\documentclass[journal,article,submit,pdftex,moreauthors]{Definitions/mdpi} 
\firstpage{1} 
\pubvolume{1}
\issuenum{1}
\articlenumber{0}
\pubyear{2025}
\copyrightyear{2025}
\datereceived{ } 

\dateaccepted{ } 
\datepublished{ } 

\hreflink{https://doi.org/} 
\Title{Symmetry Preserving Contact Interaction Approaches: An Overview of Meson and Diquark Form Factors }

\Author{Laura Xiomara Gutiérrez-Guerrero$^{1,*}$\orcidA{}, and Roger José Hernández-Pinto $^{2,*}$\orcidB{} }

\AuthorNames{L.X. Gutiérrez-Guerrero$^{1,*}$\orcidA{}, and Roger José Hernández-Pinto $^{2,*} $\orcidB{} }

\address{%
$^{1}$ \quad SECIHTI-Mesoamerican Centre for Theoretical Physics, Universidad Autónoma de Chiapas, Carretera Zapata Km. 4, Real del Bosque (Terán), Tuxtla Gutiérrez, Chiapas 29040, México; lxgutierrez@secihti.mx\\
$^{2}$ \quad Facultad de Ciencias Físico-Matemáticas, Universidad Autónoma de Sinaloa,
Ciudad Universitaria, Culiacán, Sinaloa 80000, México; roger@uas.edu.mx}

\corres{Correspondence: lxgutierrez@secihti.mx (L.X.G.-G.); roger@uas.edu.mx (R.J.H.-P.) }

\abstract{
We present an updated overview of the symmetry preserving Contact Interaction model in hadronic physics, developed a little over a decade ago to describe the mass spectrum and internal structure of mesons and diquarks composed of light and heavy quarks. Over the years, the Contact Interaction has evolved into a framework capable of treating both ground and excited states, providing a simple yet consistent approach to nonperturbative QCD. In this review, we examine the mass spectrum and elastic form factors of forty mesons with different spins and parities, together with their corresponding diquark partners. Importantly, we update the comparison of Contact Interaction  predictions using recent results from the literature, offering a fresh perspective on the model's performance, strengths, and limitations. The analysis presented here refines previous conclusions and supports the Contact Interaction as a practical tool for hadron structure studies, with potential applications to baryons and multiquark states. We also present comparisons with other theoretical models and approaches, including lattice quantum chromodynamics, and comment on future prospects in view of ongoing and planned hadron structure experimental programs.
In particular, forthcoming measurements at FAIR, together with future studies at Jefferson Lab and the Electron Ion Collider, are expected to provide key insights into hadron structure, with FAIR offering indirect constraints via hadron spectroscopy, hadronic interactions, and in-medium properties, while high-precision data on meson structure and form factors from Jefferson Lab and the Electron Ion Collider will provide valuable benchmarks to confront Contact Interaction based predictions.}


\definecolor{dark}{rgb}{0,0,0}
\definecolor{purple}{rgb}{0.5,0,0.5}
\definecolor{nblue}{rgb}{0.0,0.0,0.50}
\definecolor{scarlet}{rgb}{1.0,0.2,0}
\definecolor{darkmagenta}{rgb}{0.55, 0.0, 0.55}
\definecolor{darkolivegreen}{rgb}{0.33, 0.42, 0.18}
\definecolor{darkcandyapplered}{rgb}{0.64, 0.0, 0.0}
\definecolor{linen}{rgb}{0.98, 0.94, 0.9}
\newcommand{\tc}{\textcolor{dark}{c}}
\newcommand{\ts}{\textcolor{dark}{s}}
\newcommand{\tu}{\textcolor{dark}{u}}
\newcommand{\tb}{\textcolor{dark}{b}}

\newcommand{\td}{\textcolor{dark}{d}}
\newcommand{\be}{\begin{equation}}

\newcommand{\f}{\textcolor{dark}{f}}
\newcommand{\fu}{\textcolor{dark}{\bar{f_2}}}
\newcommand{\fd}{\textcolor{dark}{f_1}}
\newcommand{\fdu}{\textcolor{dark}{f_2}}

\newcommand{\Me}{\textcolor{dark}{V}}
\newcommand{\Meps}{\textcolor{dark}{{PS}}}

\newcommand{\Jpsi}{\textcolor{dark}{J/\Psi}}

\newcommand{\ee}{\end{equation}}
\newcommand{\bea}{\begin{eqnarray}}
\newcommand{\eea}{\end{eqnarray}}
\newcommand{\beas}{\begin{eqnarray*}}
\newcommand{\eeas}{\end{eqnarray*}}
\newcommand{\nn}{\nonumber}

\newcommand{\MeV}{\text{MeV}} 
\newcommand{\GeV}{\text{GeV}} 
\newcommand{\rmh}{\hat{\alpha}_{\mathrm {IR}}}
\newcommand{\Dps}{\textcolor{black}{{DPS}}}
\newcommand{\Mv}{\textcolor{black}{V}}
\newcommand{\Mav}{\textcolor{black}{AV}}
\newcommand{\Ms}{\textcolor{black}{S}}
\newcommand{\Ds}{\textcolor{black}{{DS}}}
\newcommand{\Dv}{\textcolor{black}{{DV}}}
\newcommand{\Dav}{\textcolor{black}{{DAV}}}

\newcommand{\eqn}[1]{Eq.~(\ref{#1})}

\newcommand{\lsim}{\mathrel{\rlap{\lower4pt\hbox{\hskip0pt $\sim$}}
\raise1pt\hbox{$<$}}}           
\newcommand{\gsim}{\mathrel{\rlap{\lower4pt\hbox{\hskip0pt$\sim$}}
\raise1pt\hbox{$>$}}}           

\begin{document}
\section{Introduction}
Since the early formulation of the quark model \cite{GellMann:1962xb,GellMann:1964nj}, understanding the structure and dynamics of hadrons, composite states of confined quarks, has remained one of the central challenges in Quantum Chromodynamics (QCD). Probing hadron structure is particularly demanding in the nonperturbative regime, where emergent phenomena such as confinement and dynamical chiral symmetry breaking (DCSB) dominate the behavior of strongly interacting matter.
In this context, meson Electromagnetic Form Factors (EFFs) play a central role, as they provide a direct window into the emergent dynamics of QCD. These observables encode detailed information about the spatial distributions of charge and current carried by quarks and establish a crucial link between theoretical descriptions and experimentally accessible quantities.\\
Calculations of meson Form Factors (FFs) are of particular importance in light of the experimental programs planned in the near future to investigate the internal structure of hadrons.
The hadron-physics program at the Facility for Antiproton and Ion Research (FAIR)~\cite{Messchendorp:2025men} is expected to play a leading role in the coming decade, particularly in the study of hadron spectroscopy, hadronic interactions, and in-medium properties of hadrons.
These efforts will be complemented by programs at the Thomas Jefferson National Accelerator Facility (Jefferson Lab), including a proposed 22~GeV energy upgrade \cite{Accardi:2023chb}, and by the U.S. Electron--Ion Collider (EIC) \cite{Aguilar:2019teb,Arrington:2021biu,Horn:2016rip,AbdulKhalek:2021gbh} and the Electron--Ion Collider in China (EicC) \cite{Anderle:2021wcy}. Additional measurements at facilities such as BESIII, LHCb, and J-PARC, spanning complementary energy regimes, will provide high precision data sets.\\
These developments strongly motivate refined theoretical analyses capable of describing meson structure in a unified and symmetry preserving manner.
The meson FFs have been investigated within a wide variety of models, including covariant, light front, and hypercentral formulations, which provide effective descriptions of electromagnetic amplitudes over broad ranges of momentum transfer $Q^{2}$ \cite{Bijker:2015gyk,Aznauryan:2015zta,Hwang:2001th}. 
In addition, light-front quark models have been extensively employed to investigate meson electromagnetic form factors, providing a relativistic framework well suited for describing hadronic structure in terms of constituent degrees of freedom (see, e.g., Refs.~\cite{Arifi:2024mff,Harjapradipta:2026kxp}).
Furthermore, extensions that combine the light-front quark model with the quark--meson coupling model offer additional insight into medium and structural effects on form factors~\cite{Arifi:2024tix}. 
Complementary to these approaches, relativistic constituent quark model formalisms have also been developed to describe hadronic transitions and decays in a fully covariant manner, enabling consistent calculations of form factors and related observables~\cite{Melikhov:1996hg}.
In parallel, lattice QCD (LQCD) offers a first principles approach to compute elastic and transition FFs from two and three point correlation functions \cite{Wang:2020nbf,Can:2012tx,Dudek:2006ej,Djukanovic:2023jag}. 
Complementary insight is provided by continuum approaches based on the coupled Schwinger-Dyson and Bethe-Salpeter equations (SDE-BSE), which enable symmetry preserving and self consistent calculations of quark propagators and interaction vertices. 
Within this framework, electromagnetic form factors are typically computed in the impulse approximation, combining Bethe--Salpeter amplitudes with dressed quark--photon vertices, thus providing a unified and covariant description of hadron structure.
These methods provide a versatile framework for the study of mesons as well as dynamical diquark correlations \cite{Miramontes:2021exi,Wilson:2011aa,Maris:2002mz,Maris:1999nt,Maris:2006ea,Maris:1998hc,Roberts:2011wy,Miramontes:2025vzb}. 
Within this picture, diquarks emerge as non pointlike dynamical correlations with internal structure encoded in their FFs, establishing them as essential building blocks for understanding baryon structure and transition mechanisms.
Alongside these dynamical formulations, effective approaches have been developed to capture the dominant nonperturbative features of QCD while avoiding the numerical complexity of full SDE-BSE calculations. 
For instance, algebraic and symmetry-based models provide simplified yet insightful descriptions of meson properties, enabling analytic control over form factors and their scaling behavior.
For example, algebraic models provide an alternative phenomenological framework in which mesons are described through symmetry based representations rather than explicit quark gluon dynamics \cite{Higuera-Angulo:2024oui,Almeida-Zamora:2023bqb,Raya:2017ggu}.
A particularly successful effective realization of continuum QCD is provided by the Contact Interaction model (CI) \cite{GutierrezGuerrero:2010md}. Originally formulated for the light quark sector, the CI can be naturally extended to heavy quarks due to the momentum independence of the quark wave function renormalization and mass function. This feature allows the quark gluon vertex to be approximated by a bare vertex while preserving both vector and axial Ward Takahashi identities. Within this framework, the CI yields a constant quark mass without a perturbative tail at large momenta. 
As the quark mass increases, the characteristic length scale of the bound state decreases, leading to more localized configurations. In this regime, hadron properties are increasingly governed by explicit quark masses and short-range dynamics, while long-range nonperturbative effects associated with dynamical chiral symmetry breaking become less dominant. Consequently, the momentum dependence of the quark mass function is significantly reduced, and approaches that approximate it as nearly constant can still provide a reliable description of heavy hadron spectra. This feature underlies the applicability of the contact interaction framework in the heavy-quark sector\cite{GutierrezGuerrero:2010md,Gutierrez-Guerrero:2019uwa,Gutierrez-Guerrero:2021rsx}.
 Beyond masses, the CI reproduces key hadronic observables such as decay constants, charge radii, and elastic FFs, demonstrating its ability to encode essential dynamical information \cite{GutierrezGuerrero:2010md,Hernandez-Pinto:2023yin,Hernandez-Pinto:2024kwg}.
The CI has been extensively applied to light and heavy mesons, including ground and excited states \cite{Paredes-Torres:2024mnz,Gutierrez-Guerrero:2024him}, as well as to diquark and baryon systems. More recently, it has also been employed to investigate medium modifications at finite temperature \cite{Wang:2013wk,Chen:2024emt}. This broad range of applications has established a solid foundation for the analysis of electromagnetic and gravitational FFs \cite{Sultan:2024hep}, positioning the CI as an efficient theoretical environment for exploring hadron structure in nonperturbative QCD.
Furthermore, the CI model has been used to compute pion twist two, twist three, and twist four generalized transverse momentum dependent parton distributions, as well as unpolarized and polarized proton parton distribution functions. These studies demonstrate that the model extends beyond static hadronic observables and is capable of describing partonic momentum correlations and transverse internal structure \cite{Zhang:2020ecj,Yu:2024qsd}.
In this review, we present a coherent synthesis of the development, theoretical foundations, and contemporary applications of CI in meson physics. We focus on the meson and diquark mass spectra in the scalar (S), pseudoscalar (PS), vector (V), and axial vector (AV) channels, as well as on the elastic form factors of S, PS, and  V mesons and their corresponding diquark partners. We outline the formulation of the model within the SDE-BSE framework, summarize its main achievements and limitations, and discuss how diquark studies encapsulate key aspects of baryon structure within the quark-diquark picture. Altogether, these elements provide an integrated perspective that situates the CI within the broader landscape of nonperturbative QCD phenomenology.


\section{Contact Interaction: Features}
\label{CI-1}
The central premise of CI approach is that the mass of a hadron is a long-wavelength, integrated observable and therefore only weakly sensitive to the fine details of its wave function. This reasoning supports the use of CI to generate realistic predictions for the ground-state spectrum of mesons and baryons, while still providing meaningful insight into selected structural features. Previous studies have confirmed the viability of this framework in SU($N_f=3$) systems and for mesons containing one or more heavy quarks, suggesting that its algebraic simplicity is well suited to uncovering systematic trends that are often obscured in numerically intensive QCD-based treatments. In this context, the CI stands out as one of the simplest yet most instructive realizations of quark dynamics within hadrons. Below, we summarize the principal features that characterize the model \cite{GutierrezGuerrero:2010md,Roberts:2010rn,Roberts:2011wy,Roberts:2011cf}.
\begin{itemize}
\item Its defining feature is the replacement of the momentum-dependent gluon propagator by one that is constant in momentum space 
\begin{eqnarray}
\label{eqn:contact_interaction}
g^{2}D_{\mu \nu}(k)&=&4\pi\hat{\alpha}_{\mathrm{IR}}\delta_{\mu \nu}\,,
\end{eqnarray}
\noindent where $\hat{\alpha}_{\mathrm{IR}}=\alpha_{\mathrm{IR}}/m_g^2$, $\alpha_{\mathrm{IR}}=0.93\pi$  can be interpreted as the interaction strength in the infrared~\cite{Binosi:2016nme,Deur:2016tte,Rodriguez-Quintero:2018wma}, and $m_g=500\,\MeV$ is a gluon mass scale generated dynamically in QCD~\cite{Boucaud:2011ug,Aguilar:2017dco,Binosi:2017rwj,Gao:2017uox}, 
an illustration of this truncation is shown in Fig.~\ref{fig:ci}.
\begin{figure}[ht]
\begin{adjustwidth}{-\extralength}{-4cm}
   \vspace{-4cm}
   \centering    \includegraphics[scale=0.4,angle=0]{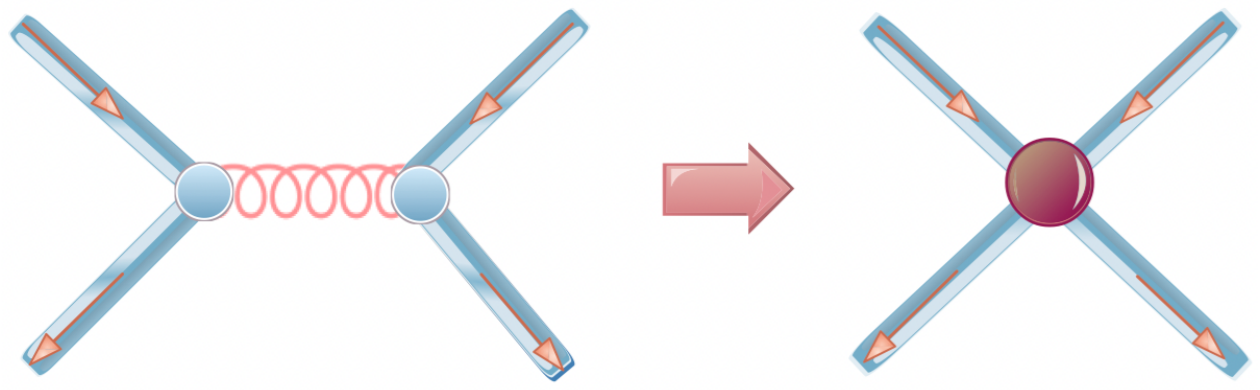}
    \end{adjustwidth}
    \vspace{-4cm}
    \caption{Diagrammatic representation of the CI, employing  the simplified interaction of Eq.~(\ref{eqn:contact_interaction}).}
    \label{fig:ci}
\end{figure}
\item The quark propagator for a quark of flavor $f$ is
\begin{eqnarray}
  && S_f^{-1}(p) =i \gamma \cdot p + m_f +  \frac{16\pi}{3}\hat{\alpha}_{\rm IR} \int\!\frac{d^4 q}{(2\pi)^4} \,
\gamma_{\mu} \, S_f(q) \, \gamma_{\mu}\,,\label{gap-1}
\end{eqnarray}
 where $m_f$ is the current-quark mass. The integral possesses quadratic and logarithmic divergences and we regularize them in a Poincar\'e covariant manner. The solution of this equation is~: 
\begin{equation}
\label{genS}
S_f^{-1}(p) = i \gamma\cdot p + M_f\,,
\end{equation}
 where $M_f$, in general, is a mass function running with a momentum scale, but within the CI it 
 is a constant dressed mass. $M_f$ is determined by
\begin{equation}
M_f = m_f + M_f\frac{4\hat{\alpha}_{\rm IR}}{3\pi}\,\,{\cal C}(M_f^2)\,,
\label{gapactual}
\end{equation}
where 
\bea
\hspace{0.75 cm}
{\cal C}(\sigma)/\sigma \equiv \overline{\cal C}(\sigma) = \Gamma(-1,\sigma \tau_{\mathrm{UV}}^2) - \Gamma(-1,\sigma \tau_{\mathrm{IR}}^2),
\eea
 with $\Gamma(\alpha,y)$ being the incomplete gamma-function and $\tau_{\rm ir, uv}$ are respectively, infrared and ultraviolet regulators. A nonzero value for  $\tau_{\mathrm{IR}}\equiv 1/\Lambda_{\mathrm{IR}}$ implements 
confinement~\cite{Roberts:2007ji}. Since the CI is a nonrenormalizable theory, 
$\tau_{\mathrm{UV}}\equiv 1/\Lambda_{\mathrm{UV}}$ becomes part of the model and therefore sets the scale for
all dimensional quantities. 
For further details on the derivation of the gap equation, see Appendix~\ref{agap}.
 %
\item For CI calculations, we use the parameters from Ref.~\cite{Gutierrez-Guerrero:2021rsx}, where a unified CI-model framework was applied to compute the masses of mesons and baryons with both light and heavy quarks. Following Ref.~\cite{Raya:2017ggu}, and guided by Refs.~\cite{Farias:2005cr,Farias:2006cs}, we introduce a dimensionless coupling $\hat{\alpha}$ defined as a function of
\begin{equation}
\hat{\alpha}(\Lambda_{\mathrm{UV}})=f(\rmh\, , \,\Lambda_{\mathrm{UV}}) , 
\label{eqn:dimensionless_alpha}
\end{equation}
which parametrizes the interaction strength in terms of the ultraviolet cutoff. In close analogy with the running of the QCD coupling with the relevant momentum scale, the dependence of $\hat{\alpha}(\Lambda_{\mathrm{UV}})$ can be described by an inverse logarithmic fit:
\bea
\label{eqn:logaritmicfit}
\hat{\alpha}(\Lambda_{\mathrm{UV}})=a\,\ln^{-1}\left(\Lambda_{\mathrm{UV}}/\Lambda_0\right),
\eea
where $a = 0.92$ and $\Lambda_0 = 0.36$ GeV~\cite{Raya:2017ggu}. Once a value of $\Lambda_{\mathrm{UV}}$ is specified, this parametrization allows for a direct determination of $\hat{\alpha}(\Lambda_{\mathrm{UV}})$, thereby providing the input needed to construct the parameter set used for the calculation of masses within the CI model,as summarized in Table~\ref{parameters}, where the first column specifies the flavor content of the constituent quarks in mesons or diquarks. The corresponding parameter set is selected according to the quark composition: light-quark systems ($u,d,s$) use the light-sector parameters (row 1); mixed heavy--light systems (e.g., $c\bar{d}$) use the corresponding mixed-sector entries; and purely heavy systems (e.g., $\eta_c$ or $\eta_b$) are described using the heavy-sector parameters (rows 3 and 6).
 \begin{table}[htbp]
 \caption{ \justifying \label{parameters} 
Parameters used for the ultraviolet regulator and the coupling strength for different quark combinations. The effective interaction is defined as $\hat{\alpha}_{\mathrm{IR}}=\hat{\alpha}_{\mathrm{IRL}}/Z_H$, with the infrared scale fixed at $\Lambda_{\rm IR} = 0.24$ GeV. 
For the light-quark sector ($u,d,s$), the choice $Z_H = 1$ corresponds to $\alpha_{\rm IR} = 0.93\pi$, which simulates the zero-momentum strength of the QCD running coupling~\cite{Chen:2012qr,Gutierrez-Guerrero:2019uwa,Gutierrez-Guerrero:2021rsx,Bedolla:2015mpa}. 
For systems involving heavy quarks, the parameters are adjusted according to Eq.~(\ref{eqn:logaritmicfit}), reflecting the change in the relevant momentum scale through a reduced effective coupling.
} 
\begin{adjustwidth}{-\extralength}{-4cm}
\begin{center}
\label{parameters1}
\begin{tabular}{@{\extracolsep{0.0 cm}} || l | c | c | c ||}
\hline \hline
 \, quarks \, &\,  $Z_{H}$ \, &\,  $\Lambda_{\mathrm {UV}}\,[\GeV] $ \,  &\,  $\hat{\alpha}_{\mathrm {IR}}$ \, [\GeV$^{-2}$]
 \\
 \hline
 \rule{0ex}{2.5ex}
$\, \tu,\td, \ts$ & 1 & 0.905 & 4.57   \\ 
\rule{0ex}{2.5ex}
$\, \tc, \td, \ts $ & \, 3.034 \, & 1.322 & 1.51 \\ 
\rule{0ex}{2.5ex}
$\, \tc $ & \, 13.122 \, & 2.305 & 0.35 \\ 
\rule{0ex}{2.5ex}
$\,  \tb,\tu$, \ts & \, 16.473 \, & 2.522 & 0.28 \\ 
\rule{0ex}{2.5ex}
$\, \tb, \tc$     &  59.056 & 4.131 & 0.08 \\
\rule{0ex}{2.5ex}
$\, \tb $ & 165.848 & 6.559 & 0.03\\
\hline \hline
\end{tabular}
\end{center}
\end{adjustwidth}
\end{table}
\item Table~\ref{table-M} presents the current quark masses $m_f$ used here and the dynamically generated dressed masses $M_f$ of $\tu$, $\ts$, $\tc$ and $\tb$ computed from the gap equation, Eq.~(\ref{gapactual}).
The similar relative contributions to the quark masses are an artifact of the momentum-independent approximation and do not indicate similar dynamical behavior between light and heavy quarks.
We assume isospin symmetry in the light-quark sector. This assumption is restricted exclusively to the light flavors and does not imply any extension to larger flavor symmetries such as SU(4) or SU(5). Heavy quarks ($c$ and $b$) are treated explicitly with their corresponding current masses, and no symmetry relations between light and heavy sectors are imposed. As a result of Eq.~(\ref{gapactual}) the generated dressed-quark masses remains constant through the whole energy range.
\begin{table}[ht]
\caption{\label{table-M}
Current ($m_{f}$) and dressed masses
($M_{f}$) for quarks in GeV, required as an input for the BSE and the EFFs.}
\begin{adjustwidth}{-\extralength}{-4cm}
 \begin{center}
\begin{tabular}{@{\extracolsep{0.0 cm}} || c | c | c | c || }
\hline 
\hline
 $m_{\tu}=0.007$ &$m_{\ts}=0.17$ & $m_{\tc}=1.08$ & $m_{\tb}=3.92$   \\
 \rule{0ex}{2.5ex}
 $ M_{\tu}=0.367$ \, & \, $  M_{\ts}=0.53$\; \, &\,   $  M_{\tc}=1.52$ \, &\,  $  M_{\tb}=4.75$   \\
 \hline
 \hline
\end{tabular}
\end{center}
\end{adjustwidth}
\end{table}
\end{itemize}
It is worth noting that, for the calculation of ground- and excited-state hadron masses at $T=0$ MeV, a fixed infrared scale of $\Lambda_{\rm IR}=0.24$ GeV is employed. This regulator effectively implements confinement by suppressing quark production thresholds in all channels. In contrast, the evaluation of screening masses requires a temperature-dependent infrared scale \cite{Chen:2024emt,Wang:2013wk}, namely
\begin{align}
\Lambda_{\rm IR} \rightarrow \Lambda_{\rm IR}(T),    
\end{align}
with the explicit form given by \cite{Chen:2024emt,Wang:2013wk,Ramirez-Garrido:2025rsu,Gutierrez-Guerrero:2026hhu}
\begin{equation}
\label{irt}
\Lambda_{\rm IR}(T) = \Lambda_{\rm IR}^{(0)}\left[\frac{M_f(T)}{M_f(0)}\right]^{1/4},
\end{equation}
where $\Lambda_{\rm IR}^{(0)} = 0.24\,\text{GeV}$ corresponds to the infrared scale fixed in $T=0$. 
In this work, we restrict our analysis to ground-state mesons with both positive and negative parity. 
Nevertheless, the extension of this framework to finite temperature remains feasible, and ongoing efforts are already addressing its application to mesons and baryons.
With these parameters specified, we now proceed to examine the resulting meson mass spectrum in the following section.
In this sense, the CI approach follows a standard strategy in hadron physics, where a minimal set of phenomenologically constrained parameters is used to generate a wide range of observables.
\section{Meson Mass Spectrum}
\label{masses-mesons}
\begin{figure}[htbp]
 \begin{adjustwidth}{-\extralength}{-4cm}
  \begin{center} 
  \hspace{5mm}
\includegraphics[scale=0.60,angle=0]{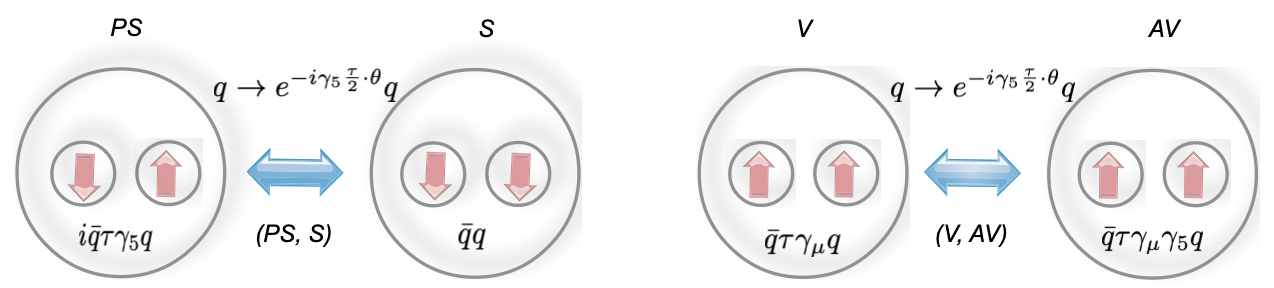}
    \end{center}
   \end{adjustwidth}
   
       \caption{\label{sca-spi}
       \justifying S mesons (such as the $\sigma$) can be interpreted as parity partners of PS mesons (e.g., the $\pi$), while V mesons (e.g., the $\rho$) form chiral partner pairs with AV mesons (e.g., the $a_1$). In this work, S states are treated exclusively as quark--antiquark systems. Consequently, S masses obtained below the $1\text{--}1.5~\mathrm{GeV}$ region correspond to hypothetical $q\bar q$ bound states. The physical nature of these low-lying S states remains an open question, owing to ongoing debates concerning their internal structure and dominant degrees of freedom~\cite{Santowsky:2020pwd,Santowsky:2021lyc,Santowsky:2021ugd}.The figure is adapted from Ref.~\cite{Hernandez-Pinto:2023yin}.}
\end{figure}

Mesons are color-singlet bound states of a quark and an antiquark. They are classified according to their total angular momentum ($J$), parity ($P$), and charge-conjugation ($C$) quantum numbers, using the standard $J^{PC}$ notation, as summarized in Table~\ref{tab:meson_classification}, which also includes the mesons analyzed in this work. We focus on PS, S, V, AV mesons composed of light and heavy quarks.
Hadrons containing top quarks are not considered here. Until recently, it was generally assumed that the top quark decays too rapidly to form bound states such as toponium, and its phenomenology was therefore largely restricted to theoretical studies~\cite{Fu:2025yft,Akbar:2024brg,Zhang:2025xxd}. More recently, however, the ATLAS and CMS collaborations have reported evidence for a state consistent with a pseudoscalar toponium interpretation~\cite{CMS:2025kzt,ATLAS:2023fsd,ATLAS:2025mvr}. In particular, ATLAS has presented measurements of $t\bar t$ production near the pair-production threshold, at an invariant mass of approximately $m_{t\bar t}\simeq 345\,\mathrm{GeV}$, in final states with two charged leptons and multiple jets. The description of such states within the CI framework remains under investigation and will be addressed in future work.\\
\begin{table}[b]
\caption{Classification of mesons according to their spin, parity, and charge-conjugation quantum numbers $J^{PC}$. For each channel, representative states are listed and organized by quark content, illustrating the PS, S, V, and AV sectors.}
\label{tab:meson_classification}
\begin{adjustwidth}{-\extralength}
{-5cm}
 \begin{center}
\begin{tabular}{@{\extracolsep{-0.1 cm}}lll}
\hline\hline
\rule{0ex}{3.5ex}
 & $J^{PC}$ & Mesons \\
\hline
\rule{0ex}{2.5ex}
PS & $0^{-+}$ &
$\pi(\tu\bar{\td}),\; K(\tu\bar{\ts}),\; h_s(\ts\bar{\ts}),\;
D^0(\tc\bar{\tu}),\; D_s^+(\tc\bar{\ts}),\;
B^+(\tu\bar{\tb}),\; B_s^0(\ts\bar{\tb}),\;
B_c^+(\tc\bar{\tb}),\; \eta_c(\tc\bar{\tc}),\; \eta_b(\tb\bar{\tb})$ \\
\rule{0ex}{3.5ex}
S & $0^{++}$ &
$\sigma(\tu\bar{\td}),\; \kappa(\tu\bar{\ts}),\; f_0(\ts\bar{\ts}),\;
D_0^\ast(\tc\bar{\tu}),\; D_{s0}^\ast(\tc\bar{\ts}),\;
B_0^\ast(\tu\bar{\tb}),\; B_{s0}^\ast(\ts\bar{\tb}),\;
B_{c0}^\ast(\tc\bar{\tb}),\; \chi_{c0}(\tc\bar{\tc}),\; \chi_{b0}(\tb\bar{\tb})$ \\
\rule{0ex}{3.5ex}
V & $1^{--}$ &
$\rho(\tu\bar{\td}),\; K^\ast(\tu\bar{\ts}),\; \phi(\ts\bar{\ts}),\;
D^{\ast0}(\tc\bar{\tu}),\; D_s^\ast(\tc\bar{\ts}),\;
B^{\ast+}(\tu\bar{\tb}),\; B_s^\ast(\ts\bar{\tb}),\;
B_c^\ast(\tc\bar{\tb}),\; J/\Psi(\tc\bar{\tc}),\; \Upsilon(\tb\bar{\tb})$ \\
\rule{0ex}{3.5ex}
AV & $1^{++}$ &
$a_1(\tu\bar{\td}),\; K_1(\tu\bar{\ts}),\; f_1(\ts\bar{\ts}),\;
D_1(\tc\bar{\tu}),\; D_{s1}(\tc\bar{\ts}),\;
B_1(\tu\bar{\tb}),\; B_{s1}(\ts\bar{\tb}),\;
B_{c1}(\tc\bar{\tb}),\; \chi_{c1}(\tc\bar{\tc}),\; \chi_{b1}(\tb\bar{\tb})$ \\
\hline\hline
\end{tabular}
\end{center}
\end{adjustwidth}
\end{table}
Within a symmetry-preserving formulation of relativistic quantum field theory, PS-S  and V-AV mesons form chiral partner pairs, as illustrated in Fig.~\ref{sca-spi}. 
 In this framework, the parity partner of a given state can be generated through a chiral rotation of the original configuration. Were chiral symmetry exact, such parity partners would be degenerate in mass. However, the observed meson and baryon spectra exhibit a clear breaking of this symmetry. In what follows, we compute the masses of mesons and their corresponding parity partners with CI, which then allows for a systematic analysis of the mass splittings between chiral partners.

\subsection{Bethe Salpeter Equation}
\label{Boundstates}
\begin{figure}[ht]
   \begin{adjustwidth}{-\extralength}{-4cm}
   \begin{center}
\includegraphics[scale=0.6,angle=0]{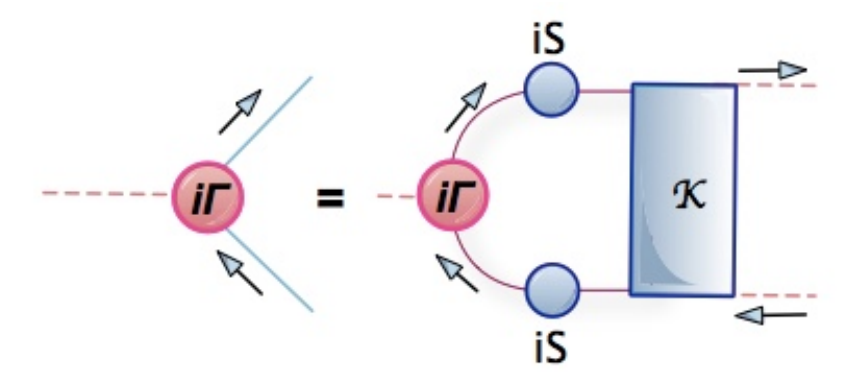}
\end{center}
\end{adjustwidth}
    \caption{\justifying Diagrammatic representation of the BSE. Blue (solid) circles represent dressed quark propagators $S$, red (solid) circle is the meson BSA $\Gamma$ while the blue (solid) rectangle is the dressed-quark-antiquark scattering kernel ${\mathcal {K}}$.}
    \label{fig:BSEfig}
\end{figure}

The BSE offers a Poincaré covariant approach to the description of bound states formed by two valence fermions, such as mesons and diquarks, within the framework of QCD. It encapsulates the dynamics of quark–antiquark systems through an integral equation that links the Bethe–Salpeter amplitude (BSA) with the dressed quark propagators and the underlying interaction kernel, as depicted in Fig.~\ref{fig:BSEfig}.
 \\
The equation takes the form~\cite{Salpeter:1951sz}
\begin{equation}
\Gamma_H(k;P)_{tu} = \int_q 
\, \chi(q;P)_{rs} \, 
{\cal K}^{rs}_{\;\;tu}(q,k;P),
\qquad
\chi(q;P) = S(q_+) \, \Gamma_H(q;P) \, S(q_-),
\label{eq:BSEnew}
\end{equation}
\begin{equation}
q_+ = q + \eta_P P,
\qquad
q_- = q - (1-\eta_P) P ,
\end{equation}
where $\Gamma_H(k;P)$ represents the Bethe--Salpeter amplitude (BSA) associated with the meson channel $H$, characterized by its flavour content.
The quark propagator matrix is written as 
$S(p) = \mathrm{diag}\big[ S_u(p), S_d(p), S_s(p) \big]$, 
and the indices $r, \ldots, u$ label colour, Dirac, and flavour components.  
The parameter $\eta_P \in [0,1]$ specifies how the total momentum is distributed between the quark and antiquark; although arbitrary, observable quantities must be insensitive to its value. 
For charge–conjugation eigenstates, one typically chooses $\eta_P = 1/2$.
In this expression, ${\cal K}^{rs}_{\;\;tu}(q,k;P)$ denotes the renormalised and fully amputated quark–antiquark scattering kernel. 
Because of the formidable complexity of ${\cal K}$, practical studies of the quark gap equation often rely on ansätze for the gluon propagator $D_{\mu\nu}(k)$ or for the vertex $\Gamma_\nu(k,p)$.
In the CI model, the momentum–independent nature of the interaction greatly simplifies the structure of the BSAs, since only a subset of the covariant components remain nonvanishing~\cite{Llewellyn-Smith:1969bcu,Maris:1999nt}.
A general decomposition of the meson BSA in this framework can be written as
\bea
\label{BSA-mesones}
\Gamma_H=A_HE_H +B_H F_H\;,
\eea
where $H={PS,V,AV,S}$ specifies the meson channel under consideration.
The explicit form of the BSA and canonical normalization for the different types of mesons and diquarks are written in Table~\ref{ff-BSE-1}. 
We adopt the notation
\bea
      \frac{d {\cal K}_{PS}(Q^2,z)}{dz}\bigg|_{Q^2=z} \equiv \frac{d {\cal K}_{PS}(z)}{dz}\bigg|_{z}  \,.
\eea
The canonical normalisation procedure ensures unit residue for the meson bound-state contribution to the quark-antiquark scattering matrix, a property of
$\Gamma_{H}^{\rm c}(P) =\frac{1}{{\cal N}} \, \Gamma_{H}(P)$.\\
With all of the above, we compute the masses and BSAs for mesons and diquarks in the next sections.
\begin{table}[h!]
\caption{\label{ff-BSE-1}
BSAs and the associated normalization conditions, $\left.{\cal N}\right|_{z=-m_H^2}=1$, for mesons (up) and diquarks (down) with CI. For a meson composed of a quark with flavor $\fd$ and an antiquark with flavor $\fu$, the reduced mass is defined as $M_R = M_{\fd} M_{\fu}/(M_{\fd} + M_{\fu})$. The explicit expressions for the kernels $K_{PS}$, $K_{S}$, $K_{V}$, and $K_{AV}$ are given in Eqs.~(\ref{bsefinalEf},\ref{Kes},\ref{KastKernel}).
}
\begin{adjustwidth}{-\extralength}{-4cm}
\centering
\begin{tabular}{@{\extracolsep{0.5 cm}}c c c c}
\hline\hline
\multicolumn{4}{c}{\textbf{Mesons}} \\
\hline
BSA & A & B & ${\cal N}$ \\
\hline
$\Gamma_{PS}$ 
& $i\gamma_5$ 
& $\dfrac{1}{2M_R}\gamma_5\gamma\!\cdot\!P$ 
& $\left.6\,\dfrac{dK_{PS}(Q^2,z)}{dz}\right|_{Q^2=z}$ 
\\[2.2ex]
$\Gamma_{V,\mu}$ 
& $\gamma^T_\mu$ 
& $\dfrac{1}{2M_R}\sigma_{\mu\nu}P_\nu$ 
& $\dfrac{9}{4\pi\hat{\alpha}_{IR}}E_V^2\dfrac{dK_V(z)}{dz}$ 
\\[2.2ex]
$\Gamma_S$ 
& $I_D$ 
& -- 
& $-\dfrac{9}{8\pi\hat{\alpha}_{IR}}E_S^2\dfrac{dK_S(z)}{dz}$ 
\\[2.2ex]
$\Gamma_{AV,\mu}$ 
& $\gamma_5\gamma^T_\mu$ 
& $\gamma_5\dfrac{1}{2M_R}\sigma_{\mu\nu}P_\nu$ 
& $-\dfrac{9}{4\pi\hat{\alpha}_{IR}}E_{AV}^2\dfrac{dK_{AV}(z)}{dz}$ 
\\[2.2ex]
\hline\hline
\multicolumn{4}{c}{\textbf{Diquarks}} \\
\hline
BSA & A & B & ${\cal N}$ \\
\hline
$\Gamma_{DS}$
& $i\gamma_5$
& $\dfrac{1}{2M_R}\gamma_5\gamma\!\cdot\!P$
& $\left.4\,\dfrac{dK_{PS}(Q,z)}{dz}\right|_{Q=z}$
\\[2.2ex]
$\Gamma_{DAV,\mu}$
& $\gamma^T_\mu$
& $\dfrac{1}{2M_R}\sigma_{\mu\nu}P_\nu$
& $\dfrac{6}{4\pi\hat{\alpha}_{IR}}E_{DAV}^2\dfrac{dK_V(z)}{dz}$
\\[2.2ex]
$\Gamma_{DPS}$
& $I_D$
& --
& $-\dfrac{3}{4\pi\hat{\alpha}_{IR}}E_{DPS}^2\dfrac{dK_S(z)}{dz}$
\\[2.2ex]
$\Gamma_{DV,\mu}$
& $\gamma_5\gamma^T_\mu$
& $\dfrac{1}{2M_R}\sigma_{\mu\nu}P_\nu$
& $-\dfrac{6}{4\pi\hat{\alpha}_{IR}}E_{DV}^2\dfrac{dK_{AV}(z)}{dz}$
\\[2.2ex]
\hline\hline
\end{tabular}
\end{adjustwidth}
\end{table}
\subsection{Pseudoscalar and Scalar  Mesons}
\begin{figure}[t!]
\begin{adjustwidth}{-\extralength}{-4cm}
\vspace{-3.29cm}
\begin{center}
\includegraphics[scale=0.45,angle=0]{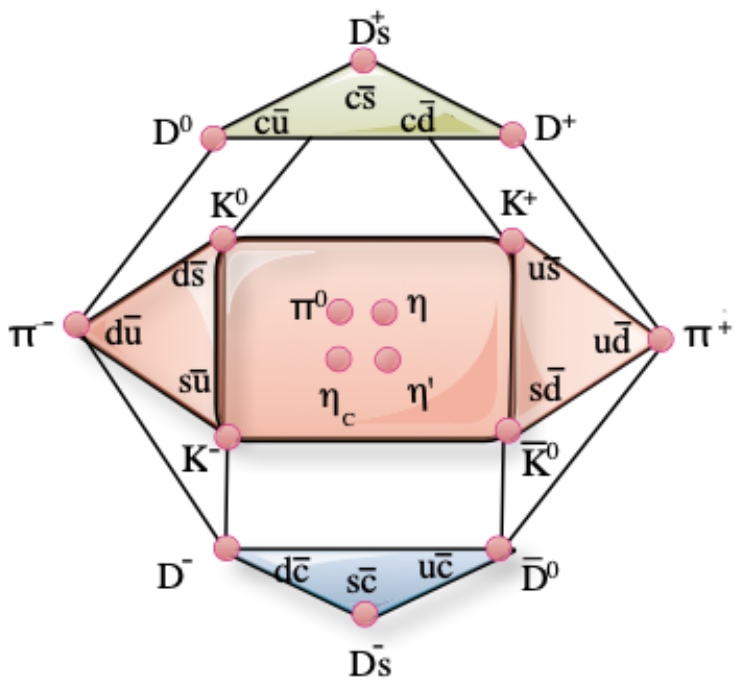}
       \vspace{-3.5cm}
       \label{meson-2}
        \end{center}
       \end{adjustwidth}
       \caption{
       \justifying The multiplet for the PS mesons made of $\tu,\;\td,\;\ts$ and $\tc$ quarks. The lightest states lie in the central orange region, while the outer vertices correspond to heavy–light mesons. The $\eta_c$ appears inside the same region but belongs to the heavy sector. 
        \vspace{0.0cm}}
       \label{meson-pse}
\end{figure}

We begin our analysis with the spin zero sector, which includes PS and S mesons. Throughout this work, we consider hadrons constructed from the five quark flavors $\tu$, $\td$, $\ts$, $\tc$, and $\tb$, corresponding to the SU(5) multiplets anticipated within the quark model~\cite{GellMann:1964nj,Zweig:1964jf,Zweig:1964ruk}.
As an illustrative example of the mesons reviewed here, Fig.~\ref{meson-pse} displays a representative set of PS mesons projected along the $\tu$, $\td$, $\ts$, and $\tc$ axes~\cite{DeRujula:1975qlm}. We note that the $\eta$ and $\eta'$ mesons are included only for completeness, since their quantitative description requires the incorporation of the axial anomaly and flavor mixing effects, which lie beyond the scope of the minimal CI framework. Nevertheless, studies of these mesons within CI can be found in Ref.~\cite{Almeida-Zamora:2023bqb}.
In Table \ref{tab:PS-S-mesons-CI} we present a consolidated overview of the studies that have investigated the masses of PS and S mesons within the CI model, encompassing systems composed of both light and heavy quarks. 
PS mesons provide a valuable window into hadronic dynamics, being the lightest quark–antiquark bound states and simultaneously emerging as the Goldstone bosons associated with dynamical chiral symmetry breaking.
As is well established, the light PS sector has been explored in far greater detail \cite{GutierrezGuerrero:2010md,Roberts:2011wy,Roberts:2011cf,Chen:2016spr,Lu:2017cln,Gutierrez-Guerrero:2019uwa,Yin:2019bxe,Gutierrez-Guerrero:2021rsx}, whereas comparatively fewer analyses have addressed heavy–light and heavy mesons. The pioneering CI studies focused primarily on the pion and the $\rho$ meson, while the extension of the model to heavy–light and heavy mesons, together with their parity partners, was carried out in Refs.~\cite{Gutierrez-Guerrero:2021rsx} and~\cite{Yin:2021uom}. This compilation illustrates the breadth of hadronic systems examined to date and provides a useful reference point for the results discussed in the sections that follow.
\begin{table}[ht]
\centering
\renewcommand{\arraystretch}{1.35}
\caption{PS $(0^-)$ and S $(0^+)$ mesons investigated within the CI , organized by flavor content. The cited references correspond to the studies in which each set of states has been analyzed.}
\label{tab:PS-S-mesons-CI}
\begin{adjustwidth}{-\extralength}{-4cm}
\begin{center}
\begin{tabular}{l l l}
\hline\hline
\multicolumn{3}{c}{\textbf{Pseudoscalar mesons $(0^-)$}} \\
\hline
 & States & Ref. \\
\hline
\rule{0ex}{3.5ex}
Light 
& $\pi(\tu\bar{\td})$, $K(\tu\bar{\ts})$, $h_s(\ts\bar{\ts})$, $\eta$, $\eta'$
& \cite{GutierrezGuerrero:2010md,Roberts:2011wy,Roberts:2011cf,Chen:2016spr,Lu:2017cln,Gutierrez-Guerrero:2019uwa,Yin:2019bxe,Gutierrez-Guerrero:2021rsx,Yin:2021uom,Almeida-Zamora:2023bqb}
\\[1.2ex]
Heavy-Light
& $D^{0}(\tc\bar{\tu})$, $D^{+}_s(\tc\bar{\ts})$, 
$B^{+}(\tu\bar{\tb})$, $B_s^{0}(\ts\bar{\tb})$
& \cite{Gutierrez-Guerrero:2019uwa,Yin:2019bxe,Yin:2021uom,Gutierrez-Guerrero:2021rsx}
\\[1.2ex]
Heavy
& $B_c^{+}(\tc\bar{\tb})$, $\eta_c(\tc\bar{\tc})$, $\eta_b(\tb\bar{\tb})$
& \cite{Bedolla:2015mpa,Raya:2017ggu,Yin:2021uom,Gutierrez-Guerrero:2021rsx}
\\
\hline\hline
\multicolumn{3}{c}{\textbf{Scalar mesons $(0^+)$}} \\
\hline
 & States & Ref. \\
\hline
\rule{0ex}{3.5ex}
Light
& $\sigma(\tu\bar{\td})$, $K_0^*(\tu\bar{\ts})$, $f_0(\ts\bar{\ts})$
& \cite{Roberts:2011cf,Chen:2016spr,Lu:2017cln,Gutierrez-Guerrero:2021rsx,Yin:2021uom}
\\[1.2ex]
Heavy-Light
& $D_0^*(\tc\bar{\tu})$, $D_{s0}^*(\tc\bar{\ts})$, 
$B_{0}^*(\tu\bar{\tb})$, $B_{s0}(\ts\bar{\tb})$
& \cite{Gutierrez-Guerrero:2021rsx,Yin:2021uom}
\\[1.2ex]
Heavy
& $B_{c0}(\tc\bar{\tb})$, $\chi_{c0}(\tc\bar{\tc})$, $\chi_{b0}(\tb\bar{\tb})$
& \cite{Gutierrez-Guerrero:2021rsx,Yin:2021uom}
\\
\hline\hline
\end{tabular}
\end{center}
\end{adjustwidth}
\end{table}

We now proceed to examine the essential aspects involved in computing the masses and BSAs of PS and S mesons. To this end, we write the explicit form of Eq.~(\ref{eq:BSEnew}) for a PS meson composed of a quark with flavour $\fd$ and an antiquark with flavour $\fu$, which reads
\begin{equation}
\label{bsefinalEf}
\left[
\begin{array}{c}
E_{\Meps}(P)\\
\rule{0ex}{3.0ex} 
F_{\Meps}(P)
\end{array}
\right]
= \frac{4 \rmh}{3\pi}
\left[
\begin{array}{cc}
{\cal K}_{EE}^{\Meps} & {\cal K}_{EF}^{\Meps} \\
\rule{0ex}{3.0ex} 
{\cal K}_{FE}^{\Meps}& {\cal K}_{FF}^{\Meps}
\end{array}\right]
\left[\begin{array}{c}
E_{\Meps}(P)\\
\rule{0ex}{3.0ex} 
F_{\Meps}(P)
\end{array}
\right],  
\end{equation}
with $P$ the total momentum of the bound state and
\begin{subequations}
\label{pionKernel}
\begin{eqnarray}
\nonumber
\nn {\cal K}_{EE}^{\Meps} &=&
\int_0^1d\alpha \bigg\{
{\cal C}(\omega)  
\bigg[ M_{\fu} M_{\fd}-\alpha (1-\alpha) P^2 - \omega\bigg]
\, \overline{\cal C}_1(\omega)\bigg\},\\
\nn {\cal K}_{EF}^{\Meps} &=& \frac{P^2}{2 M_R} \int_0^1d\alpha\, \bigg[(1-\alpha)M_{\fu}+\alpha M_{\fd}\bigg]\overline{\cal C}_1(\omega),\\
\nn{\cal K}_{FE}^{\Meps} &=& \frac{2 M_R^2}{P^2} {\cal K}_{EF}^{\Meps} ,\\
 {\cal K}_{FF}^{\Meps} &=& - \frac{1}{2} \int_0^1d\alpha\, \bigg[ M_{\fu} M_{\fd}+(1-\alpha) M_{\fu}^2+\alpha M_{\fd}^2\bigg] 
\overline{\cal C}_1(\omega)\,,
\end{eqnarray}
\end{subequations}
where $\alpha$ is a Feynman parameter and the functions $\omega\equiv\omega(M_{\fd}^2,M_{\fu}^2,\alpha,P^2)$ and ${\cal C}_1(z) $
are
\begin{eqnarray}
\label{eq:omega}
&& \hspace{-0.4cm} \omega=M_{\fu}^2 (1-\alpha) + \alpha M_{\fd}^2 + \alpha(1-\alpha) P^2\,,\\
&& \hspace{-0.4cm}  {\cal C}_1(z) \equiv z\,{\cal \overline{C}}_1(z) = - z (d/dz){\cal C}(z) = z\left[ \Gamma(0,M^2 \tau_{\mathrm {UV}}^2)-\Gamma(0,M^2 \tau_{\mathrm {IR}}^2)\right] .\rule{2em}{0ex}
\label{eq:C1}
\end{eqnarray}
The eigenvalue equation for S mesons is:
\bea
\nn 1 + {\cal K}_{S}(-m_{S}^2) = 0\,,\;\;\;\;\;
{\cal K}_{S}(P^2) = 
- \frac{4\rmh}{3\pi}
\int_0^1d\alpha\,\bigg[-{\cal L}_G
\overline{\cal C}_1(\omega^{(1)})\bigg({\cal C}(\omega^{(1)})
-{\cal C}_1(\omega^{(1)})\bigg) \bigg], \\ \label{Kes}
\end{eqnarray}
with
\bea 
{\cal L}_{G}(P^2)&=&M_{\fd} M_{\fu}+\alpha (1-\alpha)P^2.
\eea 
The equations (\ref{bsefinalEf}) and (\ref{Kes})  have a solution when $P^2=-m_{H}^2$. Then the eigenvector corresponds to the BSA of the meson.\\
It has long been known that the rainbow-ladder truncation describes V meson and flavour-nonsinglet PS meson ground-states very well but fails for their parity partners \cite{Qin:2011xq,Qin:2011dd,Maris:2006ea,Cloet:2007pi,Chen:2012qr}.
It was found that DCSB generates a large dressed-quark anomalous chromomagnetic moment and consequently the spin-orbit splitting between ground-state mesons and their parity partners is dramatically enhanced \cite{Chang:2010hb,Chang:2011ei,Chang:2010jq}. This is the mechanism responsible for a magnified splitting between parity partners; namely, there are essentially nonperturbative DCSB corrections to the rainbow-ladder kernels, which largely-cancel in the PS and V channels but add constructively in the S and AV channels.
In this context, we follow Ref.~\cite{Roberts:2011cf} and introduce spin-orbit repulsion into the S meson channels through the artifice of a phenomenological coupling $g_{SO} \leq 1$, introduced as a single, common factor multiplying the kernels defined in Eq.~ (\ref{Kes}). For S mesons we have used the value of $g_{SO}=0.32$.
The numerical results for PS and S mesons are reported in Table~\ref{par-AllFF}.
\begin{table}[ht]
\centering
\caption{PS and S meson masses and BSAs obtained within the CI model (in GeV). 
For PS states, the largest deviation with respect to Ref.~\cite{Yin:2021uom} is $9.49\%$ for the $\eta_b$ meson, 
while the difference relative to experiment remains below $3\%$. 
For S mesons, the largest deviation compared with Ref.~\cite{Yin:2021uom} is $3.29\%$ for the $D^{*}_{s0}(c\bar{s})$ state, 
and remains below $10\%$ when compared with experimental results.}
\label{par-AllFF}
\begin{adjustwidth}{-\extralength}{-4cm}
\begin{center}
\hspace{5mm}
\begin{tabular}{@{\extracolsep{0.2 cm}}l | ccccccccccc}
\hline\hline
 & $\tu\bar{\td}$ & $\tu\bar{\ts}$ & $\ts\bar{\ts}$ 
 & $\tc\bar{\tu}$ & $\tc\bar{\ts}$ 
 & $\tu\bar{\tb}$ & $\ts\bar{\tb}$ 
 & $\tc\bar{\tb}$ & $\tc\bar{\tc}$ & $\tb\bar{\tb}$ \\
\hline
\rule{0ex}{3.5ex}
$m_{PS}$ 
& 0.14 & 0.49 & 0.69 
& 1.87 & 1.96
& 5.28 & 5.37 
& 6.29 & 2.98 & 9.40 \\
$E_{PS}$
& 3.60 & 3.81 & 4.04 
& 3.03 & 3.24 
& 1.50 & 1.59 
& 0.73 & 2.16 & 0.48 \\
$F_{PS}$
& 0.47 & 0.59 & 0.74 
& 0.37 & 0.51 
& 0.09 & 0.13 
& 0.11 & 0.41 & 0.10 \\
\hline
\rule{0ex}{3.5ex}
$m_{S}$ 
& 1.22 & 1.33 & 1.45 
& 2.32 & 2.43 
& 5.50 & 5.59 
& 6.45 & 3.35 & 9.50 \\
$E_{S}$
& 0.66 & 0.65 & 0.64 
& 0.39 & 0.37 
& 0.21 & 0.20 
& 0.08 & 0.16 & 0.04 \\
\hline\hline
\end{tabular}
\end{center}
\end{adjustwidth}
\end{table}
As discussed above, a S meson can be regarded as the chiral partner of its corresponding PS state, as illustrated in Fig.~\ref{sca-spi}. In this context, our analysis of the $\pi(\tu\bar{\td})$ and $\sigma(\tu\bar{\td})$ masses reveals a mass difference of approximately $1.061$~GeV. By contrast, this mass splitting becomes significantly less pronounced for mesons composed of two heavy quarks; for instance, the $\eta_b(\tb\bar{\tb})$ and $\chi_{b0}(\tb\bar{\tb})$ states are found to have very similar masses within CI model.\\
The equal-spacing rules for mesons \cite{Okubo:1961jc,GellMann:1962xb} must also be satisfied, implying that the pseudoscalar (PS) and scalar (S) mesons obey the following relations
 \begin{align}
 \label{eq:1}\nn
    m_{D_{\ts}^{+}(\tc\overline{\ts})} - m_{D^{0}(\tc\overline{\tu})} + m_{B^{+}(\tu\overline{\tb})} - m_{B_{\ts}^{0}(\ts\overline{\tb})} &= 0 \,, \\
    m_{D_{\ts0}^{*}(\tc\overline{\ts})} - m_{D^{*}_{0}(\tc\overline{\tu})} + m_{B^{*}_{0}(\tu\overline{\tb})} - m_{B_{\ts0}(\ts\overline{\tb})} &= 0.
 \end{align}
In CI model, these equatios result in zero for PS and 0.02 for S.
The calculated masses of ground-state heavy–light mesons show a splitting relative to their chiral partners, and this difference decreases as the meson mass increases.
Although radial excitations are not included in the present analysis, it is known that their description within the CI framework requires an additional ingredient. In particular, the first radial excitation must exhibit a single node in its wave function, mirroring the behaviour of quantum–mechanical bound states. In phenomenological implementations of the CI, this feature is introduced manually by modifying the Bethe–Salpeter kernel with a factor of $(1 - q^{2} d_{F})$, which enforces a zero at $q^{2} = 1/d_{F}$, where $d_{F}$ is an extra parameter. The insertion of this node weakens the interaction within the BSE, leading to an increase in the resulting meson mass \cite{Paredes-Torres:2024mnz}.
\subsection{Vector and Axial Vector Mesons}
\begin{figure}[htbp]
\begin{adjustwidth}{-\extralength}{-4cm}
\vspace{-3.5 cm}
\begin{center}
\includegraphics[scale=0.45,angle=0]{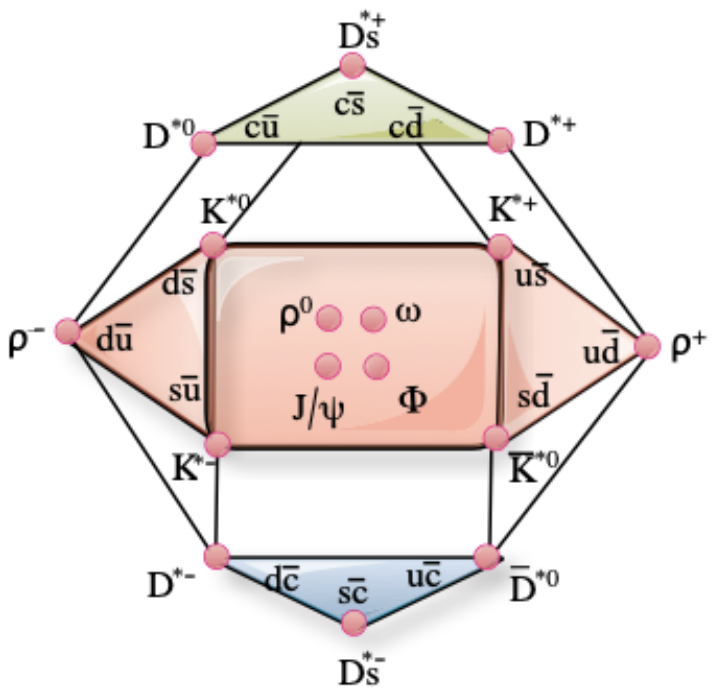}
\end{center}
\end{adjustwidth}  
       \vspace{-3.5cm}
      \caption{\label{vec-mesons}
      \justifying Structure of the V-meson multiplet composed of $\tu$, $\td$, $\ts$, and $\tc$ quarks. The diagram illustrates the quark--antiquark content of each state and highlights the organization of light, strange, and charm V mesons within the SU(4) flavor-symmetry framework.}   
\end{figure}

V and AV mesons provide important insight into the dynamical mechanisms of QCD and naturally complement the information extracted from the spin–zero sector. In the following, we present the calculated masses and BSAs for V and AV mesons in the light, heavy–light, and heavy quark sectors.
Figure~\ref{vec-mesons} displays a representative set of V mesons arranged according to their flavor content.
We further examine the characteristic trends that emerge within the CI framework, including the behavior of chiral partners and the resulting pattern of mass splittings. In parallel with CI treatment of the spin–zero sector, Table~\ref{ci-v} provides a summary of the studies that have computed V and AV meson masses using the CI.
\begin{table}[htbp]
\centering
\caption{\label{ci-v}
V $(1^-)$ and AV $(1^+)$ mesons investigated within the CI model, organized by flavor content (light, heavy--light, and heavy).}
\begin{adjustwidth}{-\extralength}{-4cm}
\begin{center}
\begin{tabular}{l | l l}
\hline\hline
 & \multicolumn{2}{c}{\textbf{Vector mesons $(1^-)$}} \\
\hline
 & States & Ref. \\
\hline
\rule{0ex}{3.5ex}
Light 
& $\rho(\tu\bar{\td})$, $K^*(\tu\bar{\ts})$, $\phi(\ts\bar{\ts})$
& \cite{Roberts:2011wy,Roberts:2011cf,Chen:2016spr,Lu:2017cln,Gutierrez-Guerrero:2019uwa,Yin:2019bxe,Yin:2021uom}
\\
\rule{0ex}{3.5ex}
Heavy--Light 
& $D^{*0}(\tc\bar{\tu})$, $D_s^{*}(\tc\bar{\ts})$, 
  $B^{*}(\tu\bar{\tb})$, $B_s^{*}(\ts\bar{\tb})$
& \cite{Gutierrez-Guerrero:2019uwa,Yin:2019bxe,Yin:2021uom}
\\
\rule{0ex}{3.5ex}
Heavy 
& $B_c^{*}(\tc\bar{\tb})$, $\Jpsi(\tc\bar{\tc})$, 
  $\Upsilon(\tb\bar{\tb})$
& \cite{Bedolla:2015mpa,Raya:2017ggu,Gutierrez-Guerrero:2019uwa,Yin:2019bxe,Yin:2021uom}
\\
\hline
\hline
 & \multicolumn{2}{c}{\textbf{Axial-vector mesons $(1^+)$} } \\
\hline
 & States & Ref. \\
\hline
\rule{0ex}{3.5ex}
Light 
& $a_1(\tu\bar{\td})$, $K_1(\tu\bar{\ts})$, $f_1(\ts\bar{\ts})$
& \cite{Roberts:2011cf,Chen:2016spr,Lu:2017cln,Yin:2021uom}
\\
\rule{0ex}{3.5ex}
Heavy--Light 
& $D_1(\tc\bar{\tu})$, $D_{s1}(\tc\bar{\ts})$, 
  $B_1(\tu\bar{\tb})$, $B_{s1}(\ts\bar{\tb})$
& \cite{Gutierrez-Guerrero:2021rsx,Yin:2021uom}
\\
\rule{0ex}{3.5ex}
Heavy 
& $B_{cb}(\tc\bar{\tb})$, $\chi_{c1}(\tc\bar{\tc})$, 
  $\chi_{b1}(\tb\bar{\tb})$
& \cite{Bedolla:2015mpa,Raya:2017ggu,Gutierrez-Guerrero:2021rsx,Yin:2021uom}
\\
\hline\hline
\end{tabular}
\end{center}
\end{adjustwidth}
\end{table}

The eigenvalue equations for V, and AV  mesons are,
\bea
\nn 0 & = & 1 - {\cal K}_{V}(-m_{V}^2)\,,\\
0 & = & 1 + {\cal K}_{AV}(-m_{AV}^2)\,,\label{eig}
\eea
where
\begin{eqnarray}\nn
{\cal K}_{\Me}(P^2)&=& \frac{2\rmh}{3\pi} \int_0^1d\alpha\,
{\cal L}_{\Me}(P^2)
\overline{\cal C}_1(\omega) \,,\\
{\cal K}_{AV}(P^2) &=&
\frac{2\rmh}{3\pi } \int_0^1d\alpha\,\bigg[
{\cal C}_1(\omega)
+ {\cal L}_{G}(P^2) \overline{\cal C}_1(\omega)\bigg] .\label{KastKernel} 
\eea 
The equations (\ref{eig} ) and (\ref{KastKernel})  have a solution when $P^2=-m_{H}^2$.  The values for the meson masses and Bethe–Salpeter amplitudes employed in this work are listed in Table \ref{tab:VAxialmasses}. These results were taken from Ref. \cite{Gutierrez-Guerrero:2019uwa}.
\begin{table}[hbt]
\caption{V and AV meson masses obtained within the CI model, expressed in GeV. The numerical values were taken from Ref.~\cite{Gutierrez-Guerrero:2019uwa} and computed using the parameter set listed in Table \ref{parameters1}. For V mesons, the largest percentage deviation between the CI results obtained with the parameters listed in Table~\ref{parameters} and those obtained using the CI parameter set of Ref.~\cite{Yin:2021uom} is is $6.32\%$, occurring for the $B^{*}_{c}(c\bar{b})$ state. 
The maximum deviation with respect to experimental values is $19.23\%$ for the $\rho(u\bar{d})$ meson. 
In the AV sector, the largest difference relative to Ref.~\cite{Yin:2021uom} appears for the $\chi_{c1}(c\bar{c})$ state, while the deviation from experiment remains below $20\%$ across all channels.}
\label{tab:VAxialmasses}
\begin{adjustwidth}{-\extralength}{-4cm}
\centering
\begin{tabular}{@{\extracolsep{0.1 cm}}c|cccccccccc}
\hline\hline
 & $\tu\bar\td$ & $\tu\bar\ts$ & $\ts\bar\ts$ & $\tc\bar\tu$ & $\tc\bar\ts$ & $\tu\bar\tb$ & $\ts\bar\tb$ & $\tc\bar\tb$ & $\tc\bar\tc$ & $\tb\bar\tb$ \\
\hline
$m_V$ 
& 0.93 
& 1.03 
& 1.12 
& 2.06 
& 2.14 
& 5.33 
& 5.41 
& 6.32 
& 3.15 
& 9.42 \\
\rule{0ex}{3.5ex}
$E_V$
& 1.53
& 1.62
& 1.73
& 1.23
& 1.32
& 0.65
& 0.67
& 0.27
& 0.61
& 0.15 \\
\hline
\rule{0ex}{3.5ex}
$m_{AV}$ 
& 1.37 
& 1.48 
& 1.58 
& 2.41 
& 2.51 
& 5.55 
& 5.64 
& 6.48 
& 3.40 
& 9.52 \\
\rule{0ex}{3.5ex}
$E_{AV}$
& 0.32
& 0.32
& 0.32
& 0.20
& 0.19
& 0.11
& 0.10
& 0.04
& 0.08
& 0.02 \\
\hline\hline
\end{tabular}
\end{adjustwidth}
\end{table}
From Tables~\ref{par-AllFF} and ~\ref{tab:VAxialmasses}, we observe that the meson spectrum exhibits a 
universal ordering pattern in all flavour channels,
$m_{PS} < m_{V} < m_{S} < m_{AV}$,
with the PS being the lightest states (Goldstone nature), followed by V, S and AV channels.
The equal spacing rules for V and AV mesons are,
\bea\label{eq:2} \nn
    m_{D_{\ts}^{*}(\tc\overline{\ts})} - m_{D^{0*}(\tc\overline{\tu})} + m_{B^{+*}(\tu\overline{\tb})} - \nn m_{B_{\ts}^{0*}(\ts\overline{\tb})} &=& 0,\\ 
    m_{D_{\ts1}(\tc\overline{\ts})} - m_{D_1(\tc\overline{\tu})} + m_{B_1(\tu\overline{\tb})} - m_{B_{\ts1}(\ts\overline{\tb})} &=& 0.
 \eea
Using the results obtained in Table~\ref{tab:VAxialmasses} we instantly infer that Eqs.~(\ref{eq:2}) are satisfied identically for V mesons while the deviation for AV mesons is less than 1\%.\\
From the calculated masses of PS, S, V and AV mesons, it follows that mesons with mixed spin and parity obey the following relations:,
\bea \label{gmo-1}
&&\hspace{-0.7cm}m_{B_c^{*}(\tc\bar{\tb})}-m_{B_s^{0*}(\ts\bar{\tb})}-m_{B_c^{+}(\tc\bar{\tb})}+m_{B_s^{0}(\ts\bar{\tb})}\approx 0\,,\\
\label{gmo-2}&& \hspace{-0.7cm}m_{B_{\ts}^{0*}(\ts\bar{\tb})}-m_{B^{+*}(\tu\bar{\tb})}-m_{B_{\ts}^0(\ts\bar{\tb})}+m_{B^{+}(\tu\bar{\tb})}= 0\,,\\
\label{gmo-3}&&\hspace{-0.7cm}m_{B_{\ts}^{0*}(\ts\bar{\tb})}-m_{B^{+*}(\tu\bar{\tb})}-m_{D_{\ts}^{+}(\tc\bar{\ts})}+m_{D^{0}(\tc\bar{\tu})}=0\,, \\
\label{gmo-4}&&\hspace{-0.7cm}m_{\eta_{\tb}(\tb\bar{\tb})}-m_{\eta_{\tc}(\tc\bar{\tc})}-2m_{B_{\ts}^{0*}(\ts\bar{\tb})}+2m_{D_{\ts}^{*}(\tc\bar{\ts})}\approx 0\,,\\
\label{gmo-5}&&\hspace{-0.7cm}m_{\eta_{\tb}(\tb\bar{\tb})}-m_{\eta_{\tc}(\tc\bar{\tc})}-2m_{B_{\ts}^0(\ts\bar{\tb})}+2m_{D_{\ts}^{+}(\tc\bar{\ts})}=0\,,\\
\label{gmo-6}&&\hspace{-0.7cm}m_{B_{\ts}^{0*}(\ts\bar{\tb})}-m_{D_{\ts}^{*}(\tc\bar{\ts})}-m_{B_{\ts}^0(\ts\bar{\tb})}+m_{D_{\ts}^{+}(\tc\bar{\ts})}=0\,,\\
\label{gmo-7}&&\hspace{-0.7cm}m_{\Upsilon(\tb\bar{\tb}) }-m_{\Jpsi(\tc\bar{\tc})}-2m_{B_{\ts}^0(\ts\bar{\tb})}+2m_{D_{\ts}^{+}(\tc\bar{\ts})}=0\,,\\
\label{gmo-8}&&\hspace{-0.7cm}m_{\Upsilon(\tb\bar{\tb}) }-m_{\Jpsi(\tc\bar{\tc})}-m_{\eta_{\tb}(\tb\bar{\tb})}+m_{\eta_{\tc}(\tc\bar{\tc})}\approx 0\,,\\
\label{gmo-9}&&\hspace{-0.7cm}m_{\Upsilon(\tb\bar{\tb}) }-m_{\Jpsi(\tc\bar{\tc})}-2m_{B_{\ts}^{0*}(\ts\bar{\tb})}+2m_{D_{\ts}^{*}(\tc\bar{\ts})}\approx 0\,.
\eea
We test the mass relations, Eqs.~(\ref{gmo-1}-\ref{gmo-9}), using both experimental data and the CI framework. 
For positive-parity states, the maximum deviation is $-0.55$ in CI and $-0.44$ experimentally, whereas for negative parity the largest deviations are $0.17$ in CI and $-0.36$ from experiment (see Ref.~\cite{Gutierrez-Guerrero:2021rsx} for details). 
Overall, the relations are well satisfied within the CI approach.
\subsection{Diquarks}
The idea of diquarks as correlated systems of two quarks was introduced in the early quark model~\cite{GellMann:1964nj}. 
A diquark may be understood as a quark–quark correlation inside a hadron, treated effectively as a compact object carrying the quantum numbers of two quarks. 
Although such an approximation is idealized, it has proven successful in reproducing a variety of hadronic properties, qualitatively and in many cases quantitatively.
Diquarks are particularly relevant in the description of multiquark states. 
Configurations such as tetraquarks and pentaquarks are commonly modelled as diquark–antidiquark bound systems~\cite{Maiani:2004vq,Shi:2021jyr,Jaffe:2003sg,MoosaviNejad:2020hhd,Liu:2019zoy}, and even baryons may be viewed as quark–diquark composites. 
Ref.~\cite{Ida:1966ev} formalized this perspective by describing baryons as bound states of a quark and an effective diquark, motivating the wide use of diquarks as phenomenological degrees of freedom in studies of nonperturbative QCD.
From the colour structure viewpoint, diquarks denoted as $[qq]$ for spin 0 and $\{qq\}$ for spin 1 arise from the combination of two quarks in the fundamental representation of $SU(3)$,
\begin{equation}
3 \otimes 3 = \bar{3}\oplus 6,
\label{eq:color-decomposition}
\end{equation}
so that only the antitriplet $\bar{3}_c$ can combine with a third quark to form a colour singlet. 
One-gluon exchange is attractive in the $\bar{3}_c$ channel and repulsive in the sextet $6_c$, making the antitriplet the physically relevant configuration.
The formalism used for mesons extends directly to diquarks, allowing computation of their BSA and masses. 
In the CI approach, the BSE for mesons and diquarks share the same algebraic structure, differing only by colour factors. 
Under ladder truncation, the colour factor for mesons is $4/3$, whereas in the $\bar{\mathbf{3}}$ diquark channel it is $2/3$, reducing the interaction strength by one half relative to the meson case.
A precise characterization of diquark masses and internal structure therefore provides a solid foundation for exploring multiquark systems, and the results presented here represent a necessary step toward a systematic description of such states within the CI framework.
Thus, mesons possess well-defined diquark partners. In particular, each mesonic channel is associated with a corresponding diquark channel, as summarized in Table~\ref{diqpart}.
\begin{table}[t!]
\caption{Correspondence between mesonic channels and their diquark partners. 
Each meson is characterized by its spin--parity $J^P$, and the associated diquark appears in the channel with opposite parity, reflecting the fermion--antifermion nature of mesons.}
\label{diqpart}
\begin{adjustwidth}{-\extralength}{-4cm}
    \begin{center}
    \begin{tabular}{@{\extracolsep{0.5 cm}}l c c l c}
    \hline \hline
    Meson & $J^P$ & & Diquark Partner & $J^P$ \\
    \hline
     \rule{0ex}{2.5ex}
    S meson        & $0^+$ & $\to$ &PS diquark & $0^-$ \\
     \rule{0ex}{2.5ex}
   PS meson & $0^-$ & $\to$ & S diquark       & $0^+$ \\ 
    \rule{0ex}{2.5ex}
    V meson       & $1^-$ & $\to$ & AV diquark & $1^+$ \\
     \rule{0ex}{2.5ex}
    AV meson & $1^+$ & $\to$ & V diquark       & $1^-$ \\
    \hline \hline
    \end{tabular}
    \end{center}
\end{adjustwidth}
\end{table}
This correspondence implies that the mass and BSA of a diquark with spin parity $J^P$ can be obtained from the equation describing a meson with opposite parity $J^{-P}$, with the only modification being a reduction of the interaction strength by a factor of one half. The reversal of parity arises from the fact that fermions and antifermions carry opposite intrinsic parity.\\
In the notation adopted for diquarks, $H = DS, DAV, DPS, DV$ denote S, AV, PS, and V diquarks, respectively. The diquark BSA has the same functional form as that given in Eq.~(\ref{BSA-mesones}). The explicit coefficients and the corresponding normalization conditions are listed in Table~\ref{ff-BSE-1}.\\
The color factor differs between mesons and diquarks because diquarks belong to the color antitriplet representation rather than being color singlets. Consequently, the canonical normalization condition for diquarks and mesons is formally identical, except for the replacement ${\cal N}_C = 3 \to 2$.
Accordingly, the eigenvalue equation for DS diquarks reads
\begin{equation}
\label{bsefinalE}
\left[
\begin{array}{c}
E_{\Ds}(P) \\
\rule{0ex}{3.0ex} 
F_{\Ds}(P)
\end{array}
\right]
= \frac{4 \rmh}{6\pi}
\left[
\begin{array}{cc}
{\cal K}_{EE}^{\Meps} & {\cal K}_{EF}^{\Meps} \\
\rule{0ex}{3.0ex} 
{\cal K}_{FE}^{\Meps}& {\cal K}_{FF}^{\Meps}
\end{array}\right]
\left[\begin{array}{c}
E_{\Ds}(P)\\
\rule{0ex}{3.0ex} 
F_{\Ds}(P)
\end{array}
\right].
\end{equation}
The equations that will give us the masses of the AV, V and PS diquarks are
\bea
\nn 0 & = & 1 + \frac{1}{2}{\cal K}_{\Mv}(-m_{\Dav}^2),\\
\nn 0 & = & 1 + \frac{1}{2}{\cal K}_{\Mav}(-m_{\Dv}^2),\\
0 & = & 1 + \frac{1}{2}{\cal K}_{\Ms}(-m_{\Dps}^2).
\label{dq}\eea
In this truncation, the diquark masses correspond to $P^2=-m_H^2$. 
We therefore present results for the masses of diquark correlations in Table~\ref{par-AllDiquarks}.
\begin{table}[htbp]
\centering
\caption{\label{par-AllDiquarks} Calculated BSA amplitudes and masses for all diquark channels within the CI model.
As in the meson sector, the values $g_{SO}=0.32$ for PS diquarks and $g_{SO}=0.25$ for V diquarks are typically employed. 
The maximum percentage deviation between the CI results obtained with the parameters listed in Table \ref{parameters} and those reported in Ref.~\cite{Yin:2021uom}, within the same model, is $4.15\%$ for S diquarks, $11.54\%$ for PS diquarks, $7.64\%$ for V diquarks, and $4\%$ for AV diquarks.
}
\begin{adjustwidth}{-\extralength}{-4cm}
\begin{center}
\begin{tabular}{@{\extracolsep{0.1cm}}l ccc|cc|cc|cc}
\toprule
 & \multicolumn{3}{c}{Scalar} & \multicolumn{2}{c}{Pseudoscalar}& \multicolumn{2}{c}{Vector} & \multicolumn{2}{c}{Axial-vector} \\
 \hline
 \rule{0ex}{2.5ex}
  & Mass & $E_{\Ms}$   &  $F_{\Ms}$  & Mass & $E_{\Meps}$  & Mass & $E_{\Me}$ & Mass & $E_{\Mav}$ \\
 \rule{0ex}{2.5ex}
$\tu\td$        & 0.77          & 2.74    & 0.31    & 1.30         &0.54     & 1.44    & 0.28  & 1.06 & 1.30  \\
\rule{0ex}{2.5ex}  
$\tu\ts$        & 0.92          & 2.88    & 0.39    &1.41          &0.54     & 1.54    & 0.28  & 1.16  & 1.36 \\
\rule{0ex}{2.5ex}
$\ts\ts$        & 1.06          & 3.03    & 0.50    &1.52           &0.53   & 1.64    & 0.27  & 1.25  & 1.42 \\
\rule{0ex}{2.5ex}
$\tc\tu$        & 2.08          & 2.00    & 0.23     &2.37          &0.32   & 2.45    & 0.17  & 2.16  & 0.93 \\
\rule{0ex}{2.5ex}
$\tc\ts$        & 2.17          & 2.11    & 0.32     &2.47          &0.31   & 2.54    & 0.16  & 2.25  & 0.95 \\
\rule{0ex}{2.5ex}
$\tu\tb$        & 5.37          & 0.99    & 0.06     &5.53          &0.18   & 5.59    & 0.09  & 5.39  & 0.48 \\
\rule{0ex}{2.5ex}
$\ts\tb$        & 5.46          & 1.00    & 0.08     &5.62          &0.14   & 5.67    & 0.05  & 5.47  & 0.48 \\
\rule{0ex}{2.5ex}
$\tc\tb$        & 6.35          & 0.42    & 0.07     &6.47          &0.07   & 6.50    & 0.04  & 6.35  & 0.20 \\
\rule{0ex}{2.5ex}
$\tc\tc$        & 3.17          & 0.96    & 0.19    & 3.38           &0.14   & 3.42    & 0.07  & 3.22  & 0.41 \\
\rule{0ex}{2.5ex}
$\tb\tb$        & 9.43          & 0.23    & 0.05    & 9.51           &0.04   & 9.53    & 0.02  & 9.44  & 0.11 \\
\hline
\hline
\end{tabular}
    \end{center}
	\end{adjustwidth}
\end{table}\\
As in the case of their meson partners, we multiply the PS and V diquark kernels by the 
spin–orbit factor $g_{SO}$, taking $g_{SO}=0.32$ for spin-zero and $g_{SO}=0.25$ for spin-one diquarks. 
In applications where these diquarks are employed to construct negative-parity baryons, this factor is further 
multiplied by $1.8$, following Ref.~\cite{Lu:2017cln}. This modification reduces the effective repulsion in the 
corresponding channels. A plausible physical interpretation is that the valence quarks inside a diquark are 
less tightly correlated than the quark–antiquark pair in a meson; hence, spin–orbit repulsion is naturally 
expected to be weaker in diquarks than in their meson partners.
With the diquark masses and amplitudes determined here, one can construct the Faddeev kernels associated with 
ground-state octet and decuplet baryons, as well as their chiral partners; see Ref.~\cite{Gutierrez-Guerrero:2021rsx}.
From Table~\ref{par-AllDiquarks}, we observe that the mass hierarchy
$m_{DS} < m_{DAV} < m_{DPS} < m_{DV}$
is satisfied in all flavour channels. The only exception is the $\tc\tb$ sector, 
where the S and AV states are nearly degenerate, while 
the overall ordering remains preserved.
These results show that diquark masses are comparable to those of mesons, but are generally larger \cite{Maris:2004bp}, supporting their interpretation as compact correlations that are not pointlike and that play a central role in quark-diquark descriptions of baryons.
\begin{equation}
m_{PS} < m_{DS},\;\;\;\;\;\;
m_{S} < m_{DPS}, \;\;\;\;\;\;
m_{AV} < m_{DV},\;\;\;\;\;\;
m_{V} < m_{DAV}.
\end{equation}
This coherent 
hierarchy provides internal consistency to the spectrum generated within the CI 
framework and establishes a solid foundation for future studies of hadronic 
structure and FFs.
\section{Elastic Form Factors}
\label{FFS-1}
The availability of high intensity and high precision electron beams at CEBAF has played a central role in advancing our understanding of hadron structure. The completion of the 12~GeV upgrade and the initiation of the corresponding physics program have opened a new stage in experimental studies of meson and baryon EFFs, allowing access to larger momentum transfers and, consequently, to a more refined resolution of nonperturbative QCD dynamics. The broad set of approved experiments planned for the coming years, together with the prospect of a future upgrade to 22~GeV \cite{Accardi:2023chb}, will significantly extend the accessible $Q^{2}$ domain and deliver high precision data that are essential for testing both effective descriptions and approaches directly rooted in QCD.
Within this experimental landscape, theoretical investigations of meson FFs are particularly timely, as these observables provide a direct connection between measured cross sections and the underlying mechanisms of confinement and dynamical chiral symmetry breaking that shape hadronic structure. FFs constitute fundamental probes of hadrons, encoding detailed information on the spatial distributions of charge, current, and momentum inside bound states, and thereby serving as a crucial interface between theory and experiment.
From the theoretical perspective, FFs also play a pivotal role within effective QCD inspired approaches. In particular, frameworks based on the SDE-BSE, and especially their realization within the CI, use EFFs as stringent tests of internal consistency and of the ability of the model to capture essential aspects of hadronic dynamics beyond static properties.
In this section, we present a coherent and comprehensive discussion of elastic EFFs within the CI framework. The mesonic FF results are taken from Refs.~\cite{Hernandez-Pinto:2023yin,Hernandez-Pinto:2024kwg}; however, we extend those analyses by providing detailed comparisons with more recent calculations obtained using alternative theoretical approaches. In this way, we offer an updated perspective on the performance and limitations of CI based predictions for meson structure.
In addition, we present results for elastic diquark FFs. These results are of particular relevance, as they provide the essential dynamical input required for subsequent calculations of elastic and transition FFs of baryons within the same framework. We begin by outlining the general considerations relevant to electromagnetic interactions of mesons, which establish the basis for the analyses presented in the following subsections.
\\\\
\noindent
\underline{Parameters}\\\\
\noindent
As in the calculation of meson and diquark masses, the parameters used in the computation of FFs are fixed according to Eq.~(\ref{eqn:logaritmicfit}), with distinct values assigned to each quark–flavor sector, following the parameter sets employed in Refs.~\cite{Hernandez-Pinto:2023yin,Hernandez-Pinto:2024kwg}. We summarize the parameters adopted for each channel in Table~\ref{parFFn}.
\begin{table}[ht]
\centering
\caption{\label{parFFn}Parameters organized by quark content for the calculation of FFs. For each quantity, the values employed in the S and PS channels, as well as those used in the V-meson sector, are listed. The parameter sets are taken from Refs.~\cite{Hernandez-Pinto:2023yin,Hernandez-Pinto:2024kwg}. It is worth noting that the adoption of the new parameter set ensures that the resulting charge radii follow the expected mass hierarchy, a desirable feature consistently satisfied across all meson channels.}
\label{tab:UV_parameters_transposed_multirow}
\small
\renewcommand{\arraystretch}{1.25}
\begin{adjustwidth}{-\extralength}{-4cm}
\begin{center}
\hspace{5mm}
\begin{tabular}{@{\extracolsep{0.1cm}}llccccccccc}
\hline\hline
 &  & $\tu,\td,\ts$ & $\ts$ & $\tc,\tu$ & $\tc,\ts$ & $\tc$ & $\tb,\tu$ & $\tb,\ts$ & $\tb,\tc$ & $\tb$ \\
\rule{0ex}{2.5ex}
\multirow{2}{*}{$Z_H$}
& S and PS & 1.00 & 3.03 & 1.32 & 1.50 & 13.12 & 11.27 & 17.54 & 30.54 & 129.51 \\
& V & 1.00 & 1.00 & 0.59 & 0.91 & 1.40 & 13.69 & 0.59 & 30.97 & 1.43 \\
\rule{0ex}{2.5ex}
\multirow{2}{*}{$\Lambda_{\rm UV}$}
& S and PS & 0.91 & 3.03 & 1.32 & 1.50 & 2.31 & 3.22 & 3.57 & 3.89 & 7.16 \\
& V  & 1.22 & 1.58 & 2.79 & 3.90 & 7.27 & 9.38 & 11.69 & 12.61 & 13.88 \\
\rule{0ex}{2.5ex}
\multirow{2}{*}{$\hat{\alpha}_{\rm IR}$}
& S and PS & 4.57 & 3.03 & 1.32 & 1.50 & 0.35 & 0.41 & 0.26 & 0.15 & 0.04 \\
& V  & 4.57 & 4.57 & 7.74 & 5.01 & 3.26 & 0.33 & 7.72 & 0.15 & 3.20 \\
\hline\hline
\end{tabular}
\end{center}
\end{adjustwidth}
\end{table}
With these parameters, the masses and BSAs of the four meson and diquark channels are modified as listed in Table~\ref{tab:unified_mesons_diquarks}. 
It is important to stress that the parameter set employed in the present framework is not arbitrarily redefined when moving from spectrum calculations to form factors. The interaction strength and regularization scales are primarily fixed by reproducing key spectroscopic observables.\\
Once this baseline is established, most quantities—including form factors and charge radii—emerge as predictions of the model.
This procedure should therefore be understood as a controlled refinement of the description, rather than an independent parameter tuning. To support this, we note that the masses obtained with the refined parameter set remain close to those of the original spectrum, with deviations below $8\%$ (see Table~\ref{tab:unified_mesons_diquarks}). This indicates that the overall spectroscopic content of the model is preserved, and that the level of flexibility remains limited.
In this sense, the framework is consistent with the QCD paradigm, where a small number of inputs constrains a wide range of hadronic observables.
For the spin–orbit coupling, $g_{SO}$, we use the values $0.32$ for S mesons and $0.25$ for AV mesons. These values are employed in the calculation of their elastic FFs.
\begin{table}[ht]
\caption{Masses and BSAs for mesons and diquarks, organized by quark content, are obtained using the parameter set listed in Table~\ref{tab:UV_parameters_transposed_multirow}. 
When meson results are compared with those reported in Tables~\ref{par-AllFF} and \ref{tab:VAxialmasses}, we find that the maximum relative difference in the PS channel is approximately $4\%$, occurring in the heavy--light $ub$ sector. 
For V mesons, the largest deviation reaches about $8\%$, again in the $ub$ channel, while in the S sector the maximum relative difference is of order $4\%$. 
Nevertheless, the use of the updated parameter set allows us to obtain charge radii that are in improved agreement with those reported in other theoretical approaches.
}
\label{tab:unified_mesons_diquarks}
\begin{adjustwidth}{-\extralength}{-4cm}
\begin{center}
\begin{tabular}{@{\extracolsep{0.1cm}}lccccccc}
\hline\hline
 & \multicolumn{7}{c}{Mesons} \\
 & $m_{\rm PS}$ & $E_{\rm PS}$ & $F_{\rm PS}$
 & $m_{\rm S}$ & $E_{\rm S}$
 & $m_{\rm V}$ & $E_{\rm V}$ \\
\hline\hline
\rule{0ex}{2.5ex}
$\tu\bar{\td}$
& 0.139 & 3.59 & 0.47
& 1.22 & 0.66
& 0.93 & 1.53 \\
\rule{0ex}{2.5ex}
$\tu\bar{\ts}$
& 0.499 & 3.81 & 0.59
& 1.38 & 0.65
& 1.03 & 1.63 \\
\rule{0ex}{2.5ex}
$\ts\bar{\ts}$
& 0.701 & 4.04 & 0.75
& 1.46 & 0.64
& 1.13 & 1.74 \\
\rule{0ex}{2.5ex}
$\tc\bar{\tu}$
& 1.855 & 3.03 & 0.37
& 2.31 & 0.39
& 2.05 & 1.23 \\
\rule{0ex}{2.5ex}
$\tc\bar{\ts}$
& 1.945 & 3.24 & 0.51
& 2.42 & 0.42
& 2.30 & 0.55 \\
\rule{0ex}{2.5ex}
$\tu\bar{\tb}$
& 5.082 & 3.72 & 0.21
& 5.30 & 1.53
& 4.93 & 2.73 \\
\rule{0ex}{2.5ex}
$\ts\bar{\tb}$
& 5.281 & 2.85 & 0.21
& 5.64 & 0.26
& 5.39 & 1.17 \\
\rule{0ex}{2.5ex}
$\tc\bar{\tb}$
& 6.138 & 2.58 & 0.39
& 6.36 & 1.23
& 6.25 & 1.06 \\
\rule{0ex}{2.5ex}
$\tc\bar{\tc}$
& 2.952 & 2.15 & 0.40
& 3.33 & 0.16
& 3.15 & 0.51 \\
\rule{0ex}{2.5ex}
$\tb\bar{\tb}$
& 9.280 & 2.04 & 0.39
& 9.57 & 0.69
& 9.51 & 0.48 \\
\hline\hline
\multicolumn{8}{c}{}\\[-1.2ex]
\hline\hline
 & \multicolumn{7}{c}{Diquarks} \\
 & $m_{\rm SD}$ & $E_{\rm SD}$ & $F_{\rm SD}$
 & $m_{\rm PSD}$ & $E_{\rm PSD}$
 & $m_{\rm AVD}$ & $E_{\rm AVD}$ \\
\hline\hline
\rule{0ex}{2.5ex}
$\tu\bar{\td}$
& 0.775 & 2.74 & 0.31
& 1.30 & 0.54
& 1.05 & 1.30 \\
\rule{0ex}{2.5ex}
$\tu\bar{\ts}$
& 0.927 & 2.88 & 0.39
& 1.41 & 0.53
& 1.15 & 1.36 \\
\rule{0ex}{2.5ex}
$\ts\bar{\ts}$
& 1.065 & 3.03 & 0.50
& 1.52 & 0.53
& 1.25 & 1.42 \\
\rule{0ex}{2.5ex}
$\tc\bar{\tu}$
& 2.063 & 2.00 & 0.22
& 2.35 & 0.32
& 2.14 & 0.92 \\
\rule{0ex}{2.5ex}
$\tc\bar{\ts}$
& 2.163 & 2.10 & 0.31
& 2.45 & 0.32
& 2.24 & 0.95 \\
\rule{0ex}{2.5ex}
$\tu\bar{\tb}$
& 5.364 & 1.64 & 0.09
& 5.56 & 0.24
& 5.45 & 0.48 \\
\rule{0ex}{2.5ex}
$\ts\bar{\tb}$
& 5.503 & 1.16 & 0.08
& 5.69 & 0.15
& 5.54 & 0.48 \\
\rule{0ex}{2.5ex}
$\tc\bar{\tb}$
& 6.361 & 0.78 & 0.12
& 6.49 & 0.13
& 6.38 & 0.32 \\
\rule{0ex}{2.5ex}
$\tc\bar{\tc}$
& 3.152 & 0.96 & 0.19
& 3.35 & 0.13
& 3.19 & 0.41 \\
\rule{0ex}{2.5ex}
$\tb\bar{\tb}$
& 9.558 & 0.31 & 0.06
& 9.63 & 0.06
& 9.56 & 0.12 \\
\hline\hline
\end{tabular}
\end{center}
\end{adjustwidth}
\end{table}
\\\\
\underline{Triangle Diagram}
\\\\
Considering a meson $M$, within the impulse approximation, the $M\gamma M$ vertex encodes the electromagnetic interaction between a meson composed of a $\fd\fu$  quark–antiquark pair and an external photon. In this framework, the coupling of the photon to the meson is described through the interaction with its constituent quarks, leading to the following expression for the $M\gamma M$ vertex:
\begin{align}
\Lambda^{M,\fd}
&= N_c \int \frac{d^{4}\ell}{(2\pi)^{4}}\,
{\rm Tr}\;\,
\Big[
 i\Gamma_{M}(k_{f})\, S(\ell+k_{i},M_{\fd})\, i\Gamma_{\lambda}(Q,M_{\fd})
\nonumber\\
&\qquad\qquad\qquad\qquad\;\times
 S(\ell+k_{f},M_{\fd})\, i\bar{\Gamma}_{M}(-k_{i})\, S(\ell,M_{\fu})
\Big] \, .
\label{General-FF}
\end{align}
The notation assumes that it is the quark $\fd$ which interacts with the photon while the antiquark $\fu$ remains a spectator. We  define 
$\Lambda^{M,\fu}$ similarly. We stress that Eq.~(\ref{General-FF}) provides a general representation of the current. The explicit form of $\Lambda^{M,\fd}$, including the identification of the Lorentz index, is fixed in each meson channel according to its quantum numbers. This is made explicit in the corresponding projections presented for pseudoscalar, scalar, vector, and axial-vector mesons in the following sections.
 Furthermore, we denote the incoming photon momentum by $Q$ while the incoming and outgoing momenta of $M$ by:
$k_{i}=k-Q/2$ and $k_f=k+Q/2$, respectively.
The assignments of momenta are shown in the triangle diagram of Fig.~\ref{vertex-1}.
\begin{figure}[htbp]
\begin{adjustwidth}{-\extralength}{-4cm}
\begin{center}
\includegraphics[scale=0.25,angle=0]{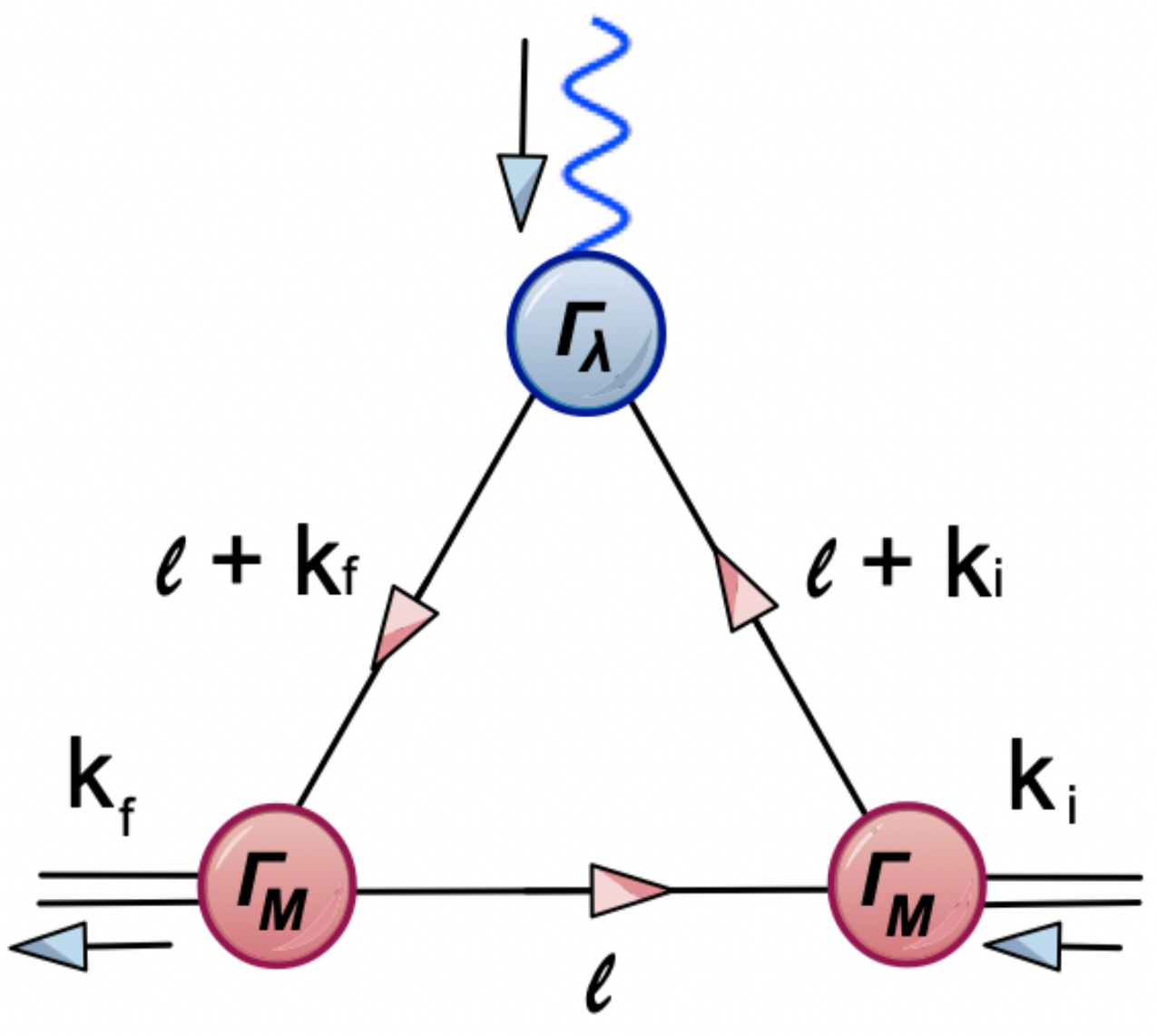}
\end{center}
    \end{adjustwidth}
    \caption{\label{vertex-1} \justifying Triangle diagram representing the impulse approximation to the $M\gamma M$ vertex.The red circles denote the Bethe--Salpeter amplitudes associated with the incoming and outgoing mesons, which encode the internal quark--antiquark correlations.
The blue circle corresponds to the dressed quark--photon vertex, describing the interaction between
the photon and the quark.
The arrows indicate the momentum flow through the diagram.}
\end{figure}\\
$\Lambda^{M,\f}$ corresponds to the EFFs of different mesons under study. The contribution from the interaction of the photon with quark $\fd$ can be represented as $F^{M,\fd} (Q^2)$ while the contribution arising from its interaction with quark $\fu$ can be represented as $F^{M,\fu} (Q^2)$. The total form factor $F^M (Q^2)$ is defined as follows~\cite{Hutauruk:2016sug}:
\begin{equation}\label{eqn:TotalMesonFF}
F^M (Q^2) = e_{\fd} F^{M,\fd} (Q^2) + e_{\fu} F^{M,\fu} (Q^2)\,,
\end{equation}
where $e_{\fd}$ and $e_{\fu}$ are the quark and the antiquark electric charges, respectively. For neutral mesons composed of same flavored quarks, the total EFF is simply $F^{M} = F^{M,\fd}$. 
With the information gathered in the previous sections, we are now able to compute the EFFs of PS, S and V mesons. In the following section, we summarize the results presented in Refs.~ \cite{Hernandez-Pinto:2023yin,Hernandez-Pinto:2024kwg}, where these EFFs were calculated for light, heavy-light, and heavy mesons.
\\\\
\underline{The Quark-Photon Vertex}
\\\\
A central ingredient in the calculation of EFFs is the quark-photon vertex \cite{Frank:1994mf},
since it governs the coupling between the electromagnetic probe and the internal quark degrees of
freedom. This vertex can be expressed in a Lorentz-Dirac tensor basis and naturally separated into two
contributions: a transverse part and a Ball-Chiu component \cite{Ball:1980ay,Miramontes:2025ofw}. The transverse sector is built from tensor
structures orthogonal to the photon momentum and encodes dynamical information relevant for bound-state
properties, while the Ball-Chiu contribution contains non-transverse terms that are uniquely fixed by
the quark propagator through the electromagnetic Ward-Takahashi identity. This construction guarantees
electromagnetic gauge invariance and provides a consistent framework for the computation of meson form
factors.
The analysis of meson observables is physically meaningful only when the vector and axial Ward--Takahashi identities (WTIs) are explicitly preserved. These relations take the form
\begin{align}
i P_{\mu}\,\Gamma_{\mu}^{\gamma}(k_{+},k_{-},M_{f}) &= S^{-1}(k_{+},M_{f}) - S^{-1}(k_{-},M_{f}), \label{eq:VWTI} \\
P_{\mu}\,\Gamma_{5\mu}(k_{+},k) &= S^{-1}(k_{+})\,i\gamma_{5} + i\gamma_{5}\,S^{-1}(k), \label{eq:AWTI}
\end{align}
where $\Gamma_{5\mu}(k_{+},k)$ denotes the dressed AV vertex, which follows from
\begin{equation}
\Gamma_{5\mu}(k_{+},k) = \gamma_{5}\gamma_{\mu}
-\frac{4}{3}\frac{1}{m_{G}^{2}}
\int\!\frac{d^{4}q}{(2\pi)^{4}}\,
\gamma_{\alpha}\,\chi_{5\mu}(q_{+},q)\,\gamma_{\alpha}.
\label{eq:axialvertex}
\end{equation}
Here, $\Gamma_{\mu}^{\gamma}$ denotes the dressed quark--photon vertex, which encodes the interaction of a photon carrying momentum $Q$ with a quark whose initial and final momenta are $k_{i}$ and $k_{f}$, respectively. This vertex is a key element in any symmetry-preserving and current-conserving calculation of EFFs~\cite{Roberts:1994hh}. A bare vertex becomes insufficient once the quark propagator acquires momentum-dependent dressing, since it no longer satisfies the Ward--Takahashi identity and consequently yields a non-conserved electromagnetic current for the meson. Therefore, the vertex must include dressing consistent with the truncation used in the Bethe--Salpeter kernel for the bound state~\cite{Maris:1997hd}. In the present framework, this constraint is implemented through the inhomogeneous BSE
 \bea
 && \hspace{-1.2cm} \Gamma_{\mu}^{\gamma}(Q,M_{\fd})= 
 \gamma_{\mu} - \frac{16 \pi \hat{\alpha}_{\rm IR}}{3} 
  \int  \frac{d^4q}{(2 \pi)^4} \gamma_{\alpha} \chi_{\mu}(q_+,q,M_{\fd})
 \gamma_{\alpha} \, ,\label{eqvertex}
 \eea
where $\chi_{\mu}(q_+,q,M_{\fd})  =  S(q+P,M_{\fd}) \Gamma_{\mu}(Q)S(q,M_{\fd})$.
Owing to the momentum-independent nature of the interaction
kernel, the general form of the solution is
  \bea \label{vertex-pq} \Gamma_{\mu}^{\gamma}(Q,M_{\fd})= \gamma_{\mu}^{L}(Q)P_{L}(Q^{2},M_{\fd}) +
 \gamma_{\mu}^{T}(Q)P_{T}(Q^{2},M_{\fd}), \;\;\;\;\;\; \gamma_{\mu}^{L} + \gamma_{\mu}^{T} = \gamma_{\mu}
 \eea
 with
  \bea \gamma_{\mu}^{T}(Q)=\gamma_{\mu}-\frac{ \gamma \cdot Q \;
 }{Q^{2}} \, Q_{\mu}  \,.
 \eea
 Inserting this general form into Eq.~(\ref{eqvertex}), one readily
 obtains (on simplifying notation)
 \bea
 P_{L} =1 \,, \quad \;\;\;\;\;\; 
 P_T= \frac{1}{1+K_\gamma(Q^2,M_{\fd})} , \;\;\;\;\;\;
 K_\gamma(Q^2,M_{\fd}) = \frac{4 \hat{\alpha}_{\rm
IR}}{3\pi} 
\int_0^1d\alpha\, \alpha(1-\alpha) Q^2\,\bar{\mathcal{C}}_1(\omega)
\,.\label{PTQ2}
\eea
One can clearly observe from Fig.~\ref{fig:rhopole} that $P_{T}(Q^{2})
\rightarrow 1$ when $Q^2 \rightarrow \infty$, yielding the
perturbative bare vertex $\gamma_{\mu}$ as expected.
\begin{figure}[ht]
\begin{adjustwidth}{-\extralength}{-4cm}
\begin{center}
\includegraphics[width=9cm]{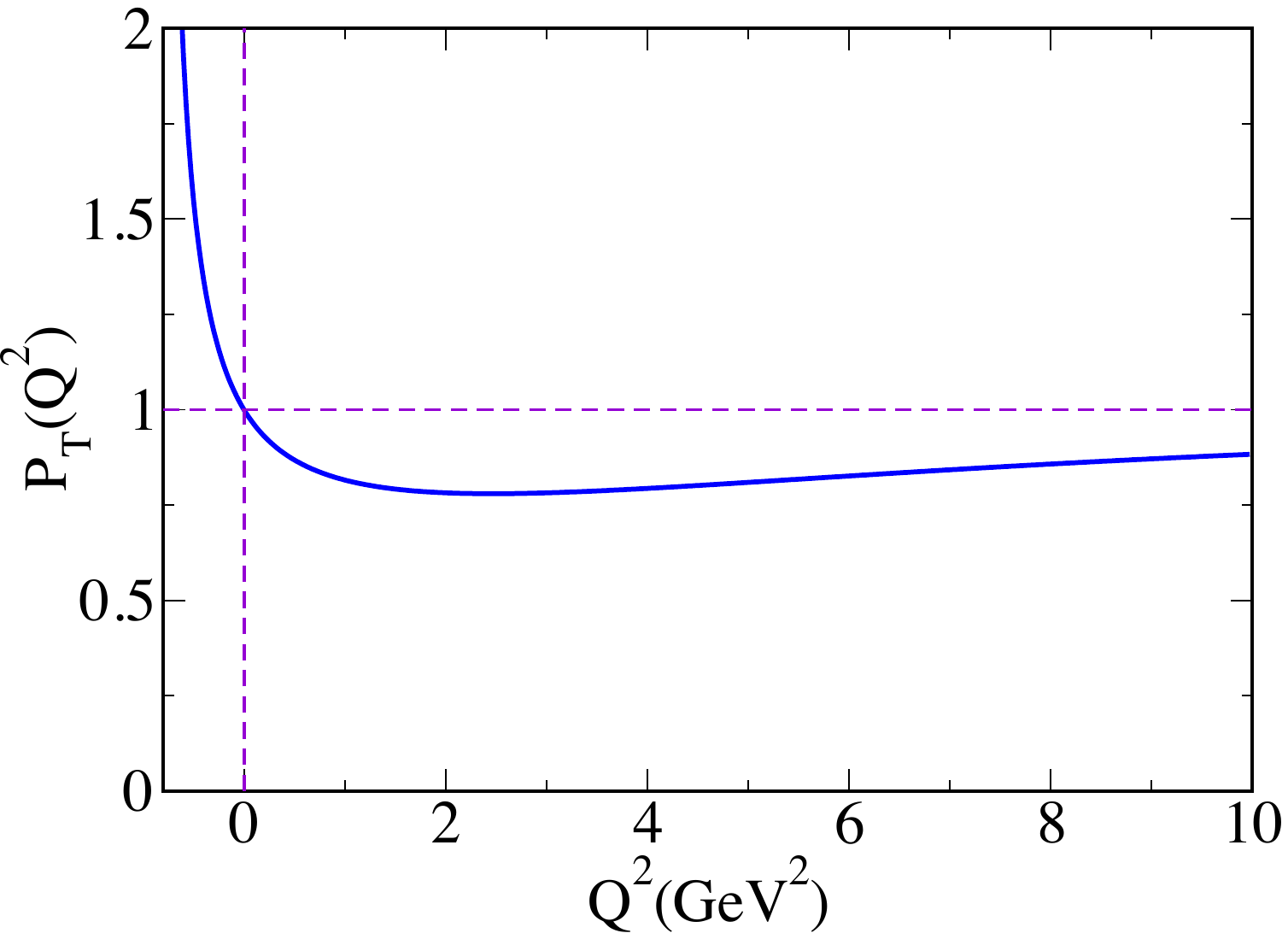}
\end{center}
\end{adjustwidth}
\caption{Dressing function associated with the transverse component of the quark–photon vertex in Eq.~(\ref{PTQ2}), evaluated using the parameter set defined in CI. The pole associated with the ground-state vector meson is clearly evident.}
\label{fig:rhopole}
\end{figure}
 We restrict our analysis to the calculation of EFFs for mesons and diquarks, using the quark-photon vertex defined in Eq.~(\ref{vertex-pq}). 
For transition FFs of mesons and baryons, however, a modified vertex is required. In such cases, the quark--photon coupling is taken as \cite{Wilson:2011aa,Xu:2024frc}
\begin{equation}
\Gamma(Q) = \gamma_{\mu}^{T} P_{T}(Q^{2}) 
+ \frac{\zeta}{2M}\,\sigma_{\mu\nu} Q_{\nu}\,
\exp\left(-\frac{Q^{2}}{4M^{2}}\right), 
\end{equation}
where \(M\) denotes the dressed-quark mass. The parameter \(\zeta = 1/2\) and the exponential damping scale follow the distribution obtained in Ref.~\cite{Chang:2010hb}.
The quark-photon vertex in Eq.~(\ref{vertex-pq}) provides the electromagnetic interaction required to probe meson electromagnetic form factors through a triangle diagram, as will be demonstrated in the following sections devoted to the calculation of the form factors using the ingredients described above.

\subsection{Form Factors of Pseudoscalar Mesons }
\label{ffps}
The EFF of the pion, the most extensively studied meson in this channel, was first investigated within CI in Ref.~\cite{GutierrezGuerrero:2010md}, with further developments presented in Ref.~ \cite{Roberts:2011wy}. The extension of this analysis to the heavy sector was explored in Ref.~\cite{Bedolla:2015mpa}. Moreover, Ref.~\cite{Hernandez-Pinto:2023yin} offers a comprehensive and unified treatment of mesons, including heavy-light systems, thereby providing a consistent framework for studies of this kind.
The PS mesons $F^{M,f_1}$ is straightforwardly related to $\Lambda^{M,\fd}$:
\begin{eqnarray}
 \Lambda^{\Meps,\fd}=-2k_\lambda F^{\Meps,\fd} \,.
\end{eqnarray}
Using the meson masses and BSAs obtained in Sec.~\ref{masses-mesons}, together with the parameters listed in Table~\ref{parFFn}, the EFFs are computed by combining the triangle diagram with the quark-photon vertex. This framework enables a systematic analysis of PS mesons composed of light-light ($qq$), light-heavy ($qQ$), and heavy--heavy ($QQ$) quark pairs. With straightforward algebraic manipulations, one obtains:
\bea 
&& \hspace{-8mm} F^{\Meps,\fd}= 
P_T(Q^2)\bigg[ 
E_{\Meps}^2 T^{\Meps}_{EE}(Q^2) +
E_{\Meps} F_{\Meps} T^{\Meps}_{EF}(Q^2) +
F_{\Meps}^2 T^{\Meps}_{FF}(Q^2) \bigg]\,,
\eea
where
\begin{align}
\nn T^{\Meps}_{EE}(Q^2) & = \frac{3}{4\pi^2}\bigg[ \int_0^1 d\alpha\, \overline{\mathcal{C}}_1(\omega_1)+ 
2\int_0^1 d\alpha\, d\beta \, \alpha \, \mathcal{A}^{\Meps}_{EE} \, \overline{\mathcal{C}}_2(\omega_2) \bigg]\,,\\
\nn T^{\Meps}_{EF}(Q^2) & 
= -\frac{3}{2\pi^2} \frac{1}{M_R}\int_0^1 d\alpha \, d\beta \, \alpha \bigg[\mathcal{A}_{EF}^{(1)} \, \overline{\mathcal{C}}_1(\omega_2)+  
(\mathcal{A}_{EF}^{(2)}-\omega_2 \mathcal{A}_{EF}^{(1)}) \, \overline{\mathcal{C}}_2(\omega_2) \bigg]\,,\\
 T^{\Meps}_{FF}(Q^2) & =  \frac{3}{4\pi^2} \frac{1}{M_R^2}
\int_0^1 d\alpha \, d\beta \, \alpha \bigg[ \mathcal{A}_{FF}^{(1)} \, \overline{\mathcal{C}}_1(\omega_2)  +
 (\mathcal{A}_{FF}^{(2)} -\omega_2 \mathcal{A}_{FF}^{(1)}) \, \overline{\mathcal{C}}_2(\omega_2)  \bigg]\,,
\end{align}
and
\begin{align}
\nn \omega_1 &= \omega_1(M_{\fd},\alpha,Q^2) = M_{\fd}^2+\alpha\, Q^2(1-\alpha) \, ,\\
\omega_2&=\omega_2(M_{\fd},M_{\fu},\alpha,\beta,m_H) = \alpha \,  M_{\fd}^2 +(1-\alpha)M_{\fu}^2 -\alpha(1-\alpha)\, m^2_{H} +\alpha^2 \, \beta \,  (1-\beta) \, Q^2\, , \nn \\
    \overline{\mathcal{C}}_2(z)&=
(\exp(-z\, \tau_{\rm UV})-\exp(-z\,\tau_{\rm IR}))/(2z)\,.
\end{align}
The coefficients ${\cal A}_i$ are given by the following expressions:
\begin{align}
\mathcal{A}^{\Meps}_{EE} &=
\alpha \left(M_{\fd}^2 + m_H^2 \right)
+ 2(1-\alpha) M_{\fd} M_{\fu}
+ (\alpha-2) M_{\fu}^2 \, , \nonumber \\[1mm]
\mathcal{A}_{EF}^{(1)} &=
M_{\fd} + M_{\fu} \, , \nonumber \\[1mm]
\nn \mathcal{A}_{EF}^{(2)} &=
2 M_{\fd}^2 M_{\fu}
- \alpha M_{\fd}
\left[ 4 (\alpha-1) m_H^2 + \alpha Q^2 \right]
\nonumber \\[0.5mm]
& \nn
+ M_{\fu}
\left[
2 (\alpha-1)^2 m_H^2
+ \alpha Q^2
\left( 2 \alpha (\beta-1)\beta + \alpha - 1 \right)
\right] \, , \\[1mm]
\mathcal{A}_{FF}^{(1)} &=
(3\alpha -2) m_H^2 + \alpha Q^2 \, , \nonumber \\[1mm]
\mathcal{A}_{FF}^{(2)} &=
2 \alpha
\Big[
(\alpha-1)^2 m_H^4
+ \alpha m_H^2 Q^2
\left(
3 \alpha \beta^2
- 3 \alpha \beta
+ \alpha
- 2 \beta^2
+ 2 \beta
- 1
\right)
\Big]
\nonumber \\[0.5mm]
&
+ 2 \alpha m_H^2 M_{\fd}^2
- 2 M_{\fd} M_{\fu}
\left[
2 (\alpha-1) m_H^2 + \alpha Q^2
\right] \, .
\end{align}
The resulting EFFs for both charged and neutral mesons have been calculated and analyzed in Ref.~\cite{Hernandez-Pinto:2023yin} and are shown in Fig.~\ref{plotPS}.
\begin{figure}[htbp]
\begin{adjustwidth}{-\extralength}{-5.0cm}
\begin{center}
\begin{tabular}{@{\extracolsep{-2.3 cm}}c}
 \renewcommand{\arraystretch}{-1.6} %
 \hspace{-5mm}
 \includegraphics[scale=0.55]{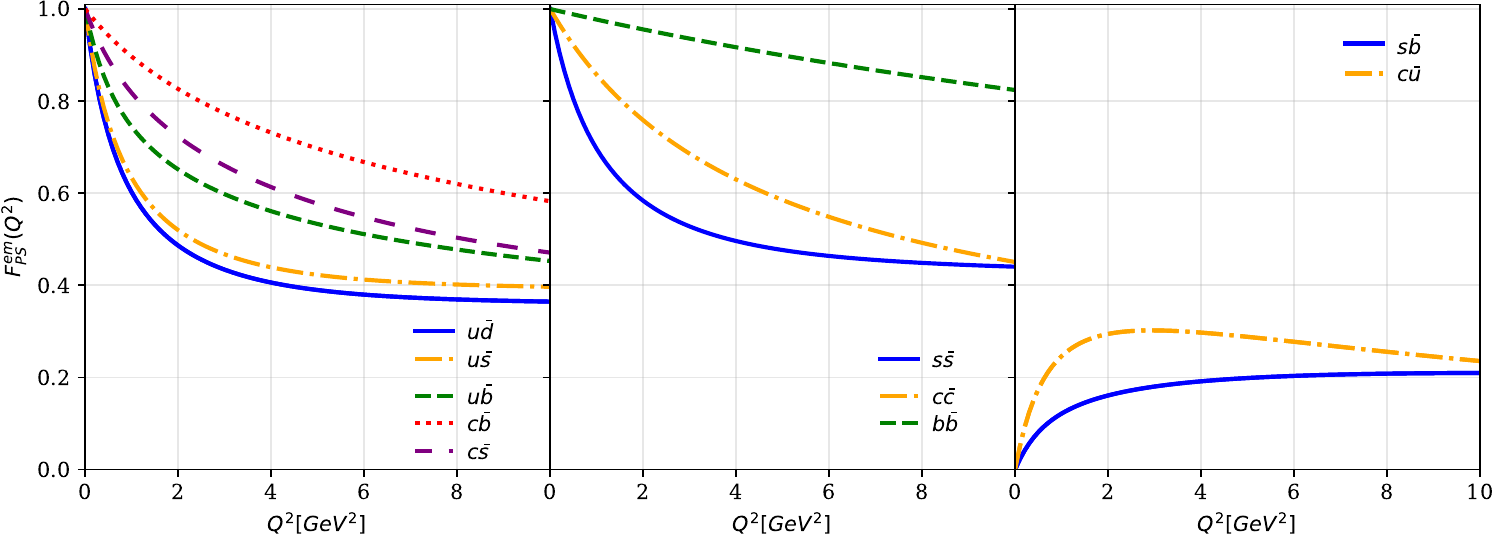}
\end{tabular}
\end{center}
\end{adjustwidth}
\vspace{-2mm}
\caption{\label{plotPS}EFFs of PS mesons computed within the CI framework.
Left panel: charged mesons formed by quarks of different flavors.
Center panel: quarkonia states, including a hypothetical ground-state strangeonium ($\ts\bar{\ts}$).
Right panel: neutral mesons composed of quarks of various flavors.
The results are taken from Ref.~\cite{Hernandez-Pinto:2023yin}.}
\end{figure}
FFs are parameterized by the following functional expression: 
\begin{align}
F^{PS}(Q^2) = \frac{e_{M} + a_{PS}\, Q^2 +b_{PS}\,Q^4}{1+c_{PS}\, Q^2 + d_{PS}\, Q^4}\,,
\label{EqfitPS}
\end{align}
where $e_M\equiv F^{PS}(Q^2=0)$ is the electric charge of the meson and $a_{PS}$, $b_{PS}$, $c_{PS}$, $d_{PS}$ are the fitted coefficients. The best fit corresponds to the values listed in Table \ref{fitParaPS}.
The fit in Eq.~(\ref{EqfitPS}) reflects the fact that PS meson EFFs approach constant values once $Q^{2}$ becomes the dominant energy scale, a natural outcome of the point-like nature of the CI interaction. Heavy and heavy--light mesons, however, reach this asymptotic behavior more slowly than light mesons, since $Q^{2}$ must exceed their larger intrinsic energy scales. \\
\begin{table}[hbt]
\caption{\label{fitParaPS}Parameters obtained from the fit of Eq.~(\ref{EqfitPS}) for PS mesons, valid in the range $Q^{2} \in [0,\, 8 m_{H}^{2}]$. This parametrization also simplifies the extraction of charge radii.
}
\begin{adjustwidth}{-\extralength}{-4cm}
\begin{center}
\begin{tabular}{@{\extracolsep{0.0 cm}} || c | c | c | c | c || }
\hline
\hline
& $a_{PS}$ & $b_{PS}$ & $c_{PS}$ & $d_{PS}$  \\ 
\hline
\rule{0ex}{2.5ex}
 $\tu\bar{\td}$ \,& \,\, 0.330\,\,   &\,\,   0.029\,\,   & \, \, 1.190\,\,   &\,\,   0.068\,\,   \\
\rule{0ex}{2.5ex}
$ \tu\bar{\ts}$ \,& 0.335 & 0.029 & 1.092 & 0.065  \\
\rule{0ex}{2.5ex}
$\ts\bar{\ts}$ \,& 0.328 & 0.040 & 0.874 & 0.092  \\
\rule{0ex}{2.5ex}
$\tc\bar{\tu}$ \,& 0.616 & $-$0.001 & 1.370 & 0.109 \\
\rule{0ex}{2.5ex}
$\tc\bar{\ts}$ \,& 0.615 & 0.028 & 0.897 & 0.111  \\
\rule{0ex}{2.5ex}
$\tu\bar{\tb}$ \,& 1.143 & 0.033 & 1.921 & 0.146  \\
\rule{0ex}{2.5ex}
$\ts\bar{\tb}$ \,& 0.218 & 0.000 & 0.840 & 0.009 \\
\rule{0ex}{2.5ex}
$\tc\bar{\tb}$ \,& 0.333 & 0.003 & 0.493 & 0.021  \\
\rule{0ex}{2.5ex}
$\tc\bar{\tc}$ \,& 1.778 & 0.057 & 1.994 & 0.334  \\
\rule{0ex}{2.5ex}
$\tb\bar{\tb}$ \,& 0.099 & 0.000 & 0.127 & 0.002   \\
\hline \hline
\end{tabular}
\end{center}
\end{adjustwidth}
\end{table}
The behavior of the FFs in the vicinity of $Q^{2} \simeq 0$ reveals the spatial distribution of charge, which makes it possible to extract the corresponding charge radii. These radii provide an essential measure of the electromagnetic size of mesons and offer a direct point of comparison with experimental results. Once the FFs are obtained, the charge radii can be computed with the following expression:
\begin{equation}
 r_{M}^2 =
-6\left.\frac{\mathrm{d}F_{M}(Q^{2})}{\mathrm{d}Q^{2}}\right|_{Q^{2}=0}
\,.\label{radii}
\end{equation}
In Table~\ref{tab:chargeradiuspsu}, we present the charge radii (in fm) of PS mesons obtained within the CI framework, taken from Ref.~\cite{Hernandez-Pinto:2023yin}, and compare them with results from other theoretical approaches and experiment~\cite{ParticleDataGroup:2022pth}. The results are contrasted with those from a Bethe–Salpeter framework (BSF)~\cite{Miramontes:2025vzb}, LQCD~\cite{Wang:2020nbf,Can:2012tx}, data-driven methods~\cite{Stamen:2022uqh,Leplumey:2025kvv}, weighted rainbow–ladder analyses~\cite{Xu:2024fun}, and an algebraic model (AM)~\cite{Almeida-Zamora:2023bqb}, as well as predictions from a hybrid model (HM)~\cite{Lombard:2000kw}, a light-front framework (LFF)~\cite{Hwang:2001th}, and a QCD potential model (PM)~\cite{Das:2016rio}.
\begin{table}[htbp]
\caption{\label{tab:chargeradiuspsu}
Charge radii (in fm) of PS mesons obtained within different theoretical approaches and from experiment~\cite{ParticleDataGroup:2022pth}. Results from CI framework are compared with those from a BSF~\cite{Miramontes:2025vzb},LQCD~\cite{Wang:2020nbf,Can:2012tx}, data-driven methods~\cite{Stamen:2022uqh,Leplumey:2025kvv}, weighted rainbow--ladder analyses~\cite{Xu:2024fun}, and an AM~\cite{Almeida-Zamora:2023bqb}, as well as predictions from a HM~\cite{Lombard:2000kw}, a LFF~\cite{Hwang:2001th}, and a PM~\cite{Das:2016rio}. We follow the notation used in Ref.~\cite{Miramontes:2025vzb}, and the imaginary factor “$i$” is employed whenever the squared charge radii are negative.}
\label{tab:ps_charge_radii_all_refs_complete}
\begin{adjustwidth}{-\extralength}{-5cm}
\begin{center}
\begin{tabular}{@{\extracolsep{-0.15 cm}}lccccccccccc}
\hline
\rule{0ex}{3.0ex}
 & $\pi$
 & $K$
 & $h_s$
 & $D^0$
 & $D_s^{+}$
 & $B^{+}$
 & $B_s$
 & $B_c$
 & $\eta_c$
 & $\eta_b$ \\
\hline
\rule{0ex}{3.0ex}
CI
& 0.45 & 0.42 & 0.36
& 0.36 & 0.26
& 0.34 & $0.24$
& 0.17 & 0.20 & 0.07 \\
\rule{0ex}{3.0ex}
BSF
& 0.656 & 0.568 & $0.270\, i$
& 0.428 & $0.542\, i$
& 0.631 & $0.330\, i$
& 0.213 & 0.267 & 0.082 \\
\rule{0ex}{3.0ex}
Experiment
& 0.659 & 0.560 & $0.277\, i$
& -- & --
& -- & --
& -- & -- & -- \\
\rule{0ex}{3.0ex}
LQCD
& 0.656 & -- & --
& 0.450(24) & --
& -- & --
& -- & -- & -- \\
\rule{0ex}{3.0ex}
Data-driven
& 0.655 & 0.599 & $0.245\, i$
& -- & --
& -- & --
& -- & -- & -- \\
\rule{0ex}{3.0ex}
Weighted-RL
& 0.646 & 0.608 & $0.253\, i$
& 0.435 & $0.556\, i$
& 0.619 & $0.337\, i$
& 0.219 & -- & -- \\
\rule{0ex}{3.0ex}
AM
& -- & -- & --
& 0.680 & --
& 0.926 & $0.345\, i$
& 0.217 & -- & -- \\
\rule{0ex}{3.0ex}
HM
& 0.66 & 0.65 & --
& 0.47 & 0.50
& -- & --
& -- & -- & -- \\
\rule{0ex}{3.0ex}
LFF
& 0.66 & 0.58 & --
& 0.55 & 0.35
& 0.61 & 0.34
& 0.20 & -- & -- \\
\rule{0ex}{3.0ex}
PM
& -- & -- & --
& 0.67 & 0.46
& 0.73 & 0.46
& -- & -- & -- \\
\hline
\end{tabular}
\end{center}
\end{adjustwidth}
\end{table}
The overall pattern in which charge radii decrease as the constituent quark mass increases is consistent with physical expectations. This behavior is evident in the following hierarchies:
\begin{eqnarray}\label{eq:hierarchies}
&& r_{\tu\bar{\td}} > r_{\tu\bar{\ts}} > r_{\tc\bar{\tu}} > r_{\tu\bar{\tb}} \,, \nn \\
&& r_{\tu\bar{\ts}} > r_{\ts\bar{\ts}} > r_{\tc\bar{\ts}} > r_{\ts\bar{\tb}} \,, \nn \\
&& r_{\tc\bar{\tu}} > r_{\tc\bar{\ts}} > r_{\tc\bar{\tc}} > r_{\tc\bar{\tb}} \,, \nn \\
&& r_{\tu\bar{\tu}} > r_{\ts\bar{\ts}} > r_{\tc\bar{\tc}} > r_{\tb\bar{\tb}} \,. 
\end{eqnarray}
The EFFs of the pion and the kaon are identified as benchmark observables for
exploring the nonperturbative dynamics of QCD. As the lightest meson and a Goldstone boson associated
with dynamical chiral symmetry breaking, the pion provides direct insight into the mechanisms
responsible for the emergence of hadronic mass. The kaon, which contains a strange quark, offers a
complementary probe by allowing the relative roles of dynamical chiral symmetry breaking and explicit
quark-mass effects to be disentangled. The experimental program at the Electron-Ion Collider is
expected to enable high-precision measurements of pion and kaon FFs over an extended range of
momentum transfer, thereby providing stringent tests for theoretical approaches and strong motivation
for unified and symmetry-preserving descriptions of meson structure~\cite{AbdulKhalek:2021gbh}.
Finally, to close this section, Figs.~\ref{pionff} and \ref{kaonff} show the elastic FFs of the 
pion and the kaon, the lightest mesons in this channel, together with experimental measurements from 
JLab and Amendolia~\cite{AMENDOLIA1986168,JeffersonLabFpi:2000nlc,JeffersonLabFpi-2:2006ysh}. 
For comparison, we also include recent calculations performed within BSE 
framework, which provide space-like EFFs for PS mesons, covering both 
light and heavy–light systems \cite{Miramontes:2025vzb}.
\begin{figure}[htbp]
\begin{adjustwidth}{-\extralength}{-4cm}
\begin{center}
\hspace{0.5cm}
\begin{minipage}[b]{0.29\linewidth}
\centering
\includegraphics[width=\textwidth]{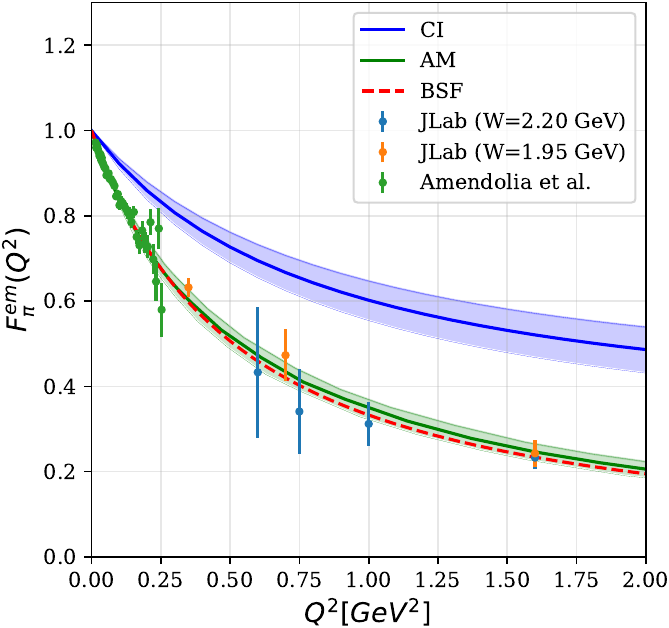}
\captionsetup{justification=justified, singlelinecheck=false}
\caption{\label{pionff}
Comparison of the pion EFF predictions within the CI framework,  AM~\cite{Higuera-Angulo:2024oui}, and BSF~\cite{Miramontes:2025vzb}, together with
experimental measurements from JLab and Amendolia~\cite{AMENDOLIA1986168,JeffersonLabFpi:2000nlc,JeffersonLabFpi-2:2006ysh}. 
}
\end{minipage}
\hspace{0.5cm}
\begin{minipage}[b]{0.28\linewidth}
\centering
\includegraphics[width=\textwidth]{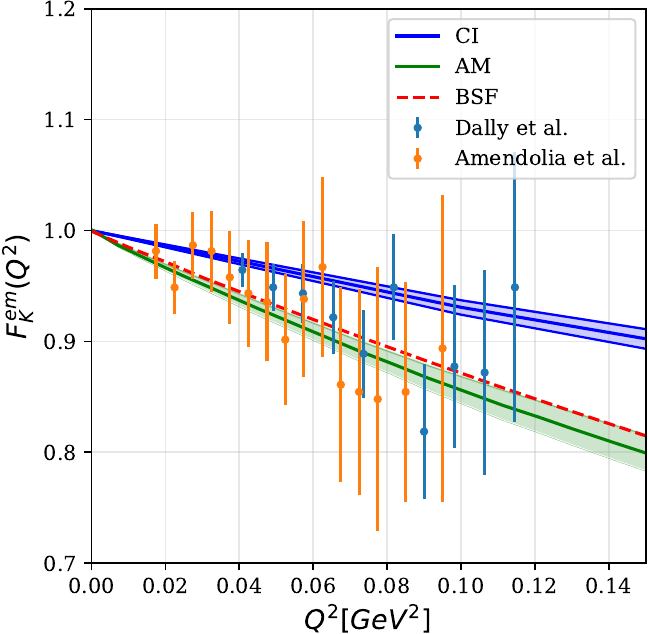}
\captionsetup{justification=justified, singlelinecheck=false}
\caption{\label{kaonff} EFF of the kaon obtained within the CI framework, compared with results from an AM \cite{Higuera-Angulo:2024oui}, BSF \cite{Miramontes:2025vzb} and with experimental data from Dally and Amendolia~\cite{AMENDOLIA1986168}.}
\end{minipage}
\end{center}
\end{adjustwidth}
\end{figure}
The central curve of the (blue) band is obtained by using the $\Lambda_{\rm UV}$ value from Table~\ref{parameters1}. The filled (blue) band allows for a $5\%$ variation in the charge radius.
From Figs.~\ref{pionff} and \ref{kaonff} it is evident that the CI reproduces the correct normalization at 
$Q^2=0$ and provides a reasonable slope in the low-momentum region, where it remains close to the experimental 
data; however, as the momentum increases, the results deviate from experiment due to the nature of the model in 
Eq.~(\ref{eqn:contact_interaction}). The FFs decrease more slowly than those obtained using BSE or algebraic models. In general, the 
CI captures the qualitative trend of the electromagnetic structure of light PS mesons despite its 
simplicity, serving as a useful reference compared with more sophisticated calculations.\\
A detailed analysis of the derivation of the pion form factor in the chiral limit is presented in Appendix~\ref{ffpice}, based on the pioneering contact-interaction study of Ref.~\cite{GutierrezGuerrero:2010md}. Subsequent works, including those presented here, follow the same methodology, with the appropriate modifications associated with the quark flavor content and spin structure. 
In the next section, we shift our focus to the parity partners of these systems, namely the S mesons, whose properties provide complementary insight into the dynamics underlying the PS sector.
\subsection{Form Factors of Scalar Mesons}
Having discussed the EFFs of PS mesons, we now turn to the S meson sector. In constituent-quark-based approaches, PS mesons are characterized by vanishing orbital angular momentum, $L=0$, and total spin $S=0$, arising from an antiparallel alignment of the constituent quark spins. By contrast, S mesons correspond to $P$-wave configurations with $L=1$ and $S=1$, where the coupling between spin and orbital degrees of freedom yields a total angular momentum $L+S=0$. Within this framework, S mesons are naturally interpreted as excitations of their PS counterparts see Fig. \ref{fig:sps}.\\
\begin{figure}[b]
\begin{adjustwidth}{-\extralength}{-4cm}
    \begin{center}
\includegraphics[width=0.35\linewidth]{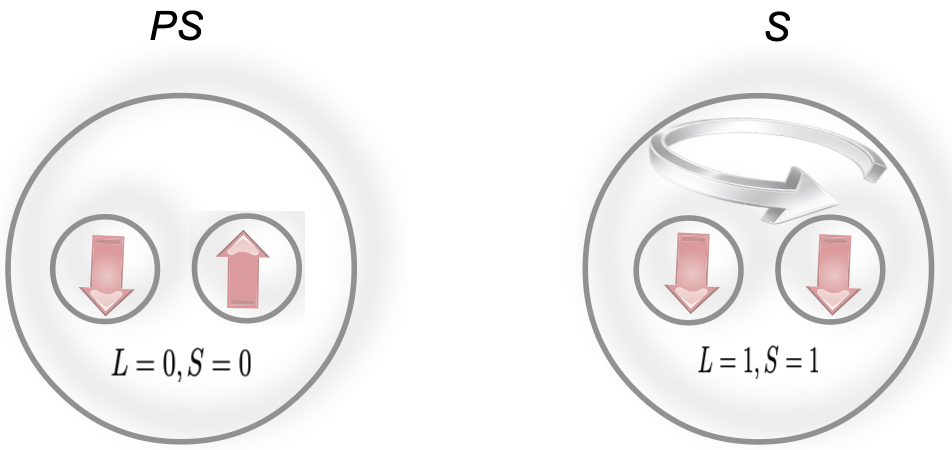}
 \end{center}
    \end{adjustwidth}
\captionsetup{justification=justified, singlelinecheck=false}
    \caption{\label{fig:sps}Within constituent-quark-based approaches, S mesons can be viewed as excited states of the corresponding PS mesons. The figure is adapted from Ref.~\cite{Bashir:2012fs}.}
\end{figure}
This structural distinction has direct consequences for their internal dynamics and electromagnetic properties, thereby motivating a dedicated analysis of S-meson FFs. 
The S mesons $F^{M,f_1}$ is straightforwardly related to $\Lambda^{M,\fd}$:
\begin{eqnarray}
\Lambda^{\Ms,\fd}=-2k_\lambda F^{\Ms,\fd} \,, \, 
\end{eqnarray}
the S channels represent a natural and meaningful extension of the analysis performed for PS mesons. In this section, we outline the main features of  S mesons FFs within the CI framework and present the corresponding results for their electromagnetic properties.
The explicit expression for the EFFs for S mesons with mass $m_H$ constituted from a quark $\fd$ and an antiquark $\bar\fdu$ is given by Eq.~(\ref{eqn:TotalMesonFF}) with
 \bea 
&& \hspace{-5mm} F^{\Ms,\fd}= 
-P_T(Q^2)
E_{\Ms}^2 \frac{3}{4\pi^2}\bigg[ \int_0^1 d\alpha\, \overline{\mathcal{C}}_1(\omega_1) + 
2\int_0^1 d\alpha\, d\beta \, \alpha \, \mathcal{A}^{\Ms}_{EE} \, \overline{\mathcal{C}}_2(\omega_2) \bigg]\,,
\eea
 with
 \begin{align}
     \mathcal{A}_{EE}^{\Ms} &= \alpha M_{\fd}-2(1-\alpha) M_{\fd} M_{\fu} 
     +(\alpha-2)M_{\fu}^2+\alpha m_H^2 \, .
 \end{align}
\begin{figure*}[t!]
\begin{adjustwidth}{-\extralength}{-4cm}
\begin{center}
\begin{tabular}{@{\extracolsep{-2.3 cm}}c}
 \renewcommand{\arraystretch}{-1.6} %
 \hspace{5mm}
 \includegraphics[scale=0.55]{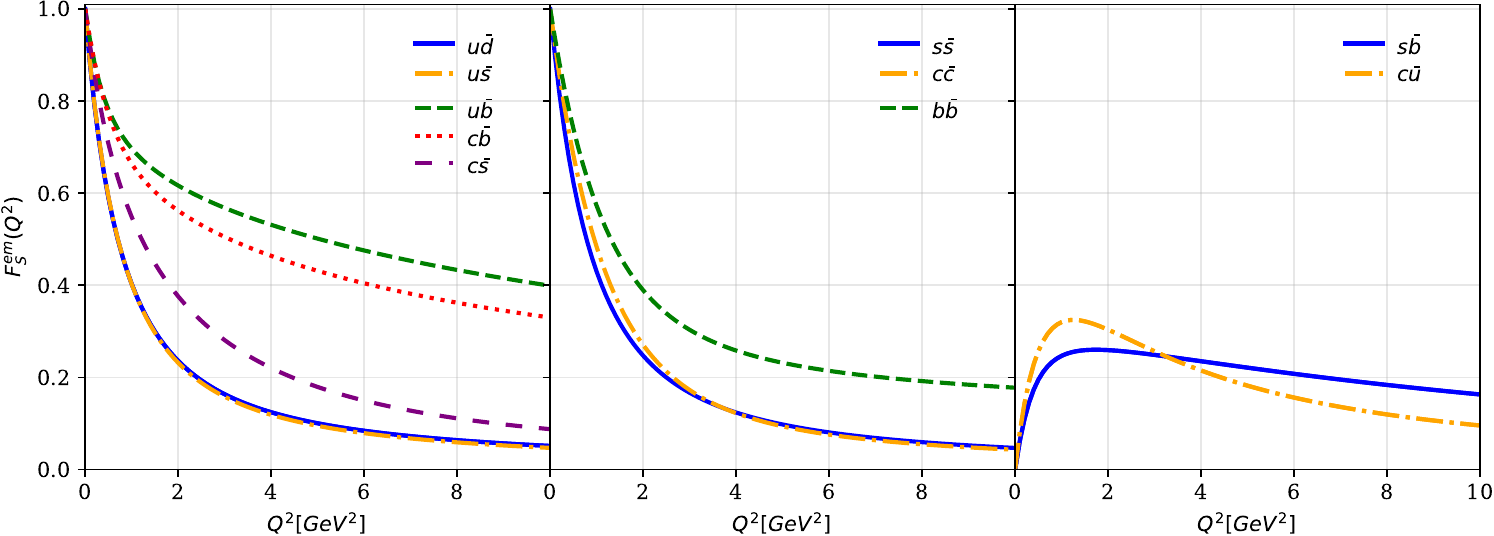}
\end{tabular}
\end{center}
\end{adjustwidth}
\vspace{-2mm}
\caption{ \justifying
EFFs for S mesons in the CI model. Left panel: electrically charges mesons composed of quarks of different flavors. Central panel: quarkonia including a hypothetical ground state {\em strangeonium} ($\ts\bar \ts$). Right panel: electrically neutral mesons composed of quarks of different flavors.
EFFs of electrically neutral but flavored mesons have been normalized to $F^S(0)=0$.}\label{plotS}
\end{figure*}
In the same spirit of PS FFs, the total EFFs of S mesons are fitted to,
\begin{align}
F^{S}( Q^2) = \frac{e_{M} + a_S\, Q^2 +b_S\,Q^4}{1+c_S\, Q^2 + d_S\, Q^4}\,,
\label{EqfitS}
\end{align}
where $e_M\equiv F^{S}(Q^2=0)$ is the electric charge of the meson and $a_S$, $b_S$, $c_S$, $d_S$ are the parameters of the fit. These values for S mesons are listed in Table \ref{fitParaS}. Based on these numbers, we can immediately infer the large $Q^2$ behavior of these EFFs. The coefficient $b_S \approx 0 $ for all S mesons under consideration. Therefore, the EFFs for S mesons fall as $1/Q^2$ for large $Q^2$. 
We also present the numerical values of the charge radii for S mesons in Table~\ref{fitParaS}. 
We must reiterate that for the S mesons there are no reported measurements of their charge radii. Theoretical results are also scarce for any direct and meaningful comparison. It is worth mentioning again that the internal structure of S mesons is not well-established. CI results are based on  considering them as effective quark-antiquark states. 
\begin{table}[hbt]
\caption{\label{fitParaS}
Parameters for the fit in Eq.~(\ref{EqfitS}) for S mesons. The last row shows the corresponding charge radii in fm.}
\begin{adjustwidth}{-\extralength}{-4cm}
\begin{center}
\hspace{0.7cm}
\begin{tabular}{@{\extracolsep{0.0 cm}}c c c c c c c c c c c}
\hline\hline
\rule{0ex}{2.5ex}
 & $\tu\bar{\td}$ & $\tu\bar{\ts}$ & $\ts\bar{\ts}$ & $\tc\bar{\tu}$ & $\tc\bar{\ts}$ & $\tu\bar{\tb}$ & $\ts\bar{\tb}$ & $\tc\bar{\tb}$ & $\tc\bar{\tc}$ & $\tb\bar{\tb}$ \\
\hline
\rule{0ex}{2.5ex}
$a_S$ & 0.29 & 0.27 & 0.22 & 0.76 & 0.00 & 0.98 & 0.21 & 0.29 & 0.22 & 0.27 \\
\rule{0ex}{2.5ex}
$b_S$ & 0.00 & 0.00 & 0.00 & $-0.01$ & 0.00 & 0.00 & 0.00 & 0.00 & 0.00 & 0.00 \\
\rule{0ex}{2.5ex}
$c_S$ & 1.54 & 1.49 & 1.27 & 0.68 & 0.78 & 1.62 & 0.18 & 0.74 & 0.86 & 1.61 \\
\rule{0ex}{2.5ex}
$d_S$ & 0.62 & 0.63 & 0.54 & 0.64 & 0.05 & 0.09 & 0.12 & 0.03 & 0.67 & 0.02 \\
\hline
\rule{0ex}{2.5ex}
$\langle r^2 \rangle^{1/2}$ [fm] 
& 0.55 & 0.54 & 0.50 & 0.47 & 0.44 & 0.42 & 0.41 & 0.40 & 0.43 & 0.39 \\
\hline\hline
\end{tabular}
\label{tab:S_fit_radii_vertical}
\end{center}
\end{adjustwidth}
\end{table}

\subsection{Form Factors of Vector Mesons}
V mesons occupy a prominent position in hadron physics, as they provide a particularly sensitive window into the nonperturbative dynamics of QCD. Owing to their spin-one nature, vector mesons are characterized by electric, magnetic, and quadrupole electromagnetic form factors, which encode complementary information on the spatial distributions of charge and current within these systems.
In this section, we discuss the main features of V mesons within the CI framework and present the corresponding results for their static and dynamical observables, thereby establishing a basis for systematic comparisons with other mesonic channels.
Within the quark model, V mesons can be interpreted as spin-flip partners of pseudoscalar mesons. This distinction leads to marked differences in their electromagnetic structure: while PS mesons are described by a single form factor, V mesons require three independent form factors. Consequently, the extraction of $F_j^{M,f_1}$ for V mesons involves suitable projection operators, reflecting the more intricate tensor structure of $\Lambda^{M,\fd}_{\lambda\mu\nu}$ associated with their spin-one character. Explicitly,  
\begin{align}\label{Eq:VFF}
     \Lambda_{\lambda \mu \nu}^{\Me,\fd} = \sum_{j=1}^3 T_{\lambda \mu \nu}^{(j)}(k,Q) \, F_j^{\Me,\fd}(Q^2) \, ,
 \end{align} with the tensors $T_{\lambda \mu \nu}^{(j)}$ expressed as\,:
%
%
 \begin{align}
 T_{\lambda \mu \nu}^{(1)}(k,Q) & =  2 k_\lambda\, {\cal
 P}^T_{\mu\alpha}(k_i) \, {\cal P}^T_{\alpha\nu}(k_f)\,, \\
 T_{\lambda \mu \nu}^{(2)}(k,Q) & =  \left[Q_\mu - k_{i\mu} \frac{Q^2}{2 m_{H}^2}\right] {\cal P}^T_{\lambda\nu}(k_f) - \left[Q_\nu + k^f_\nu \frac{Q^2}{2 m_{H}^2}\right] {\cal P}^T_{\lambda\mu}(k_i)\,, \\
 T_{\lambda\mu \nu}^{(3)}(k,Q) & =  \frac{k_\lambda}{m_{H}^2}\, \left[Q_\mu - k_{i\mu} \frac{Q^2}{2 m_{H}^2}\right] \left[Q_\nu + k_{f\nu} \frac{Q^2}{2 m_{H}^2}\right] \,.
 \end{align}
In addition, the transverse projector in the above relations is given by\,:
\begin{eqnarray} {\cal P}_{\alpha
\beta}^T(P) = \delta_{\alpha \beta} - {P_{\alpha}
P_{\beta}}/{P^2} \,.
\end{eqnarray}
%
After introducing the Feynman parameters and integrating over the internal momentum $l$, the EFFs defined in Eq.~(\ref{Eq:VFF}), $F_i^{V,f_1}$, can be projected out to be\,: 
\begin{eqnarray}
\label{eqn:vecmes}
 && \hspace{-1cm} F^{\Me,\fd}_i(Q^2) = \frac{3}{4\pi^2} E^2_{\Me}  \, P_T(Q^2)
\int_0^1 d\alpha \, d\beta \; \alpha  \bigg[
\mathcal{A}^{\Me}_i \, \overline{\mathcal{C}}_1(\omega_1)  +(\mathcal{B}^{\Me}_i -\mathcal{A}^{\Me}_i \, \omega_2) \, \overline{\mathcal{C}}_2(\omega_2) 
\bigg] \, ,
\end{eqnarray}
where the labels $i=1,2,3$ correspond to the three EFFs. Besides, $\mathcal{A}_i^{\Me}$ coefficients are given by,
\begin{align}
\nn\mathcal{A}^{\Me}_1 &= 2-\alpha \, ,\\
\nn \mathcal{A}^{\Me}_2 & = \frac{1}{2} \left( \frac{\alpha(2\beta-1)Q^2}{m_H^2}+\alpha(10\beta - 7)-4\right) \, , \\
\mathcal{A}^{\Me}_3 &= \frac{2\alpha(1-2\beta)(Q^2 + 5m^2_{H} )}{Q^2+4 m^2_{H}} \, , 
\end{align}
while $\mathcal{B}_i^{\Me}$ are,
\begin{align}
\mathcal{B}^{\Me}_1 &=
2\Big[
(\alpha-2)\alpha^2(1-\beta)\beta Q^2
+ \alpha M_{\fd}^2
+ 2(1-\alpha) M_{\fd} M_{\fu}
+ (1-\alpha)^2 \alpha m_H^2
\Big] \, , \nonumber \\[1mm]
\mathcal{B}^{\Me}_2 &=
\frac{\alpha Q^2 (2\beta-1) M_{\fd} M_{\fu}}{m_H^2}
+ \alpha^2 Q^2
\left(
2\alpha\beta^3
- 5\alpha\beta^2
+ (\alpha+2)\beta
+ \alpha - 1
\right)
\nonumber \\[0.5mm]
&
+ \alpha (6\beta-5) M_{\fd}^2
+ 2 M_{\fd} M_{\fu} (2\alpha\beta - \alpha - 2)
\nonumber \\[0.5mm]
\nn &
- (\alpha-1)\alpha m_H^2
\left(
10\alpha\beta - 7\alpha - 6\beta + 1
\right) \, , \\[1mm]
\mathcal{B}^{\Me}_3 &=
\Bigg\{
4\alpha m_H^2
\Big[
(3-6\beta) M_{\fd}^2
+ 2(1-2\beta) M_{\fd} M_{\fu}
\nonumber \\[0.5mm]
&
+ (\alpha-1) m_H^2
\left(
16\alpha\beta^2
- 6(\alpha+1)\beta
- 5\alpha + 3
\right)
\Big]
\nonumber \\[0.5mm]
&
- 4\alpha Q^2
\Big[
(2\beta-1)
\Big(
M_{\fd} M_{\fu}
+ \alpha\beta m_H^2 (\alpha(\beta-3)+2)
\Big)
\nonumber \\[0.5mm]
&
+ (\alpha-1)\alpha m_H^2
\Big]
\Bigg\}
\frac{1}{Q^2 + 4 m_H^2} \, .
\end{align}

We recall that the total EFFs must consider the participation of both constituent quarks in the process, through an analogue expression to Eq.~(\ref{eqn:TotalMesonFF}). For V mesons, the definition given in 
Eq.~(\ref{eqn:vecmes}) is important to define the three experimentally accessible and physically suggestive EFFs: electric, magnetic and quadrupole FFs, defined as
 \begin{subequations}
\begin{align}
G_E^{V}(Q^2) & =  F_1^{ V}(Q^2)+\frac{2}{3} \eta \, G_Q^{ V}(Q^2)\,,\\
G_M^{ V}(Q^2) & =  - F_2^{ V}(Q^2)\,, \label{Definemu}\\
%
G_Q^{ V}(Q^2) & =  F_1^{ V}(Q^2) + F_2^{ V}(Q^2) + \left[1+\eta\right] F_3^{ V}(Q^2)\,,
\end{align}
\label{Gsm}
\end{subequations}
where $\eta=Q^2/(4 m_{ H}^2)$ and $m_H$ is the mass of the V meson. In the limit $Q^2\to 0$, these EFFs define the charge, magnetic ($\mu_V$) and quadrupole moments ($\mathcal{Q}_V$) of the V meson; viz.,
\begin{equation}
\label{chargenorm}
G_E^{ V}(Q^2=0) =  1\,, \;\;\;\;\;\;
G_M^{ V}(Q^2=0) =  \mu_{ V},\; \;\;\;\;\;\;
G_Q^{ V}(Q^2=0) = \mathcal{Q}_{ V}\,.
\end{equation}
EFFs of light, heavy-light, and the heaviest V mesons are displayed in Fig.\,\ref{plotVE}.
\begin{figure}[ht]
\begin{adjustwidth}{-\extralength}{-1cm}
\begin{center}
\hspace{1cm}
\begin{tabular}{@{\extracolsep{-2.3 cm}}cc}
\includegraphics[scale=0.535]{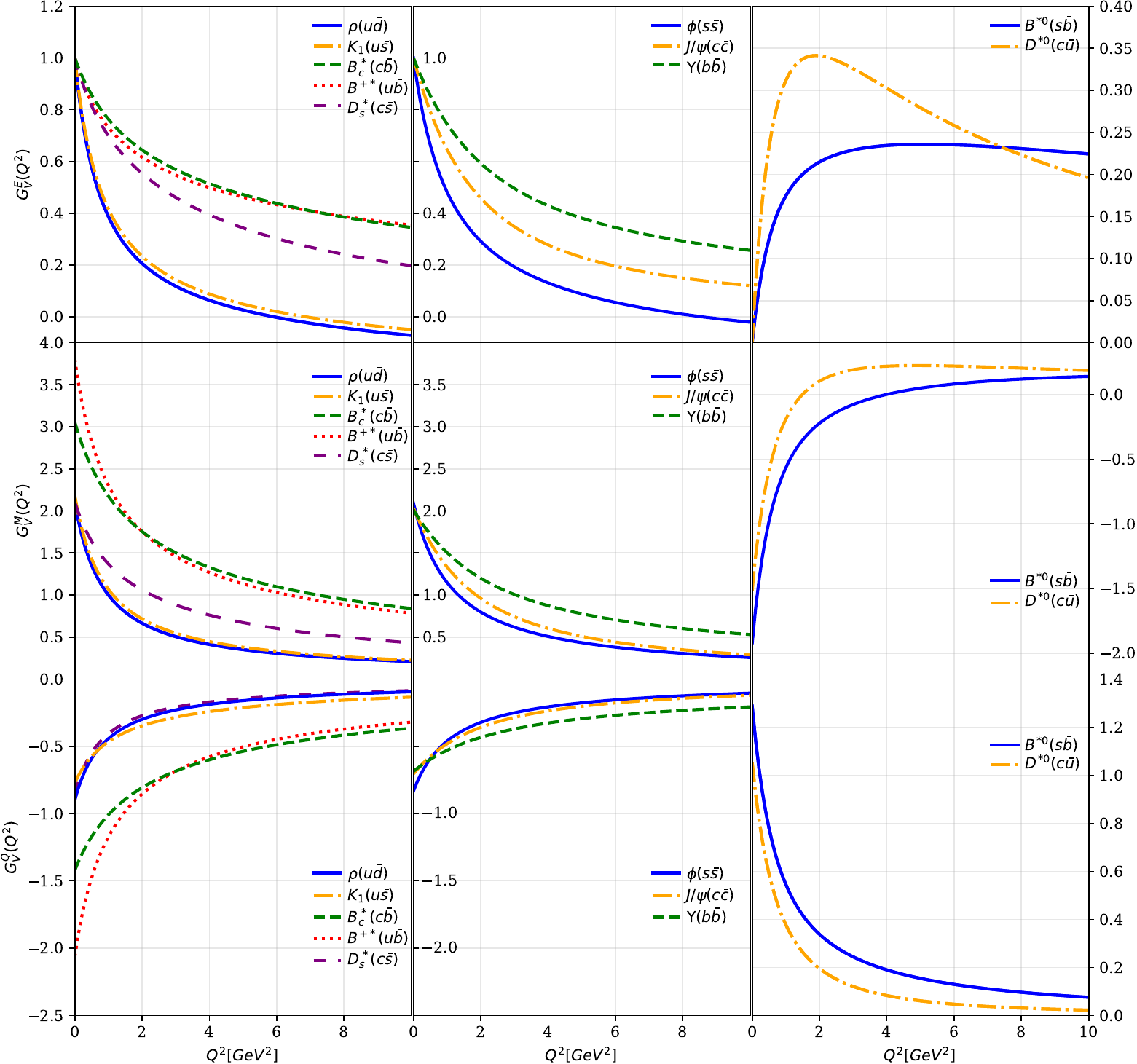}
\end{tabular} 
\end{center}
\end{adjustwidth}
\vspace{-0.2cm}
\caption{\justifying
V mesons electric, magnetic, and quadrupole FFs are displayed in the top, middle, and bottom rows, respectively. FFs are depicted on the left, center, and right columns for the charged V mesons, neutral V mesons consisting of identical flavored quarks and the neutral V mesons composed of different flavored quarks, respectively. }\label{plotVE}
\end{figure}
The asymptotic limit of QCD reported in Ref.~\cite{Brodsky:1992px} predicts 
\bea
  G_{E}(Q^2):G_{M}(Q^2):G_Q(Q^2) \stackrel{Q^2 \rightarrow \infty}{=}
  1 - \frac{2}{3} \eta : 2 : -1 \,.
 \eea
Although none of these predictions are strictly satisfied in CI computation, as noted in Ref.~\cite{Roberts:2011wy,Raya:2017ggu}, this relation is recovered for $\Lambda_{\rm UV} \rightarrow \infty$ but at the cost of a logarithmic divergence in the individual EFFs. Therefore, we conclude that a CI cannot reasonably be regularized in a manner fully consistent with the constraints of asymptotic QCD. A more involved model must be constructed to recover the predictions of Ref.~\cite{Brodsky:1992px} exactly.
As has been seen in several works including Ref.~\cite{Roberts:2011wy}, the electric FF of the $\rho$-meson might have a zero crossing, a behavior not found for S and PS mesons~\cite{Hernandez-Pinto:2023yin}. 
Similarly, we find that the electric FF of mesons composed of light quarks, the $K_1$ and the $\phi$ mesons, present a zero crosssing, {\em albeit} at increasing values of momentum transfer, $Q^2\simeq7,10$ GeV$^2$, respectively. However, in the case of mesons composed by at least one heavier quark, $c$ or $b$, we do not observe any zero crossing even for large values of momentum transfer $Q^2\simeq50$ GeV$^2$, a similar behavior was observed in Ref.~\cite{Raya:2017ggu} within the CI for $J/\psi$ and $\Upsilon$ mesons.

We can now readily compute the charge, magnetic and quadrupole radii, using 
\begin{equation}\label{fradii}
 r_{i}^2 =
-\frac{6}
{G_i(0)}\left.\frac{\mathrm{d}G_{i}(Q^{2})}{\mathrm{d}Q^{2}}\right|_{Q^{2}=0}
\,,
\end{equation}
with $i\in \{E,M,Q\}$. It is important to remark that positive slopes shall appear in some EFFs; in those cases, the definition for charge radii is opted to remove the minus sign in the equation to avoid imaginary numbers. Since the properties of V mesons are also under investigation by several other groups, we present  in Tab.~\ref{tab:all_V_mesons_explicit_refs} a comparison of charge radii with the NJL model~\cite{Luan:2015goa}, with the method based on the Bag model \cite{Simonis:2016pnh} and in a symmetry-preserving approach to the two valence-body continuum bound-state problem \cite{Bhagwat:2006pu}. Experimental data on these static properties are absent to date. Therefore, results from other theoretical models are included in the table. 
\begin{table}[ht]
\caption{ \label{tab:all_V_mesons_explicit_refs}
Electric ($r^E$), magnetic ($r^M$), and quadrupole ($r^Q$) radii (in fm), magnetic moments $\mu$,
and quadrupole moments $\mathcal{Q}$ for V mesons.
 The results are compared with the NJL model \cite{Luan:2015goa}, bag model \cite{Simonis:2016pnh} and
a symmetry-preserving continuum approach (SPCA) \cite{Bhagwat:2006pu}.}
\begin{adjustwidth}{-\extralength}{-4cm}
\setlength{\tabcolsep}{3pt}
\centering
\begin{tabular}{@{\extracolsep{-0.1cm}}lcccccccccc}
\hline\hline
 & $\rho$ & $K^{*}$ & $\phi$ & $D^{*}$ & $D_s^{*}$ & $B^{*}$ & $B_s^{*}$ & $B_c^{*}$ & $J/\psi$ & $\Upsilon$ \\
\hline
\multicolumn{11}{l}{\rule{0ex}{3.0ex}Electric radius $r^{E}$} \\
CI
& 0.56 & 0.54 & 0.47 & 0.42 & 0.37 & 0.34 & 0.30 & 0.29 & 0.35 & 0.28 \\
NJL
& 1.12 & 1.05 & -- & 0.72 & 0.49 & 0.96 & -- & -- & -- & -- \\
SPCA
& 0.54 & 0.43 & -- & -- & -- & -- & -- & -- & 0.052 &  \\
\hline
\multicolumn{11}{l}{\rule{0ex}{3.0ex}Magnetic radius $r^{M}$} \\
CI
& 0.75 & 0.73 & 0.63 & 0.82 & 0.59 & 0.82 & 0.83 & 0.56 & 0.49 & 0.40 \\
\hline
\multicolumn{11}{l}{\rule{0ex}{3.0ex}Quadrupole radius $r^{Q}$} \\
CI
& 0.47 & 0.47 & 0.40 & 0.57 & 0.36 & 0.63 & 0.60 & 0.38 & 0.28 & 0.22 \\
\hline
\multicolumn{11}{l}{\rule{0ex}{3.0ex}Magnetic moment $\mu$} \\
CI
& 2.11 & 2.18 & 2.09 & $-1.51$ & 2.10 & 3.80 & $-1.82$ & 2.95 & 2.03 & 2.01 \\
NJL
& 2.54 & 2.26 & -- & 1.16 & 0.98 & 1.47 & --& -- & -- &-- \\
Bag
& 2.10 & 2.06 & -- & $-1.21$ & 0.87 & 1.47 & $-0.48$ & 0.35 &  & -- \\
SPCA
& 2.01 & 2.23 & -- & -- & -- & -- & -- & -- & 2.13 & -- \\
\hline
\multicolumn{11}{l}{\rule{0ex}{3.0ex}Quadrupole moment $\mathcal{Q}$} \\
CI
& $-0.85$ & $-0.90$ & $-0.83$ & 1.05 & $-0.71$ & $-2.06$ & 1.20 & $-1.37$ & $-0.70$ & $-0.69$ \\
SPCA
& $-0.41$ & $-0.38$ & -- & -- & -- & -- & -- & -- & $-0.28$ & -- \\
\hline\hline
\end{tabular}
\end{adjustwidth}
\end{table}
Nonetheless, wherever possible, we compare with the available data. We highlight that for a structureless spin-$1$ particle where the magnetic and quadrupole moments take the value $ \mu = 2$ and
$\mathcal{Q}=-1$ 
\cite{Brodsky:1992px}, any deviation from these values points to the internal structure of mesons.
From the tabulated results, we can infer that the heavy mesons are nearly point-like within the CI, as expected and as was reported in Ref.~\cite{Raya:2017ggu}. However, the heavy-light system deviates from this premise.
The charge radii of V mesons are consistently larger than PS mesons.\\
To analyze the behavior of the electric FFs of V mesons at large $Q^2$, we use a  parametrization similar to the EFFs of S and PS mesons, i.e., 
\begin{align}
G_{E}^{ V}(Q^2) = \frac{e_M+a_V^E Q^2 + b_V^E Q^4}{1+ c_V^E Q^2 + d_V^E Q^4} \, .
\label{eqE}\end{align}
In analogy, in the case of magnetic and quadrupole FFs, the parametrizations adopted are 
\begin{align} 
G_{M}^{V}(Q^2) &= \frac{\mu_V + a_V^M Q^2 +b_V^M Q^4}{1+ c_V^M Q^2 + d_V^M Q^4} \, , \\
\label{eqMQ} G_{Q}^{V}(Q^2) &= \frac{\mathcal{Q}_{V} +a_V^Q Q^2 + b_V^Q Q^4}{1+ c_V^Q Q^2 +d_V^Q Q^4} \, ,
\end{align}
where $a_i^V$, $b_i^V$, $c_i^V$ and $d_i^V$ with $i\in \{ E, M, Q\}$ are the coefficients to be fitted with CI numerical results.
 In~Table \ref{tableVEMQ} we present these values for all the indicated coefficients of ground state V mesons. This parametric form is valid over the range $Q^2\in [0,8 m_H^2 ]$. From the results in Eqs.~(\ref{eqE})-(\ref{eqMQ}), one finds that all $b_V$ components of the V meson vanish. Consequently, the corresponding EFFs exhibit an asymptotic behavior proportional to $1/Q^{2}$ in the limit of large $Q^{2}$.
\begin{table}[ht]
\caption{\label{tableVEMQ}Parameters for the fits in Eqs.~(\ref{eqE})-(\ref{eqMQ}), for the electric, magnetic and quadrupole FFs of V mesons. All numerical values are rounded to two digits after the decimal point.}
\begin{adjustwidth}{-\extralength}{-4.8cm}
\begin{center}
 \renewcommand{\arraystretch}{1.6} %
\begin{tabular}{@{\extracolsep{-0.15 cm}}ccccccccccccc}
\hline \hline
 & $a_V^E$ &  $b_V^E$ &  $c_V^E$ &  $d_V^E$ & $a_V^M$ &  $b_V^M$ &  $c_V^M$ &  $d_V^M$ & $a_V^Q$ &  $b_V^Q$ &  $c_V^Q$ &  $d_V^Q$   \\ 
\hline
$\rho$ & $-0.11$ & $-0.01$ & 1.25 & 0.01 & $-0.13$ & $-0.01$ & 1.09 & $-0.11$ & 0.03 & 0.00 & 1.10 & $-0.08$ \\
$K_1$ & $-0.11$ & $-0.01$ & 1.13 & $-0.01$ & $-0.30$ & 0.00 & 0.77 & $-0.11$ & 0.06 & 0.00 & 1.01 & $-0.10$ \\
$\phi$ & $-0.22$ & 0.01 & 0.72 & $-0.08$ & $-0.17$ & $-0.00$ & 0.75 & $-0.08$ & 0.08 & 0.00 & 0.71 & $-0.09$ \\
$D^{*0}$ & 0.83 & 0.00 & 1.38 & 0.29 & 1.00 & 0.00 & 1.32 & 0.33 & 0.04 & 0.00 & 1.39 & 0.50 \\
$D^*_{\ts}$ & $-0.03$ & 0.00 & 0.51 & $-0.01$ & $-0.05$ & 0.00 & 0.68 & $-0.01$ & $-0.40$ & 0.00 & 1.41 & 0.45 \\
$\Jpsi$ & 0.97 & $-0.01$ & 1.43 & 0.72 & 0.99 & 0.01 & 0.98 & 0.38 & $-0.85$ & $-0.00$ & 1.64 & 0.71 \\
$B^{+*}$ & 0.15 & 0.00 & 0.48 & 0.01 & 0.70 & 0.00 & 0.89 & 0.04 & $-0.28$ & $-0.00$ & 0.92 & 0.06 \\
$B^{0*}_{\ts}$ & 0.15 & 0.00 & 0.37 & 0.03 & 0.45 & 0.00 & 1.20 & 0.09 & $-0.01$ & 0.00 & 1.40 & 0.04 \\
$B^{0*}_{\tc}$ & 0.07 & 0.00 & 0.30 & 0.01 & 0.43 & 0.00 & 0.54 & 0.02 & $-0.16$ & 0.00 & 0.51 & 0.02 \\
$\Upsilon$ & 0.05 & 0.00 & 0.39 & 0.01 & 0.14 & 0.00 & 0.44 & 0.01 & $-0.05$ & 0.00 & 0.37 & 0.01 \\
\hline \hline
\end{tabular}
\end{center}
\end{adjustwidth}
\end{table}

In Fig.~\ref{rho}, we plot the electric, magnetic and quadrupole FFs for the $\rho$-meson, allowing for a $5\%$ variation in the value of the charge radius of the electric FFs. It is primarily controlled by varying $\tau_{\rm UV}$.
The behavior of the $\rho$-meson EFFs is, as expected, in perfect agreement with the previous work reported with the CI~\cite{Roberts:2011wy}. We also plot the magnetic and quadrupole FFs. We emphasize that all curves exhibit a pole at $Q^2 = -m_V^2$ because of the behavior of the quark-photon vertex in the time-like region. We also display an explicit comparison with the earlier results reported by lattice groups~\cite{QCDSF:2008tjq,Shultz:2015pfa}.
It is important to emphasize that lattice QCD provides the most direct nonperturbative access to the dynamics of QCD and therefore constitutes a fundamental benchmark for continuum and effective approaches. In this context, the contact interaction framework should be regarded as a symmetry-preserving effective model that captures dominant infrared features of QCD, while relying on simplified interaction kernels.
\begin{figure}[t!]
\begin{adjustwidth}{-\extralength}{-3cm}
\begin{center}
\begin{tabular}{@{\extracolsep{-2.3 cm}}cc}
\hspace{-4mm}
\includegraphics[scale=0.53]{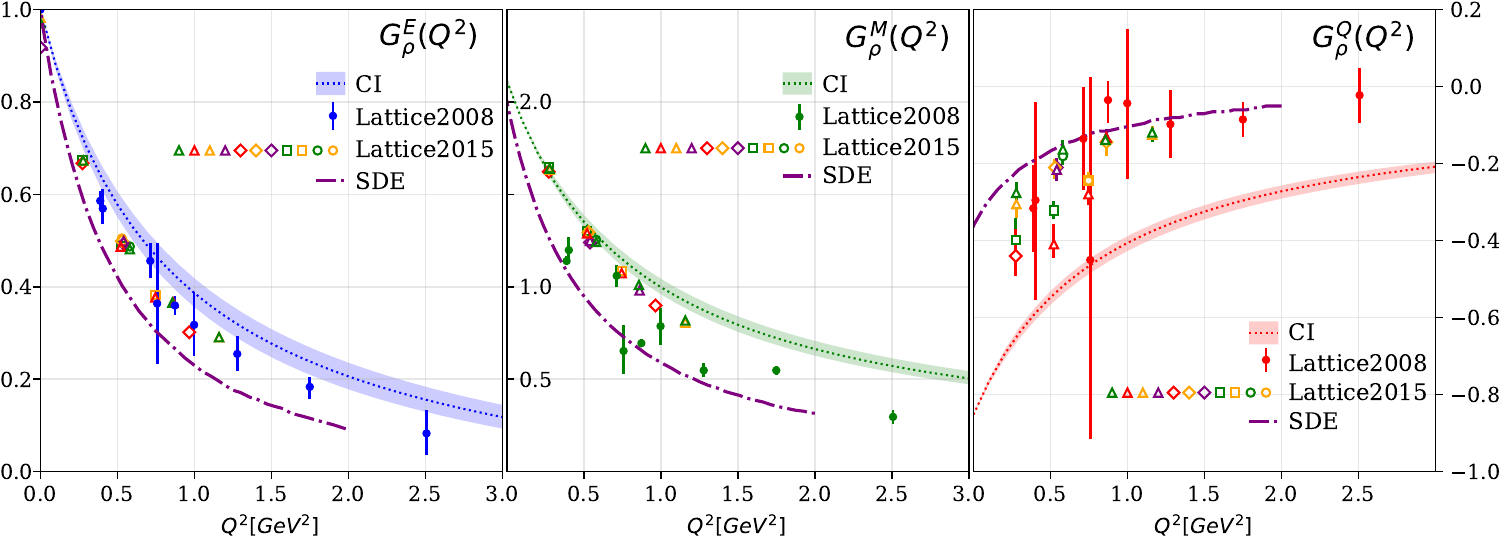}
\end{tabular} 
\end{center}
\end{adjustwidth}
\caption{\justifying
Electromagnetic (blue), Magnetic (green), and quadrupole (red) FFs for $\rho$-meson. The central curve in each case is obtained using the $\tau_{\rm UV}$ value from Tab.~\ref{parFFn}. The width of the band represents a $5\%$ variation in the charge radius. The meson life is so short that it is challenging to carry out experimental measurements of its EFFs. We compare CI results with those obtained from lattice QCD~\cite{QCDSF:2008tjq,Shultz:2015pfa} and the 
    SDE~\cite{Xu:2024fun}.}\label{rho}
\end{figure}
This completes our analysis of elastic meson FFs. The AV meson sector remains under active investigation, and, to the best of our knowledge, no corresponding elastic FF results have been reported in the literature up to the time of completion of this work. We therefore turn, in the next section, to the analysis of diquark form factors.
\subsection{Form Factors of Diquarks}
While the EIC Yellow Report does not address diquarks as explicit dynamical degrees of freedom, it
strongly motivates theoretical studies of quark correlations through high-precision measurements of
hadron structure, providing indirect constraints on diquark dynamics.
The interaction vertex that ensures current conservation for baryon FFs within the quark-diquark framework explicitly incorporates diquark FFs. In this approach, one must account for photon couplings to the dressed quark, to the diquark in elastic scattering, and for transitions between S and AV diquarks.
In order to compute the electromagnetic properties of nucleons (and, more generally, other baryons), it is necessary to understand how photons couple to diquark correlations. Since diquarks are not observable states, this information cannot be extracted directly from experiment; instead, one must rely on theoretical calculations of their electromagnetic structure. In this context, we examine the EFFs of diquarks using the CI \cite{Roberts:2011cf, Gutierrez-Guerrero:2021rsx}. CI results may provide a useful benchmark for future developments and improvements in the study of electromagnetic properties of baryons.
Since diquarks are evaluated in both the light- and heavy-flavor sectors, this framework also establishes a foundation for the systematic study of heavy-baryon FFs.\\
Within the CI framework defined in Eq.~(\ref{eqn:contact_interaction}), the relation between meson and diquark 
correlations has been extensively discussed in Ref.~\cite{Roberts:2011cf}. From those results, it follows that the 
structure of the EFFs in the diquark sector largely mirrors that of their meson partners. 
Under the impulse approximation, the photon-diquark interaction is described by a diagram analogous to Fig.~\ref{vertex-1}, 
corresponding to the photon coupling to each of the constituent quarks.
Moreover, within the rainbow-ladder truncation it becomes particularly evident that the elastic form factor of the 
S $[\tu\td]$ diquark can be obtained directly from the well-known expression for the pion form factor 
$F_\pi^{\rm em}(Q^2)$ through a simple set of substitutions in the BSAs and in the 
corresponding diquark mass.
Explicitly,
\begin{equation}
F^{0^+}_{[\tu\td]}(Q^2)
= \frac{1}{3}\,
F_\pi^{\rm em}(Q^2)\bigg|_{(E_\pi,F_\pi)\rightarrow
\sqrt{\tfrac{2}{3}}\,(E^{0^+}_{[\tu\td]},F^{0^+}_{[\tu\td]}),
\; m_\pi \rightarrow m^{0^+}_{[\tu\td]}}\, .
\label{eq:70}
\end{equation}
This relation reveals how closely the internal structure of S quark pairs resembles that of PS mesons, 
supporting the notion that diquarks may be viewed as color-antisymmetric analogues of mesons in a different kinematic 
configuration. The results for the EFFs of  S and PS diquarks are 
shown in Figs.~\ref{plotDS} and \ref{plotDPS}.

\begin{figure}[h!]
\begin{adjustwidth}{-\extralength}{-5cm}
\begin{center}
\begin{tabular}{@{\extracolsep{-2.3 cm}}c}
 \renewcommand{\arraystretch}{-1.6} %
 \hspace{-0.2cm}
 \includegraphics[scale=0.53]{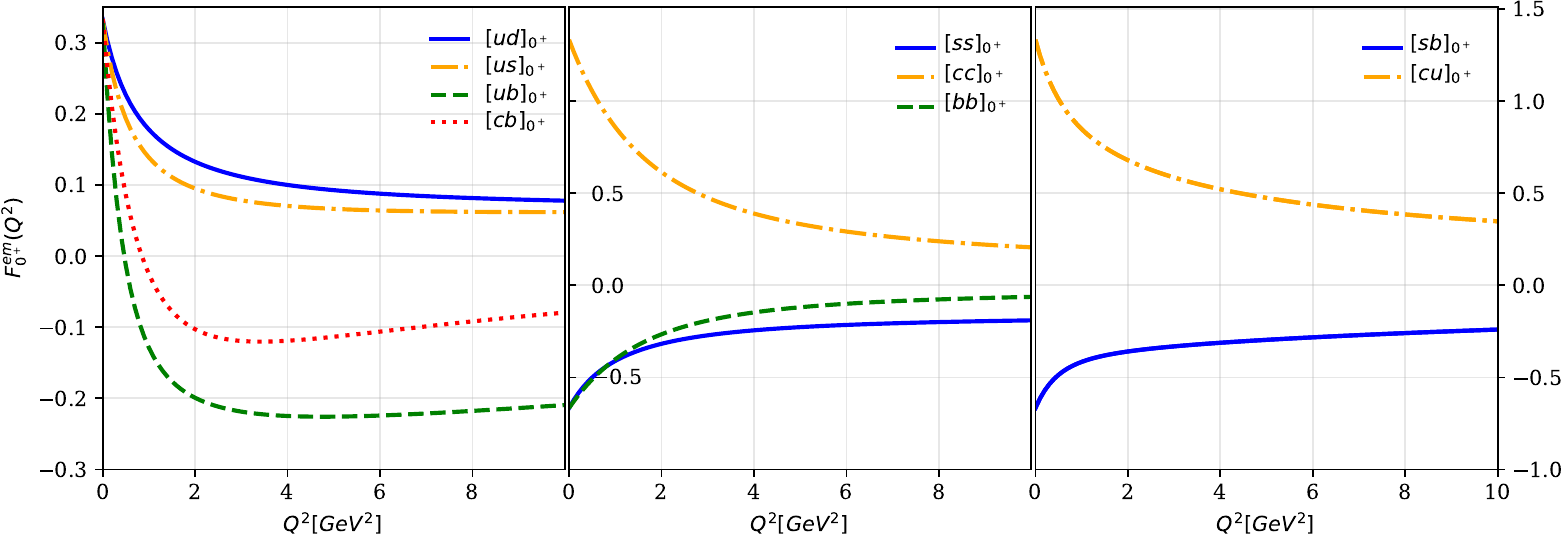}
\end{tabular}
\end{center}
\end{adjustwidth}
\caption{
\justifying
As we anticipated and expected, the CI results are harder due to its point-like interaction. This behaviour is typical of the CI model and it is irrespective of the nature of the mesons under study. 
EFFs for S diquarks in the CI model. Left panel: diquarks composed of different flavored quarks with total charge of the system of 1/3. Central panel: diquark system composed of the same flavored quarks. Right panel: diquarks composed of different flavored quarks with total charge of the system of $-$2/3 and 4/3.
\label{plotDS}}
\end{figure}
\begin{figure*}[t!]
\begin{adjustwidth}{-\extralength}{-5cm}
\begin{center}
\begin{tabular}{@{\extracolsep{-2.3 cm}}c}
 \renewcommand{\arraystretch}{-1.6} %
 \hspace{-0.5cm}
 \includegraphics[scale=0.53]{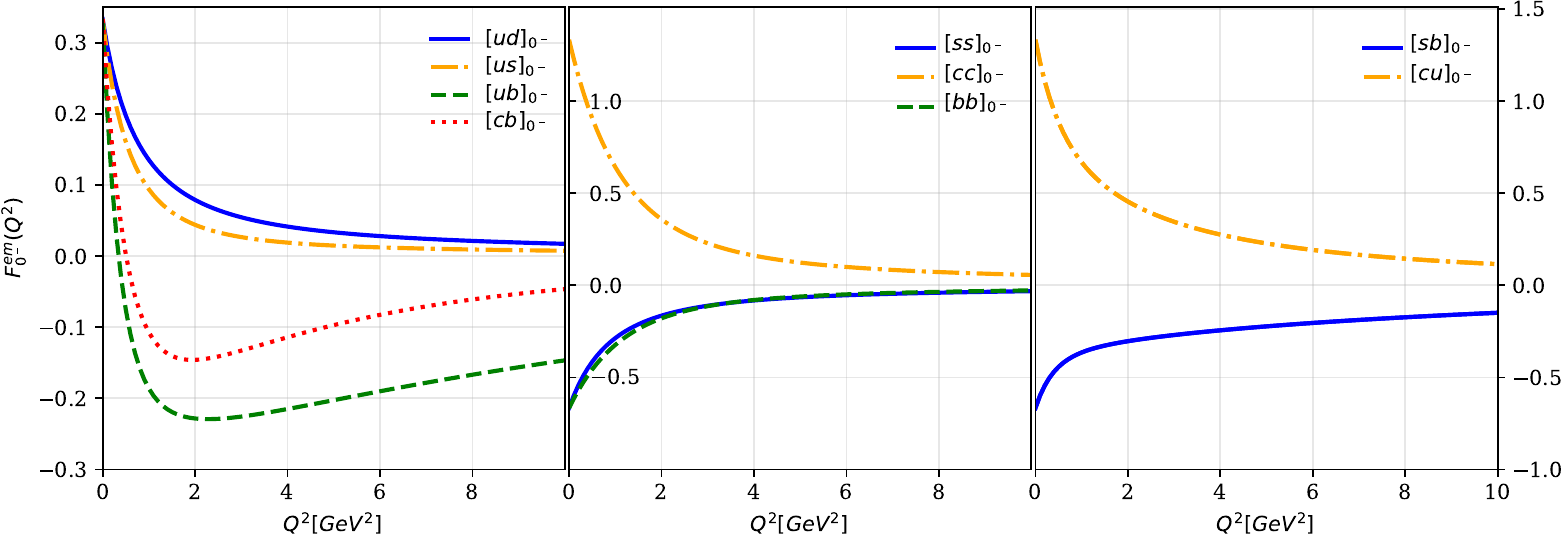}
\end{tabular}
\end{center}
\end{adjustwidth}
\caption{
\justifying
EFFs for PS diquarks in the CI. Left panel: diquarks composed of different flavored quarks with total charge of the system of 1/3. Central panel: diquark system composed of the same flavored quarks. Right panel: diquarks composed of different flavored quarks with total charge of the system of $-$2/3 and 4/3.
}\label{plotDPS}
\end{figure*}
Analogously to the meson case, the charge radii of S and PS diquarks can be obtained from their 
EFFs. Once these FFs are calculated, the mean squared charge radius is determined 
using the standard expression:
\begin{equation}
r_{0^{\pm}}^{2} =
-6\left.\frac{\mathrm{d}F^{0^{\pm}}(Q^{2})}{\mathrm{d}Q^{2}}\right|_{Q^{2}=0}
\, .
\end{equation}
This relation connects the slope of the form factor at vanishing momentum transfer with the effective spatial size 
of the bound state, allowing us to quantify the extent of the charge distribution inside the diquark.
In Table~\ref{tab:charge_radii-pss}, we present the charge radii for S and PS diquarks.
\begin{table}[htbp]
\centering
\caption{\label{tab:charge_radii-pss}
Charge radii (fm) for S and PS diquarks.}
\begin{adjustwidth}{-\extralength}{-4cm}
\begin{center}
\begin{tabular}{@{\extracolsep{0.1 cm}}lcccccccccc}
\hline\hline
\rule{0ex}{2.8ex}
Channel & $\tu\td$ & $\tu\ts$ & $\ts\ts$ & $\tc\tu$ & $\tc\ts$ & $\tu\tb$ & $\ts\tb$ & $\tc\tb$ & $\tc\tc$ & $\tb\tb$ \\
\hline
\rule{0ex}{2.8ex}
Diquark S  & 0.48 & 0.57 & 0.40 & 0.41 & 0.38 & 0.91 & 0.45 & 0.71 & 0.34 & 0.36 \\
\rule{0ex}{2.8ex}
Diquark PS & 0.31 & 0.37 & 0.50 & 0.55 & 0.24 & 0.59 & 0.42 & 0.50 & 0.43 & 0.42 \\
\hline\hline
\end{tabular}
\end{center}
\end{adjustwidth}
\end{table}
Estimates for the spatial extent of diquark correlations have been reported in the literature. 
For instance, an early Faddeev-equation analysis of nucleon form factors~\cite{Maris:2004bp} 
reported a scalar diquark charge radius of $r_{[ud]_{0^+}} = 0.71~\mathrm{fm}$, which is approximately $8\%$ larger than the pion charge radius, $r_\pi = 0.66~\mathrm{fm}$. 
Within the CI framework, we obtain $r_\pi = 0.45~\mathrm{fm}$ and $r_{[ud]_{0^+}} = 0.48~\mathrm{fm}$, indicating that the scalar diquark radius is likewise larger than that of the pion, with a comparable relative difference of about $7\%$. 
We note that both CI results and DSE predictions for diquark electromagnetic radii exhibit the same ordering~\cite{Maris:2004bp}, namely 
$r_{[ud]} > r_{\pi}$.

Similarly, the elastic FFs of AV diquarks can be obtained from those of V mesons. 
This mapping allows one to reuse well–established results in the V sector, requiring only minimal 
parameter substitutions. Consequently,
\begin{equation}
F^{1^+}_{\{\tu\td\},\,j}(Q^2)
= \frac{1}{3}\,
F_j(Q^2)\bigg|_{m_\rho \rightarrow \sqrt{\tfrac{2}{3}}\,m^{1^+}_{\{\tu\td\}}}\, .
\label{eq:fdav}
\end{equation}
The momentum dependence of the FFs for the flavour-symmetric $\{uu\}$ and $\{dd\}$ AV diquarks
is identical; however, their normalisation differs owing to charge and flavour composition, taking values
$\tfrac{4}{3}$ and $-\tfrac{2}{3}$, respectively.\\ \\
For AV diquarks, the electric, magnetic, and quadrupole FFs can be computed 
using Eq.~(\ref{eq:fdav}) together with Eqs.~(\ref{Gsm}). The resulting FFs are shown in Fig.~\ref{plotDAV}. 
We observe that their behaviour resembles that of V mesons, though remains clearly 
distinguishable, since AV diquark FFs fall off more slowly due to their larger mass.
\begin{figure}[t]
\begin{adjustwidth}{-\extralength}{-4cm}
\begin{center}
\vspace{5mm}
\includegraphics[scale=0.53]{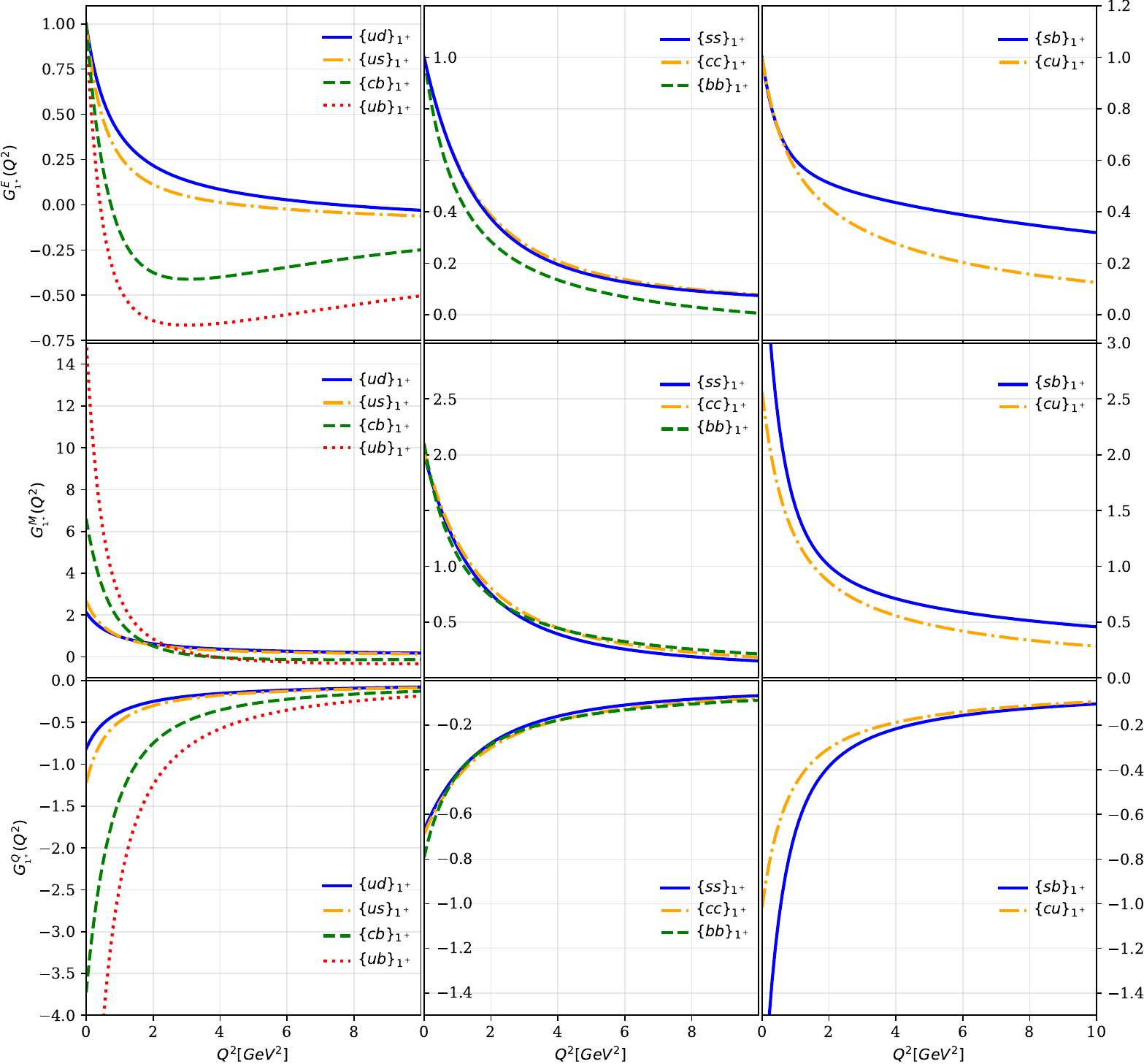}
\end{center}
\end{adjustwidth}
\caption{\justifying
AV diquarks electric, magnetic, and quadrupole FFs are displayed in the top, middle, and bottom rows, respectively. Left panel: diquarks composed of different flavored quarks
with total charge of the system of 1/3. Central panel: diquark system composed of the same flavored
quarks. Right panel: diquarks composed of different flavored quarks with total charge of the system
of $-$2/3 and 4/3.\label{plotDAV}}
\end{figure}
In the limit $Q^2 \to 0$, these EFFs define the charge, magnetic moment ($\mu_{DAV}$), 
and quadrupole moment ($\mathcal{Q}_{DAV}$) of AV diquarks, in complete analogy with the 
V-meson case as expressed in Eq.~(\ref{chargenorm}). 
In Table~\ref{tab:mesons_diquarks_radii} we present the CI results for the electric, magnetic, and quadrupole charge radii, 
as well as the magnetic and quadrupole moments of AV diquarks. 
\begin{table}[htbp]
\caption{
Electric, magnetic, and quadrupole radii (in fm), magnetic moments $\mu_{DAV}$, and quadrupole moments $\mathcal{Q}_{DAV}$
for AV diquarks composed of $\tu,\td,\ts,\tc,\tb$ quarks, obtained within the CI framework.
%
}
\label{tab:mesons_diquarks_radii}
\begin{adjustwidth}{-\extralength}{-4cm}
\begin{center}
\begin{tabular}{@{\extracolsep{0.2 cm}}lccccc}
\hline
\hline
\rule{0ex}{3.0ex}
 & $r^E$ & $r^M$ & $r^Q$ & $\mu$ & $\mathcal{Q}$ \\
\hline
\rule{0ex}{3.0ex}
$\{\tu\td\}_{1^+}$   
& 0.555 & 0.758 & 0.465 & 2.13 & $-0.81$ \\
\rule{0ex}{3.0ex}
$\{\tu\ts\}_{1^+}$   
& 0.647 & 0.962 & 0.628 & 2.68 & $-1.22$ \\
\rule{0ex}{3.0ex}
$\{\ts\ts\}_{1^+}$   
& 0.391 & 0.535 & 0.325 & 2.10 & $-0.79$ \\
\rule{0ex}{3.0ex}
$\{\tc\tu\}_{1^+}$   
& 0.445 & 0.785 & 0.512 & 2.56 & $-1.01$ \\
\rule{0ex}{3.0ex}
$\{\tc\ts\}_{1^+}$   
& 0.405 & 0.836 & 0.586 & $-0.26$ & 0.88 \\
\rule{0ex}{3.0ex}
$\{\tu\tb\}_{1^+}$   
& 0.946 & 2.665 & 1.800 & 15.1 & $-8.41$ \\
\rule{0ex}{3.0ex}
$\{\ts\tb\}_{1^+}$   
& 0.467 & 1.222 & 0.844 & 4.24 & $-2.02$ \\
\rule{0ex}{3.0ex}
$\{\tc\tb\}_{1^+}$   
& 0.743 & 1.506 & 1.015 & 6.61 & $-3.72$ \\
\rule{0ex}{3.0ex}
$\{\tc\tc\}_{1^+}$   
& 0.428 & 0.603 & 0.335 & 2.02 & $-0.68$ \\
\rule{0ex}{3.0ex}
$\{\tb\tb\}_{1^+}$     
& 0.302 & 0.427 & 0.234 & 2.00 & $-0.66$ \\
\hline
\hline
\end{tabular}
\end{center}
\end{adjustwidth}
\end{table}
In general, AV diquarks exhibit larger magnetic and quadrupole radii than their V-meson partners,
particularly in heavy-quark sectors, where the spatial extension of the correlations becomes noticeably larger.
Only in a few light-flavour channels (notably $ss$ and, to a lesser extent, $cu$) do mesons present comparable
or slightly larger radii. This pattern supports the picture in which diquarks are less compact than mesons and
their electromagnetic distributions are more spread out, especially under the presence of heavy quarks.
Magnetic and quadrupole moments display clear systematic differences between V mesons and AV diquarks. 
For light flavours, both $\mu$ and $\mathcal{Q}$ remain comparable in magnitude; however, in heavy-quark sectors the diquark 
moments increase substantially, in several cases even changing sign relative to the meson values. This behaviour 
suggests that AV diquarks possess a more pronounced and flexible internal electromagnetic structure, with stronger 
magnetic response and larger quadrupole deformation than their meson partners.
\section{Conclusions and Outlook}
In this review, we have provided a comprehensive overview of the symmetry preserving CI framework and its application to the description of meson and diquark properties across a broad range of quark masses. The defining features of the CI limit its quantitative predictive power at large momentum transfer $Q^{2}$. At the same time, these same features clearly delineate the kinematic domain in which the approach can be applied most effectively, namely the low and intermediate momentum region where nonperturbative QCD dynamics are dominant.

By examining both the mass spectrum and elastic EFFs, we have highlighted the capacity of the CI to deliver a unified and internally consistent description of hadronic observables in the nonperturbative regime of QCD. Within this framework, we have identified a set of robust and physically meaningful results obtained using the CI approach, which can be summarized as follows.

\begin{itemize}

\item
The mass spectra discussed for all mesons presented in Sec.~\ref{masses-mesons} and in Tables~\ref{par-AllFF} and~\ref{tab:VAxialmasses} demonstrate that, despite its simplicity, the CI framework characterized by a momentum independent interaction kernel, provides a coherent and reliable description of the masses of mesons composed of both light and heavy quarks. The masses predicted within the CI are in excellent agreement with those obtained from other theoretical approaches and with the available experimental data in channels where measurements exist. In particular, the relative differences between experimental values and the corresponding CI results remain below $20\%$ in all channels~\cite{Gutierrez-Guerrero:2021rsx}.

\item
In Table~\ref{par-AllDiquarks}, we present the results for the masses of all spin zero and spin one diquark channels. These results play a central phenomenological role, as they provide the essential building blocks for the description of multiquark hadrons and, in particular, form the basis for computing baryon masses within the quark-diquark approximation. 
Lattice QCD studies~\cite{Alexandrou:2006cq,Francis:2021vrr,Francis:2022fdj,Francis:2022stl} provide direct evidence for strong diquark correlations inside baryons, supporting the existence of compact and attractive diquark configurations.
In these studies, S diquarks are found to be the lightest and most tightly bound configurations, while V and negative parity diquarks appear at higher energies. Although diquarks are not physical asymptotic states, these results confirm that QCD dynamically favors diquark correlations, supporting their effective role in the description of baryon structure.

\item
The study of PS and S meson FFs is presented in Sec.~\ref{FFS-1}, where their characteristic behavior is displayed in Figs.~\ref{plotPS} and~\ref{plotS}. The EFFs of the pion and the kaon are shown in Figs.~\ref{pionff} and~\ref{kaonff}, respectively, together with comparisons to results from other theoretical approaches and to available experimental data. As the momentum transfer increases, the EFFs tend to become nearly independent of $Q^{2}$, indicating the onset of an effective pointlike behavior. It is well established that the asymptotic FFs obtained within the CI framework exhibit a harder momentum dependence than those predicted by QCD based approaches~\cite{Roberts:2011cf,Bedolla:2015mpa,Raya:2017ggu}. This feature follows directly from the momentum independent structure of the interaction kernel, as made explicit in Eq.~(\ref{eqn:contact_interaction}).

\item
The three FFs of V mesons, namely the electric, magnetic, and quadrupole FFs, are analyzed in Fig.~\ref{plotVE}.

\item
For mesons, the CI description reproduces the expected hierarchies and mass radius trends, illustrated in Fig.~\ref{jera} and supported by the results presented in Tabs.~\ref{tab:chargeradiuspsu}, \ref{tab:S_fit_radii_vertical}, and \ref{tab:all_V_mesons_explicit_refs}. In particular, Eq.~(\ref{eq:hierarchies}) is satisfied for all mesons.
The charge radii extracted within the CI using Eq.~(\ref{radii}) are consistent with expectations from DSE and lattice QCD studies, as shown in Table~\ref{tab:all_V_mesons_explicit_refs}. In particular, the PS channels preserve the ordering reported in previous works.
\begin{figure}[htbp]
\begin{adjustwidth}{-\extralength}{-5cm}
\begin{center}
\includegraphics[width=0.35\textwidth]{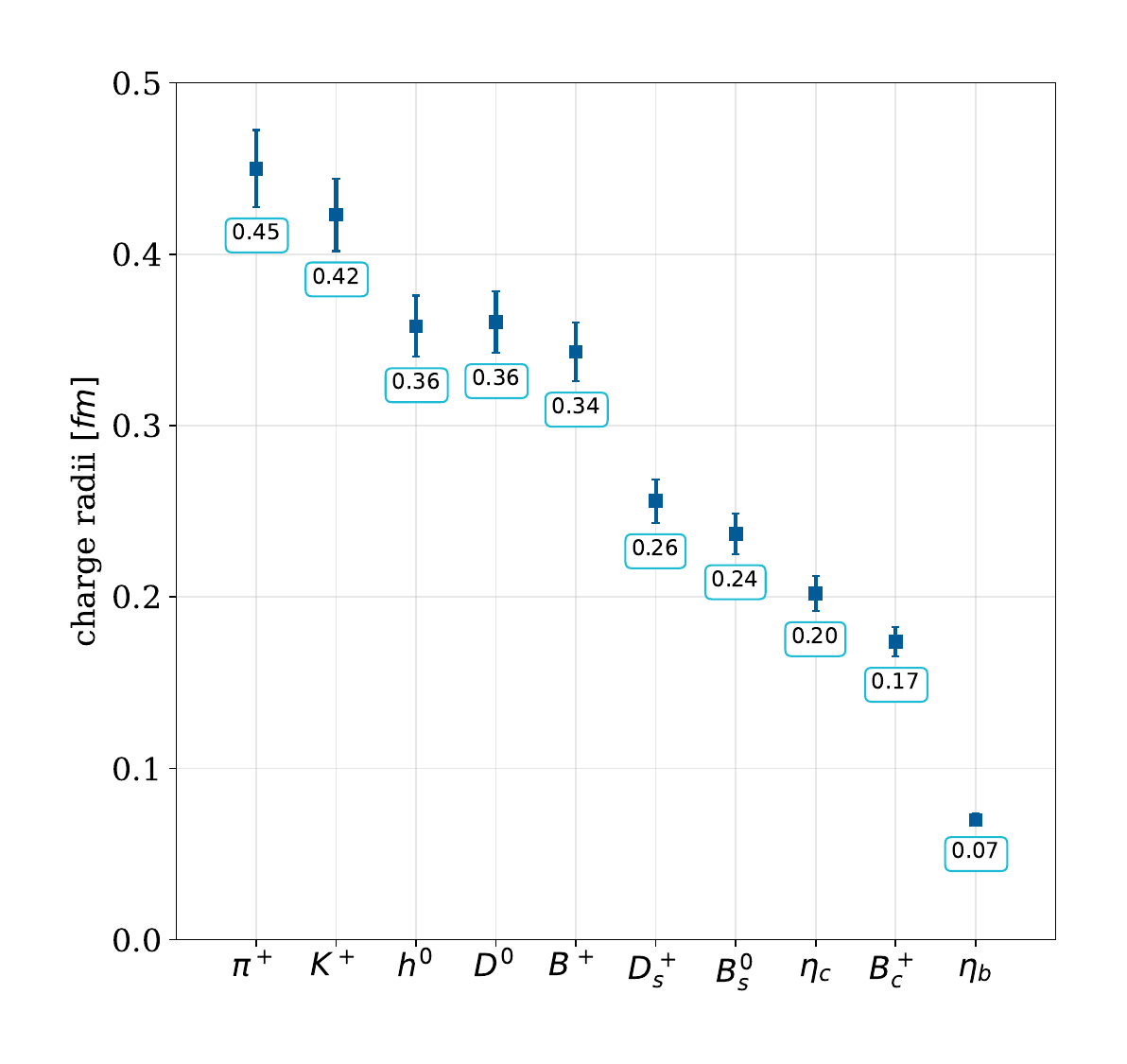}\hspace{-0.2cm}%
\includegraphics[width=0.35\textwidth]{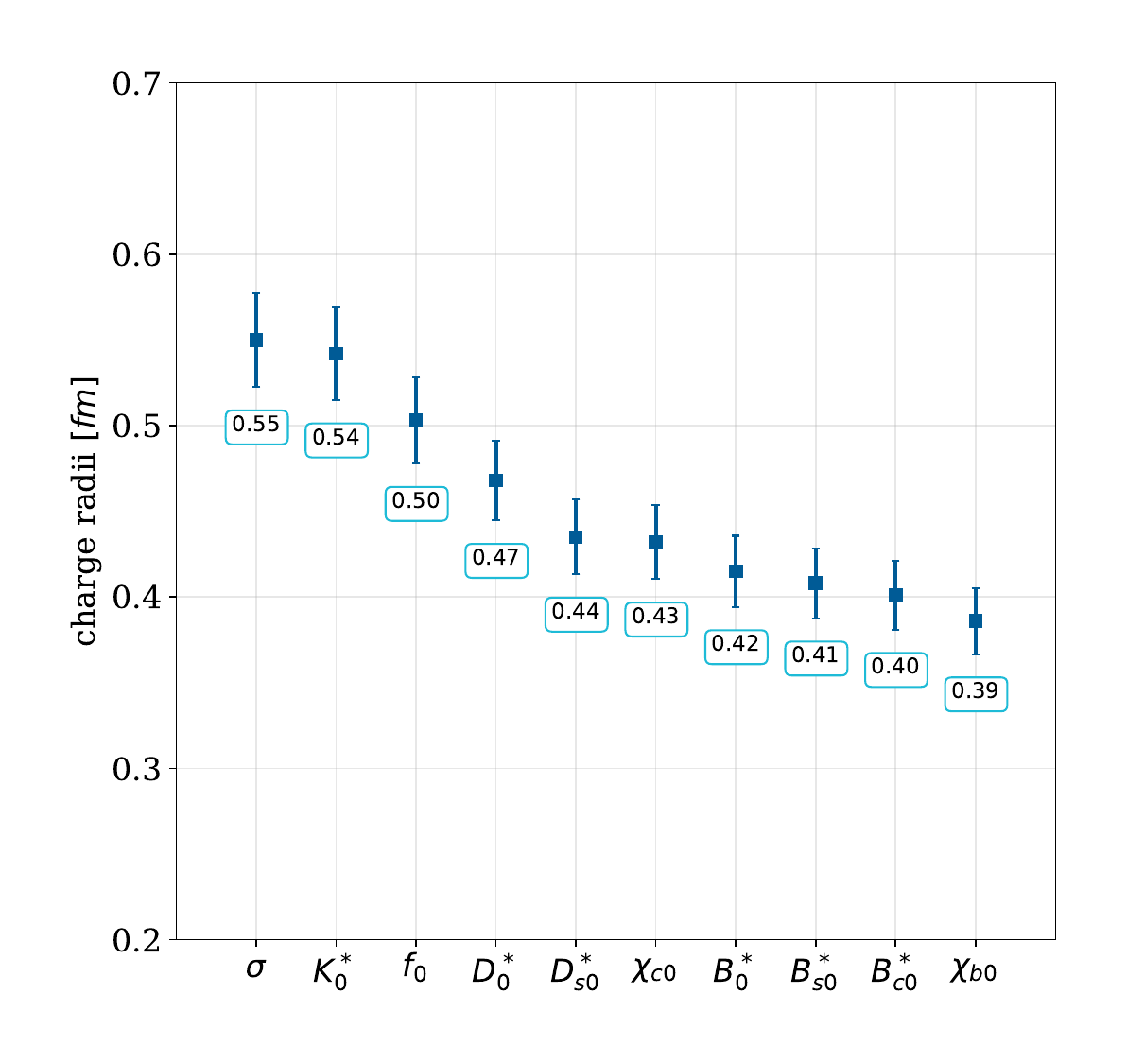}\hspace{-0.2cm}%
\includegraphics[width=0.35\textwidth]{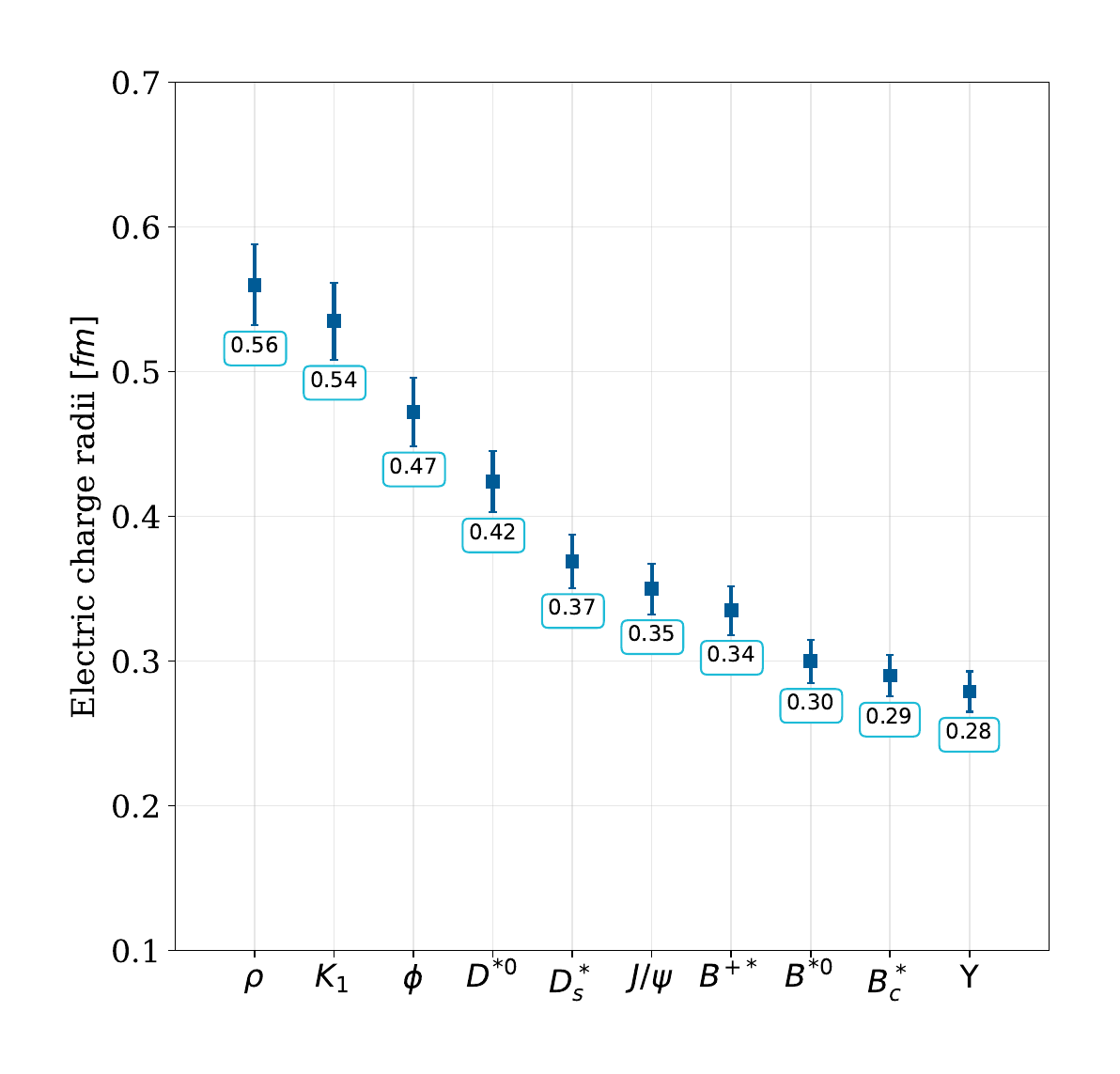}
\end{center}
 \end{adjustwidth}
    \caption{\label{jera}
    \justifying
    Charge radii of PS (left), S (center) and V (right) ground state mesons in the CI framework. It is clear from this plot that the radii tend to decrease as the constituent quarks' masses increase.}
    \label{fig:threefigsher}
\end{figure}

\item
In contrast to mesons, diquarks exhibit a more extended spatial structure. Our analysis shows that S and PS diquarks display larger charge radii than their meson partners in most flavor configurations, with differences becoming more pronounced in systems containing heavy quarks. For AV diquarks, the electric, magnetic, and quadrupole FFs can be obtained from the V meson case through simple parameter substitutions. Nevertheless, their behavior remains clearly distinct, as AV diquark FFs decrease more slowly with increasing $Q^{2}$, reflecting the larger masses and weaker binding of quark-quark correlations compared to quark-antiquark mesons.

\item
The quantitative comparison presented here reinforces the role of diquark correlations as fundamental effective degrees of freedom in Faddeev equation based descriptions of baryons. These results provide a consistent and testable input for future studies of hadron structure at finite temperature, at high density, and in phenomenological applications connected to experiments at Jefferson Lab and in heavy ion physics.
Lattice QCD studies~\cite{Alexandrou:2006cq} provide direct evidence for strong diquark correlations inside baryons, with S diquarks emerging as the most tightly bound configurations. Relativistic quark-diquark models further indicate that diquarks cannot be treated as pointlike constituents when describing baryon electromagnetic properties. Instead, diquarks possess their own EFFs, which encode their internal structure and effective dynamics. These FFs are essential for achieving a realistic description of nucleon electromagnetic observables over a wide range of momentum transfer, emphasizing the role of diquarks as extended and effective degrees of freedom within baryons rather than elementary objects~\cite{DeSanctis:2011zz}.
\end{itemize}
It is important to emphasize that the contact interaction (CI) framework should be understood as a symmetry-preserving effective approach that captures the dominant infrared features of QCD while employing a momentum-independent interaction kernel. As such, it does not incorporate the full running behavior of the interaction and, consequently, has limitations in describing the transition between infrared and ultraviolet regimes. In this context, lattice QCD provides a fundamental nonperturbative benchmark, and, whenever available, we have compared CI results with lattice data, finding qualitative agreement for a variety of low-energy observables. This agreement should not be interpreted as a proof of completeness of the approach, but rather as an indication that the CI model encodes essential aspects of the underlying dynamics within its domain of applicability. Future developments aimed at incorporating momentum-dependent effects and strengthening connections with lattice QCD are expected to further clarify both the scope and limitations of this framework.\\
In a broader context, the study of hadron structure, including masses and form factors, is part of a coordinated effort within the hadron-physics community aimed at understanding the emergence of mass and structure from QCD. In this endeavor, different theoretical approaches—such as lattice QCD, continuum SDE-BSE frameworks, and effective models—play complementary roles, each being applicable in distinct kinematic regimes and for different classes of observables. \\
Within this landscape, the contact interaction (CI) should be regarded as a symmetry-preserving and computationally efficient framework that captures the dominant infrared dynamics of QCD, particularly in the description of static properties and low--to--moderate momentum observables. At the same time, its intrinsic limitations highlight the importance of systematic comparisons with other approaches, especially lattice QCD, as well as the need to confront theoretical predictions with experimental data.\\
Future progress in the field will rely on strengthening these connections, extending calculations to heavy and heavy--light systems, and exploring a wider class of observables. In this regard, form factors remain key quantities, providing direct insight into the internal structure of hadrons and serving as essential benchmarks for assessing the reliability and domain of applicability of different theoretical frameworks.\\
Finally, the present work establishes a solid foundation for future studies based on the CI framework. These include the computation of FFs at finite temperature, the investigation of transition FFs, the analysis of processes involving transitions between excited hadronic states, and the exploration of the mass spectrum of hadrons containing top quarks.\\
This is a particularly stimulating period for hadron physics, shaped by the breadth and precision of current and forthcoming experimental programs. The continued development of high precision measurements at facilities such as FAIR, Jefferson Lab, and the Electron Ion Collider will provide increasingly stringent benchmarks for theoretical descriptions of hadron structure. Within this context, the CI offers a transparent and computationally economical framework in which qualitative features and systematic trends can be clearly identified and critically compared with experimental data, as well as with results from lattice QCD and other continuum approaches. In this way, the CI is expected to remain a valuable complementary tool in ongoing efforts to understand the emergent phenomena of QCD and the internal structure of hadrons.\\

\acknowledgments{
L.~X.~Guti\'errez-Guerrero acknowledges the {\em Secretaría de Ciencia, Humanidades, Tecnología e Inovación} (SECIHTI) for the support provided to her through the {\em Investigadores e Investigadoras por México} SECIHTI program and Project CBF2023-2024-268, Hadronic Physics at JLab: Deciphering the Internal Structure of Mesons and Baryons, from the 2023-2024 frontier science call. The work of R.~J.~Hern\'andez-Pinto is partly supported by SECIHTI (Mexico) through {\em Sistema Nacional de Investigadores}.}
\appendixtitles{no} 
\appendixstart
\appendix
\section[\appendixname~\thesection]{The Gap Equation}
\label{agap}

In this Appendix, we briefly summarize the derivation of the quark gap equation within the symmetry-preserving contact interaction (CI) framework. This provides the basis for determining the dressed quark mass, which constitutes a key input for all subsequent calculations of meson and diquark properties.
The SDE for the quark propagator is given by
\begin{eqnarray}
S_f^{-1}(p)  = i\gamma \cdot p + m_f + \Sigma(p) \, ,
\end{eqnarray}
where $m_f$ is the current quark mass and $\Sigma(p)$ denotes the quark self-energy. The latter encodes the interaction of the quark with the gluon field and is given by
\begin{eqnarray}
\Sigma(p) = \int \frac{d^4 q}{(2\pi)^4} \,
g^2 D_{\mu\nu}(p-q)\,\frac{\lambda^a}{2}\gamma_\mu \, S(q)\, \Gamma^a_\nu(q,p) \, ,
\end{eqnarray}
with $D_{\mu\nu}(k)$ the dressed gluon propagator and $\Gamma^a_\nu(q,p)$ the dressed quark-gluon vertex.
The quark propagator can be expressed in the general form
\begin{eqnarray}
S_f(p) = -i\gamma \cdot p \, \sigma_V(p^2) + \sigma_S(p^2)
= \frac{Z(p^2)}{i\gamma \cdot p + M_f(p^2)} \, ,
\end{eqnarray}
which defines the quark mass function $M_f(p^2)$ and the wave-function renormalization $Z(p^2)$. In particular,
\begin{eqnarray}
\sigma_S(p^2) = \frac{ M_f(p^2)}{p^2 + M_f^2(p^2)} \, .
\end{eqnarray}
Within the CI framework, the interaction kernel is simplified by employing the ansatz in \eqn{eqn:contact_interaction}, together with a bare quark--gluon vertex
\begin{eqnarray}
\label{eqn:contact_interaction-2}
\Gamma^a_\nu(p,q) &=& \frac{\lambda^a}{2} \gamma_\nu \, ,
\end{eqnarray}
which preserves the relevant symmetries while rendering the equations tractable. Using the identity $\sum_a \lambda^a \lambda^a = \frac{16}{3} I$, the gap equation becomes
\begin{eqnarray}
  && S_f^{-1}(p) =i \gamma \cdot p + m_f +  \frac{16\pi}{3}\hat{\alpha}_{\rm IR} \int\!\frac{d^4 q}{(2\pi)^4} \,
\gamma_{\mu} \, S_f(q) \, \gamma_{\mu}\,,\label{gap-2}
\end{eqnarray}
where $\hat{\alpha}_{\rm IR}$ encodes the effective interaction strength.\\
 This simplification allows for an analytic treatment of the gap equation.
We introduce the notation
$\int_q^\Lambda = \frac{1}{4\pi^2} \int d^4 q$. 
We then multiply the resulting equation by $\gamma \cdot p$ and take the Dirac trace
\begin{eqnarray}\label{eq:gap_mass}
{\rm Tr}\left[M_f(p)\right] &=& m_f 
+ \frac{16\pi}{3} \hat{\alpha}_{\rm IR}\int_q^\Lambda 
\frac{M_f(q)\,(\gamma_\mu \gamma_\mu = \delta_{\mu\mu} = 4)}
{-q^2 + M_f^2(q)} \, ,
\end{eqnarray}
where we have dropped the terms with odd number of $\gamma$ matrices, as they vanish under the Dirac trace.
To proceed, we carry out the angular integration. Since the integrand depends only on $q^2$, the angular part can be performed analytically. Writing the integration measure as
$d^4 q = dq\, q^3 \sin^2\theta\, d\theta \sin\phi\, d\phi\, d\psi$, one obtains
\begin{eqnarray}
d^4 q &=& dq\, q^3 \int_0^\pi \sin^2\theta\, d\theta \left(=\frac{\pi}{2}\right)
\int_0^\pi \sin\phi\, d\phi \left(=2\right)
\int_0^{2\pi} d\psi \left(=2\pi\right) \nonumber \\
&=& dq^2\, q^2 \pi^2 \, .
\end{eqnarray}
 To regulate the ultraviolet divergence and implement confinement, a proper-time regularization is introduced. This procedure removes the pole in the quark propagator, thereby ensuring that quarks do not appear as asymptotic states.Then we note
\begin{eqnarray}
\label{eq:proper_time_substitution1}
\int_a^b dx \, e^{-x(s+M)} 
= \frac{e^{-a(s+M)} - e^{-b(s+M)}}{M + s} \, .
\end{eqnarray}
$M(p)=M$ is a solution of the gap equation,Substituting Eq.~(\ref{eq:proper_time_substitution1}) into 
Eq.~(\ref{eq:gap_mass}), we obtain
\begin{eqnarray}
M_f = m_f + M_f\frac{4\hat{\alpha}_{\rm IR}}{3\pi}
\int_0^\infty ds \, \frac{s}{s+M_f^2}
\left[ e^{-(s+M_f^2)r_{\rm UV}^2} - e^{-(s+M_f^2)r_{\rm IR}^2} \right] \, .
\label{eq:gap_mass_regularized}
\end{eqnarray}
Making the change of variables $s+M_f^2 = s'$, we have
\begin{eqnarray}
\int_0^\infty ds \, \frac{s}{s+M_f^2}
\left[ e^{-(s+M_f^2)r_{\rm UV}^2} - e^{-(s+M_f^2)r_{\rm IR}^2} \right]
&=& \int_{M_f^2}^\infty ds' 
\left[ e^{-s'r_{\rm UV}^2} - e^{-s'r_{\rm IR}^2} \right] \nonumber \\
- M_f^2 \int_{M_f^2}^\infty \frac{ds'}{s'} e^{-s'r_{\rm UV}^2}
+ M_f^2 \int_{M_f^2}^\infty \frac{ds'}{s'} e^{-s'r_{\rm IR}^2} \, .
\label{eq:change_variable}
\end{eqnarray}
With another change of variables $s' r_{\rm UV}^2 = t$, and denoting the integral as 
$C(M_f^2; r_{\rm IR}^2, r_{\rm UV}^2)$, we have
\begin{eqnarray}
C(M_f^2; r_{\rm IR}^2, r_{\rm UV}^2) &=& 
- M_f^2 \Gamma\left(0, M_f^2 r_{\rm UV}^2\right)
+ M_f^2 \Gamma\left(0, M_f^2 r_{\rm IR}^2\right) \nonumber \\
&& + \frac{1}{r_{\rm UV}^2} e^{-M_f^2 r_{\rm UV}^2}
- \frac{1}{r_{\rm IR}^2} e^{-M_f^2 r_{\rm IR}^2} \, ,
\label{eq:C_function}
\end{eqnarray}
where
\begin{eqnarray}
\Gamma(\alpha,y) = \int_y^\infty dt \, t^{\alpha-1} e^{-t} \, .
\label{eq:gamma_function}
\end{eqnarray}
With partial integration of the first term in Eq.~(\ref{eq:C_function}), we can also write
\begin{eqnarray}
C(M_f^2; r_{\rm IR}^2, r_{\rm UV}^2) &=& 
M_f^2 \Gamma\left(-1, M_f^2 r_{\rm UV}^2\right)
- M_f^2 \Gamma\left(-1, M_f^2 r_{\rm IR}^2\right)\, .\equiv C(M_f^2)
\label{eq:C_function_alt}
\end{eqnarray}
Therefore, the gap equation reduces to the algebraic expression
\begin{equation}
M_f = m_f + M_f\frac{4\hat{\alpha}_{\rm IR}}{3\pi}\,\,{\cal C}(M_f^2)\,,
\label{gapactual2}
\end{equation}
which determines the dressed quark mass self-consistently. This quantity serves as a fundamental input for the calculation of bound-state properties within the CI framework.
\section{Pion electromagnetic form factor}
\label{ffpice}

We now evaluate the pion electromagnetic form factor in the chiral limit, detailing the calculation of form factors within the present contact-interaction framework. Consider an incoming pion with momentum $p_1$ that absorbs a spacelike photon carrying momentum $q$, such that the outgoing pion has momentum $p_2 = p_1 + q$. Working in the Breit frame, we define $p_1 = K - q/2$. In the chiral limit, this implies
\begin{eqnarray}
p_1^2 = (K - q/2)^2 = 0 = p_2^2 = (K + q/2)^2 \Rightarrow K \cdot q = 0 , \quad K^2 = -\frac{1}{4} q^2 .
\label{eq:breit_kinematics}
\end{eqnarray}
A convenient representation of Eq.~(\ref{eq:breit_kinematics}) is given by
\begin{eqnarray}
q = (0, 0, Q, 0), \quad K = (0, 0, 0, iQ/2).
\label{eq:euclidean_choice}
\end{eqnarray}
Within the impulse approximation, one obtains
\begin{eqnarray}
2K_\mu F_\pi(Q^2) =
\frac{3}{2\pi^2} \int^\Lambda_t
\mathrm{tr}_D \left[
i\Gamma_\pi(-p_2) S(t+p_2) i\gamma_\mu S(t+p_1) i\Gamma_\pi(p_1) S(t)
\right].
\label{eq:impulse_ff}
\end{eqnarray}
Employing the relation $\sigma_S(s) = M\sigma_V(s)$, one identifies representative contributions such as
\begin{eqnarray}\nn
Q^2 F_\pi(Q^2) =
\frac{3}{2\pi^2} \int^\Lambda_t
4t^2 [Q^2 - 2K\cdot t]
\sigma_V(t^2)\sigma_V((t+p_1)^2)\sigma_V((t+p_2)^2)
E_\pi(-P)E_\pi(P).\\
\label{eq:ff_intermediate}
\end{eqnarray}
Introducing Feynman parameters and performing a shift of the integration variable, Eq.~(\ref{eq:ff_intermediate}) can be rewritten as
\begin{eqnarray}
F_\pi(Q^2) =
\frac{3}{2\pi^2} \int_0^1 dx_1 dx_2\, 2x_1
\int_0^\infty ds\, s
\frac{s(1-\tfrac{1}{4}x_1) - \rho^2(1-\tfrac{x_1}{2})}
{(s+\mu^2)^3}
E_\pi(-P)E_\pi(P),
\label{eq:ff_feynman}
\end{eqnarray}
where
\begin{eqnarray}
\rho^2 = Q^2 x_1^2 x_2(1-x_2), \quad \mu^2 = M^2 + \rho^2 .
\label{eq:def_rho_mu}
\end{eqnarray}
After applying proper-time regularization, Eq.~(\ref{eq:ff_feynman}) becomes
\begin{eqnarray}\nn
F_\pi(Q^2) =
\frac{3}{2\pi^2} \int_0^1 dx_1 dx_2\, 2x_1
\left[
\left(1 - \frac{1}{4}x_1\right)\frac{1}{\mu^2}(C_1 - C_2)
- \frac{\rho^2}{\mu^4}\left(1-\frac{x_1}{2}\right)C_2
\right]
E_\pi(-P)E_\pi(P),\\
\label{eq:ff_regularized}
\end{eqnarray}
with
\begin{eqnarray}
C_2(M^2) = \frac{1}{2} M^4 \left(\frac{d}{dM^2}\right)^2 C(M^2).
\label{eq:def_C2}
\end{eqnarray}
The dominant contribution at low and moderate momentum transfer can be expressed as
\begin{eqnarray}
F_{\pi,E}(Q^2) =
\frac{3}{4\pi^2} \int_0^1 dx\,
\frac{1}{M^2 + Q^2 x(1-x)} C_1(M^2 + Q^2 x(1-x))
\left[E_\pi(P)\right]^2.
\label{eq:ff_E_term}
\end{eqnarray}
Defining
\begin{eqnarray}
\sigma(x) := M^2 + Q^2 x(1-x),
\label{eq:def_sigma}
\end{eqnarray}
Eq.~(\ref{eq:ff_E_term}) can be recast as
\begin{eqnarray}
F_{\pi,E}(Q^2) =
\frac{3}{4\pi^2} \int_0^1 dx\,
\frac{1}{\sigma(x)} C_1(\sigma(x)) \left[E_\pi(P)\right]^2.
\label{eq:ff_E_sigma}
\end{eqnarray}
The interference between scalar and pseudovector components yields
\begin{eqnarray}\nn
F_{\pi,EF}(Q^2) =
-\frac{3}{4\pi^2} \int_0^1 dx\,
\frac{2}{\sigma(x)} C_1(\sigma(x)) E_\pi F_\pi
+ \frac{3}{4\pi^2} \int_0^1 dx_1 dx_2\,
\frac{2x_1^2 Q^2}{\omega^2(x_1,x_2)} C_2(\omega) E_\pi F_\pi,\\
\label{eq:ff_mixed}
\end{eqnarray}
where
\begin{eqnarray}
\omega(x_1,x_2) = M^2 + Q^2 x_1^2 x_2(1-x_2).
\label{eq:def_omega}
\end{eqnarray}
Finally, the purely pseudovector contribution is
\begin{eqnarray}
F_{\pi,F}(Q^2) =
\frac{3}{4\pi^2} \int_0^1 dx_1 dx_2\,
\frac{x_1^2 Q^2}{M^2}
\left[
\frac{1}{\omega} C_1(\omega)
- \frac{\omega + 2M^2}{\omega^2} C_2(\omega)
\right]
\left[F_\pi(P)\right]^2.
\label{eq:ff_F_term}
\end{eqnarray}
Combining Eqs.~(\ref{eq:ff_E_sigma})--(\ref{eq:ff_F_term}), the full electromagnetic form factor is obtained:
\begin{eqnarray}
F_\pi(Q^2) =
F_{\pi,E}(Q^2)
+ F_{\pi,EF}(Q^2)
+ F_{\pi,F}(Q^2).
\label{eq:ff_total}
\end{eqnarray}
By construction, the normalization condition is satisfied:
\begin{eqnarray}
F_\pi(Q^2=0) = 1.
\label{eq:ff_normalization}
\end{eqnarray}
A complete numerical analysis of the pion electromagnetic form factor in the chiral limit can be found in Ref.~\cite{GutierrezGuerrero:2010md}. This study was later extended to the massive pion in Ref.~\cite{Roberts:2011wy}, whose results are presented in Sec.~\ref{ffps}.
Once this calculation is established, the procedure for evaluating form factors of other mesons follows in a similar manner, requiring only the appropriate modifications associated with the quark flavor content and spin structure of each meson.\\\\
\reftitle{References}

\end{document}